\documentclass[reqno,11pt]{amsart}
\usepackage{amsmath, amssymb, amsthm,amsfonts, bm} 
\usepackage[english]{babel}
\usepackage{bbm}
\usepackage{graphicx}
\usepackage{url}
\usepackage{epstopdf}
\usepackage[a4paper,bindingoffset=0.5cm,left=2cm,right=2cm,top=2.5cm,bottom=2cm,footskip=.8cm]{geometry}
\usepackage[pdfencoding=auto]{hyperref}
\usepackage{kbordermatrix}
\usepackage{nicematrix}
\usepackage{amsfonts}
\usepackage{tikz}
\usetikzlibrary{arrows, automata,positioning,calc,shapes,decorations.pathreplacing,decorations.markings,shapes.misc,petri,topaths}
  \usepackage{pgfplots}
  \pgfplotsset{compat=newest}
  \usetikzlibrary{plotmarks}
  \usepackage{grffile}
\newlength\figureheight
  \newlength\figurewidth
\pgfplotsset{%
    tick label style={font=\scriptsize},
    label style={font=\footnotesize},
    legend style={font=\footnotesize},
         every axis plot/.append style={very thick}
}
 
\usepackage{rotating}
\usepackage{amsbsy,enumerate}
\usepackage{graphicx}
\usepackage{ccaption}
\usepackage{comment}
\usepackage{mathrsfs} 
\usepackage{mathtools}
\usepackage{slashbox}
\usepackage{xcolor}
\usepackage{algorithm}
\usepackage{algpseudocode}
\usepackage{subfig}

\makeatletter
\newcommand{\specialcell}[1]{\ifmeasuring@#1\else\omit$\displaystyle#1$\ignorespaces\fi}
\makeatother

\newcommand{\dy}{{{\rm d}}y}
\newcommand{\dx}{{{\rm d}}x}

\newcommand{\vb}{\vspace{3.2mm}}
\renewcommand{\hat}{\widehat}

\newcommand{\TC}{\theta_{\text{crit}}}
\newcommand{\TTC}{\theta_{\text{crit},t}^*}

\allowdisplaybreaks

\usepackage[normalem]{ulem}

\begin{document}
\title[Opinions beyond social influence]{Opinion dynamics beyond social influence}

\author{Benedikt V Meylahn\textsuperscript{$\star$} and Christa Searle\textsuperscript{$\dagger$}}

\begin{abstract}
We present an opinion dynamics model framework discarding two common assumptions in the literature: (a) that there is direct influence between beliefs of neighbouring agents, and (b) that agent belief is static in the absence of social influence. Agents in our framework learn from random experiences which possibly reinforce their belief. Agents determine whether they switch opinions by comparing their belief to a threshold. Subsequently, influence of an alter on an ego is not direct incorporation of the alter's belief into the ego's but by adjusting the ego's decision making criteria. We provide an instance from the framework in which social influence between agents generalises majority rules updating. We conduct a sensitivity analysis as well as a pair of experiments concerning heterogeneous population parameters. We conclude that the framework is capable of producing consensus, polarisation and fragmentation with only assimilative forces between agents which typically, in other models, lead exclusively to consensus.
\vb

\noindent
{\sc AMS Subject Classification (MSC2020).} Primary: 91D30 (Social networks; opinion dynamics); Secondary: 91D15 (Social learning)
\vb

\vb

\noindent
{\sc Affiliations.} 
\textsuperscript{$\star$}Korteweg-de Vries Institute for Mathematics, University of Amsterdam; Amsterdam; The Netherlands ({\it contact}: {\tt\scriptsize  b[dot]v[dot]meylahn[at]uva[dot]nl}).

\noindent
\textsuperscript{$\dagger$}Edinburgh Business School, Heriot-Watt University; Edinburgh, Scotland

\vb

\noindent
{\sc Funding.} This research was supported by the European Union’s Horizon 2020 research and innovation programme under the Marie Skłodowska-Curie grant agreement no. 945045, and by the NWO Gravitation project NETWORKS under grant no. 024.002.003. \includegraphics[height=1em]{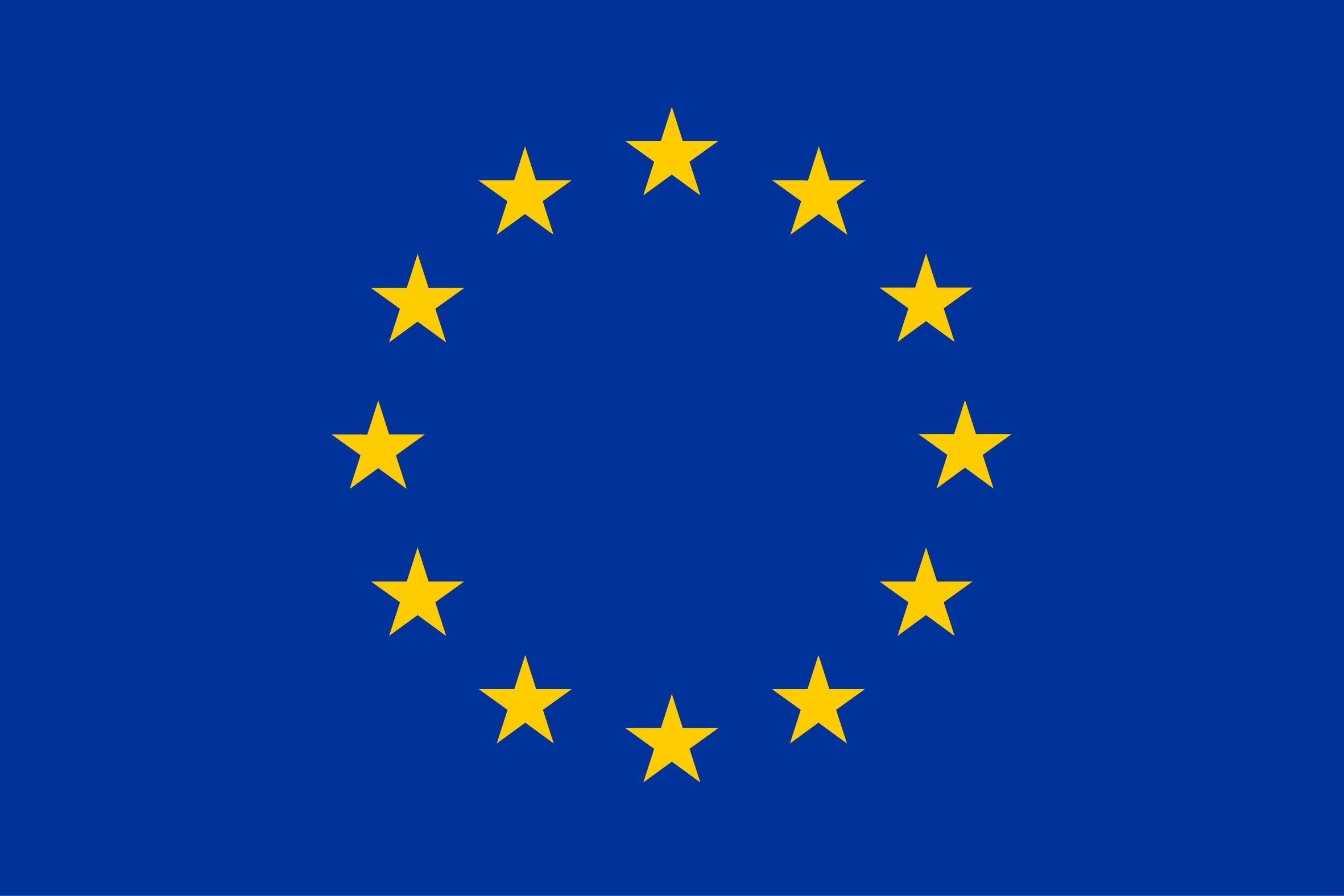} Version: \today.

\vb

\noindent{\sc Declaration of interest:} None.
\end{abstract}

\maketitle
\section{Introduction}
The opinions held by an agent may be of crucial importance to their expressed behaviour. Models that consider opinion formation tend to focus exclusively on social influence mechanisms of opinion change. This leaves agent-based modellers wanting in terms of models they might include as part of a larger agent-based model. The topic of such models may be a system in which opinions influence behaviour of the agents pertaining to other parts of the system. The framework we suggest includes a calculating, rational component in terms of how the agent incorporates information resulting from life's experience, as well as an affective component in terms of the effect of the opinions held by alters on that of the agent. Our model thus allows for opinion change even in the absence of social influence as well as opinion retention in the presence of social influence. This gives an explicit formulation of an agent's internal thought process which we believe should not be governed exclusively by social influence in a somewhat realistic model.

\subsection{Relation to the literature}
The literature in the field of opinion dynamics is expansive which is attested to by the abundance of review papers aiming to capture a moment of the state of the art of the field (see e.g.,\ Castellano \textit{et al.}\,\cite{Castellano2009}, Flache \textit{et al.}\,\cite{Flache2017}, Noorazar \textit{et al.}\,\cite{Noorazar2019}, and Zha \textit{et al.}\,\cite{Zha2021}). As such, an exhaustive review of the literature is beyond the scope of this paper. The discussion that follows focuses on the commonalities between the models in the field and why these may be seen to fall short. Furthermore, we restrict ourselves to literature pertinent to this paper in particular.

There is a stream of literature in which the agents incorporate a (possibly) weighted average of their neighbours' beliefs into their own (see e.g.\ the seminal works of French~\cite{French1956}, Harary~\cite{Harary1959}, and DeGroot~\cite{DeGroot1974}, and more recently Altafani~\cite{Altafini2013}, Proskurnikov \textit{et al.}\,\cite{Proskurnikov2016}, Liu \textit{et al.}\,\cite{Liu2017} and Chan \textit{et al.}\,\cite{Chan2022}). A second stream of literature follows the voter model~\cite{Clifford1973,Holley1975} in which agents directly copy the opinion held by someone in their neighbourhood. Castellano \textit{et al.}\,\cite{CastellanoQ2009} made a significant improvement by means of the $q$-voter model in which instead of copying a random neighbour, agents copy the opinion held by at least $q$ of their neighbours. For an overview the interested reader may consult Redner~\cite{Redner2019}. 

The models within these two streams can further be categorised according to modeling decisions:
\begin{itemize}
    \item Opinion representation being continuous or discrete;
    \item Opinion updating happening simultaneously or asynchronously; 
    \item Forces between neighbouring agents consisting only of attractive forces or including repulsive forces.
\end{itemize}
Despite these differences, the common thread is that agents are initialised with an opinion, the evolution of which is governed only by inter-agent communication. Another similarity of these models is the assumption that an agent's opinion is influenced directly by their neighbours. 

Giardini \textit{et al.}\,\cite{Giardini2015} present a model which does away with the assumption that agent opinions directly influence each other. In their model an agent's opinion is a combination of; subjective truth value, a level of confidence therein and a perceived sharedness. Subsequently these three variables may change (rather than the opinion directly) as result of interactions between agents.

More recently Baccelli \textit{et al.}\,\cite{Bacelli2017} challenge the assumption that agents change their opinions only as consequence of their network. They present a model in which noisy signals between agents represent the possible endogenous evolution of an opinion within an agent. Though the agents have the possibility (by means of noise) to change their opinion without social influence, the influence of another agents' opinion is still directly on their own opinion. 

Flache \textit{et al.}\,\cite{Flache2017} highlight the need for closer inspection of the assumptions underlying social influence and the modelling decisions that take place as a result of these assumptions. Flache \textit{et al.}\,\cite{Flache2017} acknowledge that Giardini \textit{et al.}\,\cite{Giardini2015} make such an effort and call upon researchers in the field to follow suit.  Furthermore, Noorazar \textit{et al.}\,\cite{Noorazar2019} recognize that the work of Baccelli \textit{et al.}\,\cite{Bacelli2017} questions the assumption that opinions should evolve exclusively as a result of social interaction. They mention the need for more models in which agent opinions may evolve outside the confines of social interaction. This evolution may be characteristic of sophisticated agents who have internal thought processes beyond copying their neighbours or behaving as an average of their social connections.

The gap in the existing literature is evident: There is a clear need for a model of opinion dynamics in which simultaneously (a) social influence between agents is not acting directly on the belief of agents and (b) agents have an internal cognitive process by which opinion change may occur beyond the effects of social influence. Such a model would bring researchers one step closer to a semblance of reality. It would also provide a useful tool for the modelling of complex systems which are concerning more than simply the evolution of opinions. We believe that such a model should also be computationally light in order to feasibly be applied to simulation models by practitioners without fear of a large slowdown.

\subsection{Contribution}
 In order to address the gap identified above, we present a framework in which an opinion is modelled as a lens through which experiences are interpreted. This constitutes a random process by which an opinion sometimes successfully aligns with an experience had by an agent and sometimes fails to do so. The agent's opinion then is a choice of lens, hoping for alignment between their opinion and experiences which creates cognitive harmony. In our model this is a cognitive decision making process by which each agent asks themself if the opinion they hold aligns with a sufficient portion of life's experiences. This lends some sophistication to the agents capable of some form of reasoning about the opinion they hold. We model the influence from one agent to another by means of adjusted decision making criteria rather than a direct incorporation of a neighbours opinion into one's own. That is, an agent is inclined to require a lower reliability from an opinion they share with a large portion of their neighbourhood in order to maintain that opinion.

The result is a lightweight framework which may easily be implemented on top of other agent-based simulations. We showcase the framework by means of a model instance. The model instance (and therefore the framework) has desirable properties which we confirm by a sensitivity analysis as well as a pair of experiments concerning heterogeneous population parameters: The framework instance enables polarisation, consensus and fragmentation as steady state outcomes, all without repulsive forces between agents (though these may also be included if desired). The framework features agents with sophistication in their view of the world yet does not require a large computational load.

\subsection{Organisation of paper}
The remainder of the paper is presented in two parts. The first part details the framework: In \S\ref{sec:solo} we describe an opinion dynamics model framework for a single agent which we extend to many agents in \S\ref{sec:many}. We end part one with a summary of the framework in \S\ref{sec:p1_summary}. The second part of the paper entails a framework instance of the framework with a sensitivity analysis and a set of experiments. Specifically, \S\ref{sec:ExModel} deals with the framework instance from the framework. In \S\ref{sec:SensAn} we describe the process and results of a sensitivity analysis of the framework instance. In \S\ref{sec:Experiments} we discuss experiments conducted on the framework instance. We close with a discussion of our work and possible avenues of future research in \S\ref{sec:Discussion}.

\section{Solo agent opinion dynamics model}\label{sec:solo}
For ease of exposition we first present a solo agent opinion dynamics model. We believe that agents should be able to adjust their opinion also in the absence of social influence. This model grew out of, and therefore closely follows, the model of Meylahn \textit{et al.}\,\cite{denBoerWIP} that investigates the problem of trusting institutions as a learning problem. Specifically we present a generalised framework which covers the single agent model of Meylahn \textit{et al.}\,\cite{denBoerWIP} as a special case. We posit that holding an opinion is akin to trusting that this opinion provides a good enough lens through which to interpret experiences and therefore may be modelled to have a reliability.

\subsection{Definition of opinions}
The dynamics evolve over rounds indexed $t=1,2,\ldots$. At the start of each round our agent holds an opinion $a$ from the set of possible opinion $\mathcal{A}$. We refer to the opinion held by the agent in round $t\in\mathbb{N}$ as $a_t\in \mathcal{A}$. The agent is subsequently exposed to an experience which they try to interpret using their opinion. We call the outcome of the interpretation of an experience in round $t$ using opinion $a\in\mathcal{A}$: $X_a^t\in\{0,1\}$. When $X_{a}^t$ takes the value one, the agent's opinion $a$ aligns with the experience in round $t$. Conversely, when $X_a^t$ takes the value zero, the agent's opinion $a$ does not align with the experience in round $t$. Specifically $X_a^t$ for all $a\in\mathcal{A}$ and $t\in\mathbb{N}$ are random variables:
\begin{equation}
    X_a^t = \begin{cases}
        1, \quad&\text{with probability } \theta_a\\
        0, &\text{with probability }1-\theta_a,
    \end{cases} \quad \text{with }\theta_a \in (0,1), \forall a\in\mathcal{A}.
\end{equation}
We call the probability that opinion $a$ aligns with an experience, opinion  $a$'s \textit{reliability}, $\theta_a\in (0,1)$ for all $a\in\mathcal{A}$. Note that interpretations of experiences $\{X_a^t:t\in\mathbb{N}\}$ are i.i.d.\,for each opinion $a\in\mathcal{A}$. The agent only interprets experiences using the opinion they hold. This means they do not observe $X_b^t$ for $b\neq a_t$. Furthermore, they do not know the respective opinion's reliabilities $\theta_a$ for $a\in\mathcal{A}$. We suppose that the agent receives utility $p\in\mathbb{N}$ when an experience aligns with their opinion ($X_a^t = 1$) and loses utility $l\in\mathbb{N}$ when it doesn't ($X_a^t = 0$). This formulation of an experience with an opinion is analogous to the formulation of an interaction between a truster agent and the trustee institution of Meylahn \textit{et al.}\,\cite{denBoerWIP}.

\subsection{Agent belief}
The agent holding opinion $a\in\mathcal{A}$ has a belief on the probability that $[\theta_a=x]$. In the absence of other evidence, the agent uses their prior belief. After any number of experiences, the agent adjusts their belief accordingly.
\subsubsection{Prior belief}
We model the agent to start with a prior belief distribution which is what they believe the density of $\theta_a$ to be for all opinions $a\in\mathcal{A}$ that they have no other information about. The agent has prior belief $B_0(x)$ meaning that they initially believe that the probability density relating to the reliability of opinion $a_{0}$ is:
\begin{equation}
B_0(x):=\mathbb{P}(\theta_{a_{0}} = x) \quad x\in [0,1].
\end{equation}
If the agent switches their opinion at some time $t_0$ to opinion $a_{t_0}\in\mathcal{A}$ then they revert to their prior belief $B_0(x)$ regardless of which opinion they are switching to. This models the agent's forgetting of experiences with an opinion they may have held in the past. This generalises the formulation of belief in Meylahn \textit{et al.}\,\cite{denBoerWIP} by allowing general belief distributions instead of only the Beta-distribution.
\subsubsection{Belief update}

We call the consecutive rounds in which the agent held opinion $a$, a \textit{run} with opinion $a$. We refer to the most recent switching time as $S_t$, which identifies the round in which the current run started:
\begin{equation}
    S_t:=\min\{n:a_{m}=a_{t}, \forall m\in[n,t]\}.
\end{equation}
We model the agent to `forget' previous runs with an opinion. In constructing their belief of an opinion, they only use their most recent run's history which started at time $S_t$. We define the agent's current experience history until the end of round $t\in\mathbb{N}$ as $H_t$. That is the set of experiences observed up until (and including) round $t$ during their current run:
\begin{equation}
    H_t = \{X_{a_{n}}^n : n \in[ S_t,t] \} \quad \text{for }t\in\mathbb{N}.
\end{equation}
This information along with their prior belief is used by the agent to hold a belief at time $t\in\mathbb{N}$:
\begin{equation}
B_0\times H_t\to B_t(x).
\end{equation}
We model the agent to use this belief to attain a point estimate $\hat{\theta}_{t}\in[0,1]$ of the reliability, $\theta_{a_{t}}$. A common example is the mean value of the belief distribution (using the Riemann-Stieltjes integral to allow point mass belief distributions),
\begin{equation}
    \hat{\theta}_{t} =\int_0^1 x \text{d}B_t(x), \quad \in [0,1]. 
\end{equation}
Alternatives to the mean value are the upper or lower confidence intervals and many more. The way in which the history is incorporated is part of the modeller's choice. A logical choice for a Beta-distributed prior belief is Bayesian updating. Alternatively if the prior is a point mass on the estimate, simple exponential smoothing might better serve the task.

Note that each agent has only one belief at any time which is their belief on the opinion they currently hold. This is a consequence of the agent's threshold decision making.

\subsection{Threshold decision making (choosing an opinion)}
The agent is faced with deciding whether to place their trust in the opinion which they held in the previous round. They do so by a satisficing procedure; checking if the current opinion is good enough. In choosing an opinion to hold the agent asks themselves whether they expect positive utility from the opinion they are currently holding. In other words they check the truth of the inequality:
\begin{equation}
    p\hat{\theta}_{t}-l(1-\hat{\theta}_{t})\geq 0.
\end{equation}
This inequality may be rearranged and so equivalently the agent asks themselves whether:
\begin{equation}\label{eq:defTC}
    \hat{\theta}_{t}\geq \frac{l}{p+l} =: \TC\in(0,1).
\end{equation}
Here we have defined $\TC$, the minimum reliability point estimate the agent requires an opinion to have in order to continue holding that opinion. If the agent chooses to switch opinions, they choose a new one from the set of opinions excluding the opinion they are switching from. The choice of the agent at time $t\in\mathbb{N}$ can now be defined:
\begin{equation}
    a_{t} = \begin{cases}
    a_{t-1},\quad &\text{if }\hat{\theta}_{t}\geq \TC,\\
    b \in \mathcal{A}\setminus a_{t-1}, &\text{otherwise}.
    \end{cases}
\end{equation}
 This generalises the decision making in the single agent model presented by Meylahn \textit{et al.}\,\cite{denBoerWIP} from trusting or not trusting to holding one of the opinions in $\mathcal{A}$. The protocol used to choose which of the alternative opinions the agent chooses is up to the modeller. For a set $\mathcal{A}$ of only two opinions the choice is straightforward; simply the other opinion. If there are three or more opinions the choice could be made uniformly at random.

This is a simple satisficing decision making model in which the agent is not concerned with comparisons between options but has a desired level of reliability and retains the opinion they are holding if  they believe it to satisfy this level.

\section{Many agent opinion dynamics framework}\label{sec:many}
In this section we extend the solo agent framework by placing $N$ solo agents into a network. The reason for delaying this exposition is clarity. There is interdependence of the actions taken by agents and the actions taken by their neighbours. By first introducing a solo agent model which contains the basic elements of the many agent model, all that remains is for us to describe how these are influenced by the interplay between agents. The agents in the network communicate their opinion with their neighbours which is how we induce social influence between them. The crucial difference between this framework and other opinion dynamics models is that the effect of the inter-agent communication is not on the agents' belief distributions but rather on their decision making threshold $\TC$.

\subsection{A network of solo agents}
The framework does not prescribe a population structure but assumes that one is given. That is, either the modeller uses a network empirically sourced or makes use of an appropriate random graph model to generate a network. The interested reader may consult the work of Robbins \textit{et al.}\,\cite{Robins2001} relating to models of network generation. We suppose that there exists a population of agents of finite size $N\in\mathbb{N}$. The population is embedded in a network $G=(V,E)$, with $|V| = N$ vertices representing agents and a set of social ties represented by the edges, $E$. Importantly for the framework we do not place restrictions on the form that the network may take. Edges may be directed or bidirectional according to the need within the application. 

We define the opinion held by agent $j\in V$ in round $t$ as $a_{t}(j)\in\mathcal{A}$. In each round every agent holds an opinion, has an experience which corroborates or contradicts their opinion and observes the opinions held by their neighbours. The experience had by agent $j\in V$, holding opinion $a\in\mathcal{A}$ at time $t\in \mathbb{N}$ is the random variable:
\begin{equation}
    X_a^t(j) = \begin{cases}
        1\quad &\text{with probability }\theta_a,\\
        0&\text{with probability } 1-\theta_a,
    \end{cases}\quad\text{with }\theta_a\in (0,1), \forall a\in\mathcal{A}.
\end{equation}
Similarly to the solo agent model, agent $j\in V$  receives utility $p_j\in\mathbb{N}$ when an experience agrees with their opinion and loses utility $l_j\in\mathbb{N}$ when an experience disagrees with their opinion.

Now each agent $j\in V$ is equipped with a prior belief relating to the reliability of opinion $a_{0}(j)$ as before:
\begin{equation}
 B_0^j(x):=\mathbb{P}(\theta_{a_{t}(j)} = x)\quad x\in [0,1].
\end{equation}
Furthermore, agents may be switching between opinions and so we refer to agent $j$'s most recent switching time:
\begin{equation}
    S_t(j):=\min\{n:a_{m}(j)=a_{t}(j), \forall m\in[n,t]\}, \text{for }j\in V \text{ and }t\in\mathbb{N}.
\end{equation}
Subsequently, agent $j$'s current experience history until the end of round $t\in\mathbb{N}$ is $H_t(j)$. That is the set of experiences observed up until (and including) round $t$ during their current run:
\begin{equation}
    H_t(j) = \{X_{a_{n}(j)}^n(j) : n \in[ S_t(j),t] \} \quad \text{for }j\in V,\text{and }t\in\mathbb{N}.
\end{equation}
This information along with their prior belief is used by the agent to hold a belief at time $t\in\mathbb{N}$:
\begin{equation}
B_0^j\times H_t(j)\to B_t^j(x).
\end{equation}
This belief, in turn gives an updated estimate $\hat{\theta}_{t,j}$ (assuming use of the mean value),
\begin{equation}
    \hat{\theta}_{t}(j)= \int_0^1\theta B_t^j(x)\dx.
\end{equation}

\subsection{Social influence}\label{sec:Social_influence}
An agent $u\in V$ influences agent $j\in V$ if there is an edge $(u,j)\in E$. Each agent $j\in V$ has a set of social ties which we call the neighbourhood of agent $j$:
\begin{equation}
N(j):= \left\{u: (u,j)\in E\right\},
\end{equation}
that is the set of agents who are said to influence agent $v$. For ease of notation we assume that the edges between agents are unweighted. Agent $j\in V$ observes the opinions held by the agents in their neighbourhood $N(j)$. This provides agent $j\in V$ with their network influence set for time $t\in\mathbb{N}$:
\begin{equation}
    I_t(j) := \{a_{t}(k): k\in N(j)\}.
\end{equation}

 In formulating the influence of $N(j)$ on agent $j\in V$ we aim to follow empirical literature in marketing which shows that peer-to-peer influence has its effect not in cognitive elements of belief but rather affective elements influencing decision making~\cite{Johnson2005,Ozdemir2020}. In doing so, we draw a connection between the actions of brand loyalty and the expression of opinions. The affective elements may of course still influence decision making, but the channel they follow does not effect the rational calculating elements associated with decision making. Instead the agent's threshold is adjusted based on the information gained from their network. Consider an agent's thoughts: `If it is good enough for my neighbours, why should it not be good enough for me?'

Define $\TTC(j)$ as the \textit{network adjusted critical reliability} of agent $j\in V$ at time $t\in\mathbb{N}$:
\begin{equation}
    \TTC(j): p_j\times l_j\times I_t(j) \mapsto [0,1].
\end{equation}
The ingredients this function may use are thus contained in $p_j$, $l_j$ and $I_t(j)$. We define the number of agents in $j$'s neighbourhood expressing the same opinion as agent $j\in V$ at time $t\in\mathbb{N}$ as:
\begin{equation}\label{eq:net_Agree}
    m_t(j) := |\{b: \left(b\in I_t(j)\right) \land \left(b=a_{t}(j)\right)\}|,
\end{equation}
\textit{i.e.}\,the number of agents in agent $j$'s neighbourhood \textit{matching} agent $j$'s opinion. Similarly we define the number of agents in agent $j$'s neighbourhood \textit{not matching} agent $j$'s opinion as:
\begin{equation}\label{eq:net_Disagree}
    n_t(j):= |N(j)|-m_t(j).
\end{equation}
Note that if it is desirable to have weighted connections between agents, the above may be replaced with total weight within agent $j$'s neighbourhood in agreement with agent $j$'s opinion and the total weight remaining respectively.
The functions $\TTC(i)$ can be chosen in various ways. We suggest the following properties of the network influence function regarding the neighbourhood of an agent holding an opinion:
\begin{itemize}
    \item If there is equal support and opposition ($m_t(j)=n_t(j)$), the effect is null, $\TTC(j)=\TC(j)$.
    \item If there is more support than opposition ($m_t(j)>n_t(j)$), then the threshold is lowered, $\TTC(j)<\TC(j)$.
    \item If there is more opposition than support ($m_t(j)<n_t(j)$), then the threshold is increased, $\TTC(j)>\TC(j)$. 
\end{itemize}
The belief updating of the agents thus has not changed, yet their decision making is affected by the communication of their neighbours. This means that $\hat{\theta}_{t}(j)$ is still constructed in a way which is akin to isolation and instead of comparing this to a constant threshold $\TC(j)$, agent $j\in V$ uses the truth of the inequality $\hat{\theta}_{t}(j)>\TTC(j)$ to determine whether they keep holding their opinion. Note that there are two possible triggers for an agent to switch their opinion. Their estimate may drop below their critical reliability as a result of one too many experiences which did not align with their opinion. Alternatively, a change in the opinions held by their neighbours may increase their critical reliability above their current estimate. 

This formulation of social influence is in stark contrast with the mechanism of social influence in the dual agent model presented by Meylahn \textit{et al.}\,\cite{denBoerWIP}. Where in their study, agents use the observation of their neighbours' action rationally to adjust their belief, our agents simply use the heuristic of changing the threshold which they use for the decision making. This saves a lot of computation making the model tractable for more than two agents. As mentioned before, the fact that the agent communication does not change beliefs but thresholds also follows empirical literature on social influence which distinguishes between cognitive and affective components of decision making~\cite{Johnson2005,Ozdemir2020} thereby taking a step closer to reality.

\subsection{Summary of framework}\label{sec:p1_summary}
\begin{figure}[htbp]
\centering
    \includegraphics{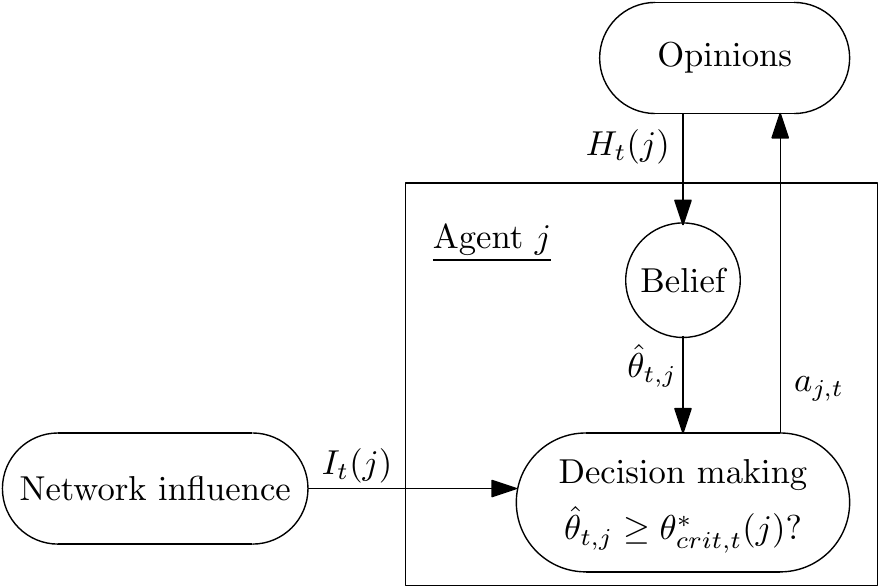}
    \caption{A graphical illustration of the opinion dynamics framework proposed.}
    \label{fig:frameW}
\end{figure}

Agents choose an opinion to hold in each round $a_{t}(j)\in\mathcal{A}$. Subsequently they experience agreement or disagreement. The effect of this experience is an updated belief $B_t^j$ based on $H_t(j)$. In order to choose whether to switch opinion in the following round, they compare a point estimate from their belief, $\hat{\theta}_{t}(j)$ with a critical reliability $\TTC(j)$. This critical reliability is adjusted according to the opinions expressed by the agent's neighbours captured in $I_t(j)$. Their own opinion expression is also how the agent influences their neighbours. Figure~\ref{fig:frameW} serves to illustrate how the elements of the framework fit together. 

\section{Framework instance}\label{sec:ExModel}
 We consider an opinion set of two opinions $\mathcal{A}:=\{0,1\}$. The opinions $0$ and $1$ have reliability $\theta_0, \theta_1\in(0,1)$ respectively. The agents in our framework instance are embedded in a Watts-Strogatz~\cite{Watts1998} generated network. The workings of the Watts-Strogatz model are illustrated in the Appendix.
\subsection{Belief in the opinion}
 The agents in our model have a prior belief distribution in the form of a Beta-distribution with shape parameters $\alpha,\beta\in \mathbb{N}$,
\begin{equation}
    B_0(x) := \frac{x^{\alpha-1} (1-x)^{\beta-1}}{\int_0^1 y^{\alpha-1}(1-y)^{\beta-1}\dy}.
\end{equation}
At time $t\in\mathbb{N}$, each agent $j\in V$ is only aware of the most recent history $H_t(j)$, which pertains to the opinion they currently hold. As such the agent keeps track of the number of \textit{confirming} experiences during their most recent history $H_t(j)$ up until time $t\in \mathbb{N}$ using:
\begin{equation}
    c_t(j) = \sum _{n=S_t(j)}^t X_n.
\end{equation}
Similarly they use 
\begin{equation}
    r_t(j)=\sum _{n=S_t(j)}^t (1-X_n),
\end{equation}
to keep track of the number of experiences during their most recent history $H_t(j)$ up until time $t\in \mathbb{N}$ which \textit{refute} their current opinion. Furthermore, the agents make use of Bayesian belief updating to keep:
\begin{equation}
    B_t^j(x) := \mathbb{P}\left(\theta_{a_{t}(j)}=x\mid H_t(j)\right), \quad \text{for }t\in \mathbb{N}.
\end{equation}
With the convention that if $H_t(j) = \emptyset $, then $\mathbb{P}(\theta_{a_{t}(j)}=x) = B_0^t(x)$.
The agent  $j$ thus holds a belief $B_{t}^j(\theta)$ at time $t\in\mathbb{N}$ given by,
\begin{equation}
    B_{t}^j(x)=\frac{x^{c_t(j)}(1-x)^{r_t(j)}B_0^j(x)}{\int_0^1 y^{c_t(j)}(1-y)^{r_t(j)}B_0^j(y) \dy}, \quad  x\in[0,1].
\end{equation}
As point estimate the agents use the mean of their belief distribution at time $t\in\mathbb{N}$. Conveniently, for the combination of Beta-distributed prior belief and Bayesian updating the mean of the belief distribution is given by,
\begin{equation}
    \hat{\theta}_{t}(j) =\frac{\alpha + c_t(j)}{\alpha+\beta+c_t(j)+r_t(j)},\quad \forall t\in\mathbb{N}, \text{ and all }j\in V.
\end{equation}

\subsection{Threshold and network influence}
In equations (\ref{eq:net_Agree}) and (\ref{eq:net_Disagree}) in \S\ref{sec:Social_influence} we have defined the number of agents in agent $j$'s neighbourhood in agreement with agent $j\in V$ at time $t\in\mathbb{N}$ as $m_t(j)$ and the number of agents in disagreement as $n_t(j)$ respectively. We define the network influence function $\TTC(j)$ representing the influence of agent $j\in V$'s neighbourhood on their decision making as a mapping $\TTC(j):p_j\times l_j\times I_t(j)\mapsto [0,1]$ by,
\begin{equation}
    \TTC (j) = \frac{p_j +f_j(I_t(j))}{p_j+l_j}, \quad \text{for }t\in\mathbb{N},
\end{equation}
with the crucial element $f_j(I_t(j))$ defined:
\begin{equation}\label{eq:netF}
    f_j(I_t(j)):=\frac{n_t(j)-m_t(j)}{\kappa_j |N(j)|},\quad \text{for }t\in\mathbb{N}, j\in V,
\end{equation}
where $\kappa_j\in(0,\infty)$ is agent $j$'s \textit{stubbornness} parameter. A large $\kappa$ represents 
 agents who are not very influenced by their neighbourhood. In fact if $\kappa$ is large enough for an agent, their opinion switching happens independently from their neighbourhood\footnote{Yildiz \textit{et al.}\,\cite{Yildiz2013} studied an extension of the voter model including stubborn agents who were unable to adjust their opinion. We consider our stubbornness parameter a generalisation as an agent might be anywhere between the two extremes: completely unaffected by their neighbourhood, or completely susceptible to the majority opinion in their neighbourhood.}. It should be noted that even the most stubborn agents in our model may change their opinion based on their individual belief updating. It is worth mentioning that if $\kappa$ is small enough for all agents then the model reduces to one in which agents adopt the opinion being held by the majority of their neighbourhood\footnote{Majority rules models have received attention in their own right by~\cite{Mossel2013,Tamuz2015,Benjamini2016} and more recently~\cite{Nguyen2020}.}.

We use a natural starting point for utility parameters. By setting $l_j=p_j=1$ for all $j\in V$ we have that every agent's starting threshold is $\TC = 1/2$. This means that an agent in the absence of a network influence will continue to hold their current opinion if their belief satisfies: $\hat{\theta}_{t,j}>0.5$. 
The resulting decision making is made by checking the inequality of $\hat{\theta}_{t}(j)>\TTC(j)$:
\begin{equation}\label{eq:net}
    \frac{\alpha + c_t(j)}{\alpha+\beta +c_t(j)+r_t(j)}\geq \left({1+\frac{n_t(j) - m_t(j)}{\kappa_j|N(j)|}}\right)/{2}.
\end{equation}

\begin{figure}[ht]
    \centering
    \includegraphics{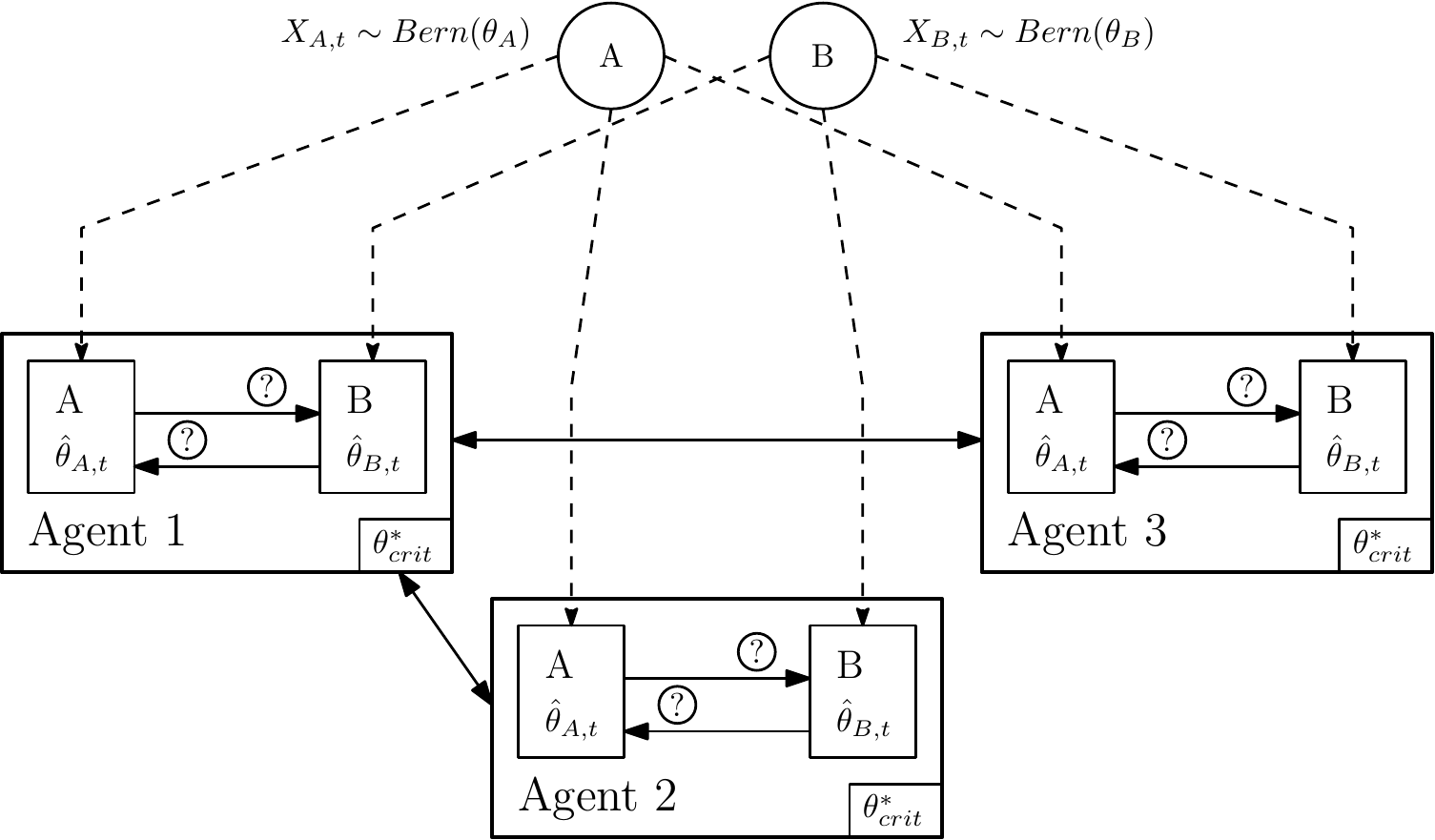}
    \caption{A graphical illustration of an example network agent model.}
    \label{fig:nets}
\end{figure}

A graphical representation of an example of this extension is presented in Figure~\ref{fig:nets}. This example has three agents $V=\{1,2,3\}$ with connections $E=\{(1,2),(2,1),(1,3),(3,1)\}$. The question mark icons represent a check of the inequality (\ref{eq:net}). A transition from trusting one opinion to trusting the other should take place whenever this inequality is not true.

\subsection{Observable outcomes of a simulation}
In the model we have presented, agents are able to converge onto an opinion.
For an agent $j\in V$ holding opinion $a\in \mathcal{A}$, with enough time (and thus sampling of experiences) the agent's estimated reliability may tend towards the true reliability, $\hat{\theta}_{t}(j)\to \theta_a$ and if $\theta_a>\TTC(j)$ it is possible that the agent holds this opinion indefinitely. If at some time $t_0\in\mathbb{N}$ this is true for all the agents in the network, the process has reached a steady state, as anymore switching of opinions becomes more unlikely. For the purposes of the simulation, we use a proxy for steady state: 100 simulated rounds in which no agent switches their opinion. We are interested in whether there is consensus in such a steady state or if there is some level of discordance. We define the probability of consensus as the probability of each agent holding the same opinion at a steady state time $t_0\in\mathbb{N}$:
\begin{equation}
    C:=\mathbb{P}\left( a_{t_0}(j) = a_{t_0}(1), \forall j \in V \right).
\end{equation}
The choice of reference to the first agent's opinion $a_{t_0}(1)$ is arbitrary as all agents are required to be in agreement.

We define the proportion of discordance as the number of discordant edges divided by the total number of edges. A discordant edge $(u,v)$ is one in which the opinions of the agents are different: $a_u\neq a_v$. We label the set containing the discordant edges at time $t\in\mathbb{N}$, $E_D(t)$:
\begin{equation}
E_D(t):= \{(u,v)\mid a_{t}(u)\neq a_{t}(v),\text{with } (u,v)\in E\}.
\end{equation}
As such we can define the asymptotic proportion of discordance as:
\begin{equation}
    D:= \mathbb{E}\left[\liminf_{t\to\infty}\frac{|E_D(t)|}{|E|}\right].
\end{equation}
Considering that we are performing a simulation study we can only approximate the quantities of interest with empirical measures at the end of simulation runs. To this end we consider the empirical value:
\begin{equation}
\hat{C} = \frac{z_c}{Z_{\text{sim}}},
\end{equation}
where $z_c$ is the number of simulation runs in which $a_{t_0}(j) = a_{t_0}(1), \forall j \in V$ at termination time $t_0$, and $Z_{\text{sim}}$ is the total number of simulations that were run. Similarly, for the proportion of discordance:
\begin{equation}
    \hat{D} = \frac{\sum_{n=1}^{Z_{\text{sim}}}|E_D(t_0, n)|}{Z_{\text{sim}}\times|E|},
\end{equation}
where $E_D(t_0, n)$ is the set of discordant edges at the time when the $n$-th simulation iteration terminates $t=t_0$.

\section{Sensitivity analysis}\label{sec:SensAn}
We are interested in the effect of the model parameters on the probability of consensus and the level of discordance. To study this we conduct a sensitivity analysis of which we now describe the setup and the results. The base network consists of $N=20$ agents in the Watts-Strogatz model~\cite{Watts1998} with $l=4$ nearest neighbours and a rewiring probability of $w=0.20$. Note that we generate a new network for each simulation run. All the agents in the population hold a uniform prior belief distribution which is a Beta-distribution with shape parameters $\alpha = 4$, and $\beta = 2$. We reiterate that whenever an agent switches their opinion they restart their learning process from their prior belief. The value of the stubbornness is varied along with the parameters being tested for the sensitivity though is kept homogeneous between agents, $\kappa_j=\kappa$, $\forall j\in V$. We take $\kappa = 0.5$ to $\kappa = 5$ in increments of $0.3$. In general the trend we observe is that as $\kappa$ increases the probability of consensus decreases and the level of discordance increases. 

The opinions in the sensitivity analysis have identical reliability $\theta_0=\theta_1=0.6$. Keeping these the same allows us to focus on the agent dynamics rather than questions of convergence to a `better' opinion. The simulation starts with a warm-up phase $t_s = 5$ rounds. In these rounds the agents follow the solo agent model. Thereafter they start communicating with neighbours and take this communication into consideration changing their $\TTC$ accordingly. We run the simulation model $5\,000$ times under each of the parameter settings in order to obtain 95\% confidence intervals for the quantities of interest.

\subsection{Number of agents}
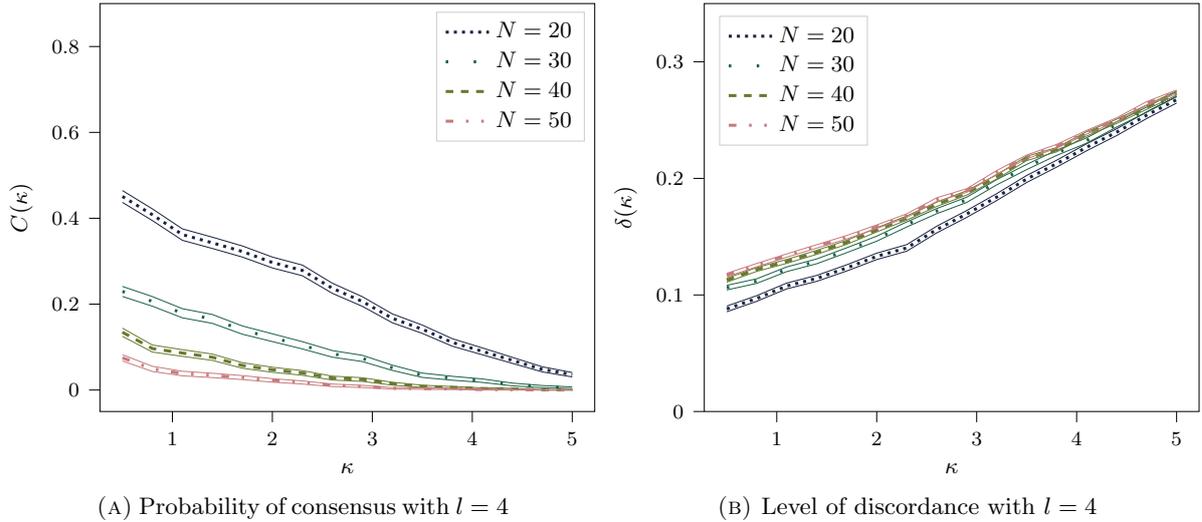
\begin{figure}
\centering
\subfloat[Probability of consensus with $l=4$]{
\begin{tikzpicture}[scale = 0.95]

\definecolor{darkgray176}{RGB}{176,176,176}
\definecolor{darkslategray279768}{RGB}{27,97,68}
\definecolor{lightgray204}{RGB}{204,204,204}
\definecolor{midnightblue263565}{RGB}{26,35,65}
\definecolor{olivedrab10412248}{RGB}{104,122,48}
\definecolor{rosybrown199122124}{RGB}{199,122,124}

\begin{axis}[
legend cell align={left},
legend style={fill opacity=0.8, draw opacity=1, text opacity=1, draw=lightgray204},
tick align=outside,
tick pos=left,
x grid style={darkgray176},
xlabel={\(\displaystyle \kappa\)},
xmin=0.275, xmax=5.225,
xtick style={color=black},
y grid style={darkgray176},
ylabel={\(\displaystyle C(\kappa)\)},
ymin=-0.05, ymax=0.9,
ytick style={color=black}
]
\addplot [line width=0.24pt, midnightblue263565, forget plot]
table {%
0.5 0.437007904052734
0.799999952316284 0.39457631111145
1.10000002384186 0.348679065704346
1.39999997615814 0.329643487930298
1.70000004768372 0.309642314910889
2 0.284334421157837
2.29999995231628 0.266173481941223
2.59999990463257 0.225212931632996
2.90000009536743 0.194985747337341
3.20000004768372 0.156661629676819
3.5 0.131936192512512
3.79999995231628 0.102099537849426
4.69999980926514 0.0422630310058594
5 0.0304640531539917
};
\addplot [very thick, midnightblue263565, dotted]
table {%
0.5 0.450799942016602
0.799999952316284 0.408200025558472
1.10000002384186 0.361999988555908
1.39999997615814 0.34280002117157
1.70000004768372 0.32260000705719
2 0.296999931335449
2.29999995231628 0.278599977493286
2.59999990463257 0.236999988555908
2.90000009536743 0.206200003623962
3.20000004768372 0.16700005531311
3.5 0.141600012779236
3.79999995231628 0.11080002784729
4.09999990463257 0.0901999473571777
4.69999980926514 0.0482000112533569
5 0.035599946975708
};
\addlegendentry{$N = 20$}
\addplot [line width=0.24pt, midnightblue263565, forget plot]
table {%
0.5 0.464591979980469
0.799999952316284 0.421823740005493
1.10000002384186 0.375320911407471
1.39999997615814 0.355956554412842
1.70000004768372 0.335557699203491
2 0.309665679931641
2.29999995231628 0.291026592254639
2.59999990463257 0.24878716468811
2.90000009536743 0.217414259910583
3.20000004768372 0.177338361740112
3.5 0.151263833045959
3.79999995231628 0.119500398635864
4.09999990463257 0.0981404781341553
4.69999980926514 0.0541369915008545
5 0.0407359600067139
};
\addplot [line width=0.24pt, darkslategray279768, forget plot]
table {%
0.5 0.217942237854004
0.799999952316284 0.195181727409363
1.10000002384186 0.168178677558899
1.39999997615814 0.155296444892883
1.70000004768372 0.129993557929993
2.29999995231628 0.0957313776016235
2.59999990463257 0.0765029191970825
2.90000009536743 0.0663621425628662
3.20000004768372 0.0456569194793701
3.5 0.0293481349945068
3.79999995231628 0.0228750705718994
4.09999990463257 0.0177522897720337
4.40000009536743 0.00986015796661377
4.69999980926514 0.0055307149887085
5 0.00320637226104736
};
\addplot [very thick, darkslategray279768, dash pattern=on 1pt off 10pt]
table {%
0.5 0.229599952697754
0.799999952316284 0.206400036811829
1.10000002384186 0.178799986839294
1.39999997615814 0.165600061416626
1.70000004768372 0.139600038528442
2.29999995231628 0.104200005531311
2.59999990463257 0.0842000246047974
2.90000009536743 0.0736000537872314
3.20000004768372 0.051800012588501
3.5 0.0343999862670898
3.79999995231628 0.027400016784668
4.09999990463257 0.0218000411987305
4.40000009536743 0.0130000114440918
4.69999980926514 0.00800001621246338
5 0.00520002841949463
};
\addlegendentry{$N = 30$}
\addplot [line width=0.24pt, darkslategray279768, forget plot]
table {%
0.5 0.241257786750793
0.799999952316284 0.217618227005005
1.10000002384186 0.18942129611969
1.39999997615814 0.175903558731079
1.70000004768372 0.149206519126892
2.29999995231628 0.112668514251709
2.59999990463257 0.0918971300125122
2.90000009536743 0.0808378458023071
3.20000004768372 0.0579431056976318
3.5 0.0394518375396729
3.79999995231628 0.0319249629974365
4.09999990463257 0.0258477926254272
4.40000009536743 0.0161397457122803
4.69999980926514 0.0104693174362183
5 0.00719356536865234
};
\addplot [line width=0.24pt, olivedrab10412248, forget plot]
table {%
0.5 0.125139713287354
0.799999952316284 0.0884115695953369
1.10000002384186 0.0784205198287964
1.39999997615814 0.0690369606018066
1.70000004768372 0.0507631301879883
2 0.041321873664856
2.29999995231628 0.0343812704086304
2.59999990463257 0.0232430696487427
2.90000009536743 0.0201233625411987
3.20000004768372 0.0114529132843018
3.5 0.00604057312011719
4.09999990463257 0.00178647041320801
4.40000009536743 0.00118851661682129
4.69999980926514 -7.87973403930664e-05
5 -0.000191926956176758
};
\addplot [very thick, olivedrab10412248, dashed]
table {%
0.5 0.134600043296814
0.799999952316284 0.0966000556945801
1.10000002384186 0.0861999988555908
1.39999997615814 0.0764000415802002
1.70000004768372 0.0571999549865723
2 0.0471999645233154
2.29999995231628 0.0398000478744507
2.59999990463257 0.0277999639511108
2.90000009536743 0.024399995803833
3.20000004768372 0.014799952507019
3.5 0.00859999656677246
4.09999990463257 0.00339996814727783
4.40000009536743 0.00259995460510254
4.69999980926514 0.000599980354309082
5 0.000200033187866211
};
\addlegendentry{$N = 40$}
\addplot [line width=0.24pt, olivedrab10412248, forget plot]
table {%
0.5 0.144060254096985
0.799999952316284 0.104788422584534
1.10000002384186 0.0939794778823853
1.39999997615814 0.0837631225585938
1.70000004768372 0.0636368989944458
2 0.0530781745910645
2.29999995231628 0.0452187061309814
2.59999990463257 0.0323569774627686
2.90000009536743 0.0286766290664673
3.20000004768372 0.0181471109390259
3.5 0.0111594200134277
3.79999995231628 0.00837576389312744
4.09999990463257 0.00501346588134766
4.40000009536743 0.00401151180267334
4.69999980926514 0.00127875804901123
5 0.00059199333190918
};
\addplot [line width=0.24pt, rosybrown199122124, forget plot]
table {%
0.5 0.0676991939544678
0.799999952316284 0.0432049036026001
1.10000002384186 0.0330735445022583
1.39999997615814 0.0289765596389771
1.70000004768372 0.0243486166000366
2 0.0182981491088867
2.29999995231628 0.0139552354812622
2.59999990463257 0.00810885429382324
2.90000009536743 0.00587022304534912
3.20000004768372 0.00225043296813965
3.79999995231628 0.00104367733001709
4.09999990463257 0.000123858451843262
5 0
};
\addplot [very thick, rosybrown199122124, dash pattern=on 3pt off 5pt on 1pt off 5pt on 1pt off 5pt]
table {%
0.5 0.0750000476837158
0.799999952316284 0.0492000579833984
1.10000002384186 0.0384000539779663
1.39999997615814 0.034000039100647
1.70000004768372 0.0290000438690186
2 0.0224000215530396
2.29999995231628 0.0176000595092773
2.59999990463257 0.0110000371932983
2.90000009536743 0.00839996337890625
3.20000004768372 0.00399994850158691
3.79999995231628 0.00240004062652588
4.09999990463257 0.0010000467300415
5 0
};
\addlegendentry{$N = 50$}
\addplot [line width=0.24pt, rosybrown199122124, forget plot]
table {%
0.5 0.0823007822036743
0.799999952316284 0.0551950931549072
1.10000002384186 0.0437264442443848
1.39999997615814 0.0390233993530273
1.70000004768372 0.0336513519287109
2 0.0265017747879028
2.29999995231628 0.0212447643280029
2.59999990463257 0.0138911008834839
2.90000009536743 0.0109297037124634
3.20000004768372 0.00574958324432373
3.79999995231628 0.00375628471374512
4.09999990463257 0.0018761157989502
4.40000009536743 0.00127875804901123
4.69999980926514 0
5 0
};
\end{axis}

\end{tikzpicture}\label{fig:con_l4}}
\subfloat[Level of discordance with $l=4$]{
\begin{tikzpicture}[scale = 0.95]

\definecolor{darkgray176}{RGB}{176,176,176}
\definecolor{darkslategray279768}{RGB}{27,97,68}
\definecolor{lightgray204}{RGB}{204,204,204}
\definecolor{midnightblue263565}{RGB}{26,35,65}
\definecolor{olivedrab10412248}{RGB}{104,122,48}
\definecolor{rosybrown199122124}{RGB}{199,122,124}

\begin{axis}[
legend cell align={left},
legend style={
  fill opacity=0.8,
  draw opacity=1,
  text opacity=1,
  at={(0.03,0.97)},
  anchor=north west,
  draw=lightgray204
},
tick align=outside,
tick pos=left,
x grid style={darkgray176},
xlabel={\(\displaystyle \kappa\)},
xmin=0.275, xmax=5.225,
xtick style={color=black},
y grid style={darkgray176},
ylabel={\(\displaystyle \delta(\kappa)\)},
ymin=0, ymax=0.35,
ytick style={color=black}
]
\addplot [line width=0.24pt, midnightblue263565, forget plot]
table {%
0.5 0.0855966806411743
0.799999952316284 0.0945550203323364
1.10000002384186 0.105224609375
1.39999997615814 0.111667990684509
1.70000004768372 0.12055504322052
2 0.130353808403015
2.29999995231628 0.137535572052002
2.59999990463257 0.153218507766724
2.90000009536743 0.166508436203003
3.20000004768372 0.18100118637085
3.5 0.196465730667114
3.79999995231628 0.209680676460266
4.09999990463257 0.223692536354065
4.40000009536743 0.236708760261536
4.69999980926514 0.251335620880127
5 0.264644742012024
};
\addplot [very thick, midnightblue263565, dotted]
table {%
0.5 0.0880199670791626
0.799999952316284 0.0969849824905396
1.10000002384186 0.107715010643005
1.39999997615814 0.114225029945374
1.70000004768372 0.123214960098267
2 0.133100032806396
2.29999995231628 0.140354990959167
2.59999990463257 0.156100034713745
2.90000009536743 0.169450044631958
3.20000004768372 0.1839599609375
3.5 0.199489951133728
3.79999995231628 0.212650060653687
4.09999990463257 0.226690053939819
4.40000009536743 0.2396399974823
4.69999980926514 0.254215002059937
5 0.267509937286377
};
\addlegendentry{$N= 20$}
\addplot [line width=0.24pt, midnightblue263565, forget plot]
table {%
0.5 0.0904433727264404
0.799999952316284 0.0994150638580322
1.10000002384186 0.110205411911011
1.39999997615814 0.116781949996948
1.70000004768372 0.125874996185303
2 0.135846257209778
2.29999995231628 0.143174529075623
2.59999990463257 0.158981561660767
2.90000009536743 0.172391653060913
3.20000004768372 0.18691873550415
3.5 0.202514290809631
3.79999995231628 0.215619325637817
4.09999990463257 0.229687452316284
4.40000009536743 0.242571353912354
4.69999980926514 0.257094383239746
5 0.27037525177002
};
\addplot [line width=0.24pt, darkslategray279768, forget plot]
table {%
0.5 0.104263782501221
0.799999952316284 0.109957456588745
1.10000002384186 0.120035409927368
1.39999997615814 0.126763820648193
1.70000004768372 0.136280179023743
2 0.146132469177246
2.29999995231628 0.158272266387939
2.59999990463257 0.169706225395203
2.90000009536743 0.179336309432983
3.20000004768372 0.194427967071533
3.5 0.207531809806824
3.79999995231628 0.219729661941528
4.09999990463257 0.230224251747131
4.40000009536743 0.244024276733398
4.69999980926514 0.256739020347595
5 0.270044565200806
};
\addplot [very thick, darkslategray279768, dash pattern=on 1pt off 10pt]
table {%
0.5 0.106256723403931
0.799999952316284 0.111916661262512
1.10000002384186 0.122036695480347
1.39999997615814 0.12882661819458
1.70000004768372 0.138363361358643
2 0.148273348808289
2.29999995231628 0.160463333129883
2.59999990463257 0.17192006111145
2.90000009536743 0.181583285331726
3.20000004768372 0.196689963340759
3.5 0.20976996421814
3.79999995231628 0.221969962120056
4.09999990463257 0.232506632804871
4.40000009536743 0.246283292770386
4.69999980926514 0.258946657180786
5 0.272279977798462
};
\addlegendentry{$N= 30$}
\addplot [line width=0.24pt, darkslategray279768, forget plot]
table {%
0.5 0.108249664306641
0.799999952316284 0.113875865936279
1.10000002384186 0.124037861824036
1.39999997615814 0.130889534950256
1.70000004768372 0.140446424484253
2 0.150414228439331
2.29999995231628 0.162654399871826
2.59999990463257 0.174133777618408
2.90000009536743 0.183830261230469
3.20000004768372 0.198952078819275
3.5 0.212008118629456
3.79999995231628 0.224210381507874
4.09999990463257 0.23478901386261
4.40000009536743 0.248542308807373
4.69999980926514 0.261154294013977
5 0.274515390396118
};
\addplot [line width=0.24pt, olivedrab10412248, forget plot]
table {%
0.5 0.110912680625916
0.799999952316284 0.120238304138184
1.10000002384186 0.127100944519043
1.39999997615814 0.134828329086304
1.70000004768372 0.143414378166199
2 0.154046773910522
2.29999995231628 0.164202690124512
2.59999990463257 0.175919532775879
2.90000009536743 0.185676693916321
3.20000004768372 0.199383735656738
3.5 0.215279102325439
3.79999995231628 0.2227623462677
4.09999990463257 0.236336946487427
4.40000009536743 0.247118234634399
5 0.271345138549805
};
\addplot [very thick, olivedrab10412248, dashed]
table {%
0.5 0.112627506256104
0.799999952316284 0.121875047683716
1.10000002384186 0.128785014152527
1.39999997615814 0.136564970016479
1.70000004768372 0.145137548446655
2 0.155807495117188
2.29999995231628 0.165997505187988
2.59999990463257 0.177732467651367
2.90000009536743 0.18754243850708
3.20000004768372 0.201247453689575
3.5 0.217144966125488
3.79999995231628 0.224612474441528
4.09999990463257 0.238197565078735
4.40000009536743 0.248975038528442
5 0.273169994354248
};
\addlegendentry{$N= 40$}
\addplot [line width=0.24pt, olivedrab10412248, forget plot]
table {%
0.5 0.114342331886292
0.799999952316284 0.123511672019958
1.10000002384186 0.130468964576721
1.39999997615814 0.138301730155945
1.70000004768372 0.146860599517822
2 0.157568216323853
2.29999995231628 0.167792320251465
2.59999990463257 0.179545402526855
2.90000009536743 0.189408302307129
3.20000004768372 0.203111171722412
3.5 0.219010829925537
3.79999995231628 0.226462602615356
4.09999990463257 0.240058064460754
4.40000009536743 0.250831842422485
5 0.274994850158691
};
\addplot [line width=0.24pt, rosybrown199122124, forget plot]
table {%
0.5 0.115586400032043
0.799999952316284 0.124169111251831
1.10000002384186 0.132388472557068
1.70000004768372 0.147989630699158
2 0.156962513923645
2.29999995231628 0.166484236717224
2.59999990463257 0.180207133293152
2.90000009536743 0.188012003898621
3.20000004768372 0.203548789024353
3.5 0.216831803321838
3.79999995231628 0.225934624671936
4.09999990463257 0.238007068634033
4.40000009536743 0.248708724975586
4.69999980926514 0.262653589248657
5 0.272580981254578
};
\addplot [very thick, rosybrown199122124, dash pattern=on 3pt off 5pt on 1pt off 5pt on 1pt off 5pt]
table {%
0.5 0.117086052894592
0.799999952316284 0.125594019889832
1.10000002384186 0.133811950683594
1.70000004768372 0.149502038955688
2 0.158506035804749
2.29999995231628 0.168050050735474
2.59999990463257 0.181789994239807
2.90000009536743 0.189587950706482
3.20000004768372 0.205153942108154
3.5 0.218446016311646
3.79999995231628 0.227568030357361
4.09999990463257 0.239599943161011
4.40000009536743 0.250344038009644
4.69999980926514 0.264258027076721
5 0.274204015731812
};
\addlegendentry{$N= 50$}
\addplot [line width=0.24pt, rosybrown199122124, forget plot]
table {%
0.5 0.118585586547852
0.799999952316284 0.127018928527832
1.10000002384186 0.135235548019409
1.70000004768372 0.15101432800293
2 0.160049438476562
2.29999995231628 0.169615745544434
2.59999990463257 0.183372855186462
2.90000009536743 0.191164016723633
3.20000004768372 0.206759214401245
3.5 0.220060110092163
3.79999995231628 0.229201316833496
4.09999990463257 0.241192936897278
4.40000009536743 0.251979351043701
4.69999980926514 0.265862345695496
5 0.275827050209045
};
\end{axis}

\end{tikzpicture}\label{fig:dis_l4}}
\caption{Results of the sensitivity analysis inspecting the number of agents in the system with 4 nearest neighbours.}
\end{figure}

\begin{figure}
\centering
\subfloat[Probability of consensus with $l=6$]{
\begin{tikzpicture}[scale = 0.95]

\definecolor{darkgray176}{RGB}{176,176,176}
\definecolor{darkslategray279768}{RGB}{27,97,68}
\definecolor{lightgray204}{RGB}{204,204,204}
\definecolor{midnightblue263565}{RGB}{26,35,65}
\definecolor{olivedrab10412248}{RGB}{104,122,48}
\definecolor{rosybrown199122124}{RGB}{199,122,124}

\begin{axis}[
legend cell align={left},
legend style={fill opacity=0.8, draw opacity=1, text opacity=1, draw=lightgray204},
tick align=outside,
tick pos=left,
x grid style={darkgray176},
xlabel={\(\displaystyle \kappa\)},
xmin=0.275, xmax=5.225,
xtick style={color=black},
y grid style={darkgray176},
ylabel={\(\displaystyle C(\kappa)\)},
ymin=-0.05, ymax=0.9,
ytick style={color=black}
]
\addplot [line width=0.24pt, midnightblue263565, forget plot]
table {%
0.5 0.704514026641846
0.799999952316284 0.660804986953735
1.10000002384186 0.627300024032593
1.39999997615814 0.62044620513916
1.70000004768372 0.580588817596436
2 0.555072546005249
2.29999995231628 0.484740734100342
2.59999990463257 0.431623578071594
2.90000009536743 0.37111485004425
3.20000004768372 0.312017321586609
3.5 0.235045909881592
3.79999995231628 0.180691123008728
4.09999990463257 0.146526455879211
4.40000009536743 0.0984318256378174
4.69999980926514 0.0696104764938354
5 0.0458457469940186
};
\addplot [very thick, midnightblue263565, dotted]
table {%
0.5 0.717000007629395
0.799999952316284 0.673799991607666
1.10000002384186 0.640599966049194
1.39999997615814 0.633800029754639
1.70000004768372 0.594200015068054
2 0.56879997253418
2.29999995231628 0.498600006103516
2.59999990463257 0.44539999961853
2.90000009536743 0.384600043296814
3.20000004768372 0.325000047683716
3.5 0.246999979019165
3.79999995231628 0.19159996509552
4.09999990463257 0.156599998474121
4.40000009536743 0.10699999332428
4.69999980926514 0.0770000219345093
5 0.0520000457763672
};
\addlegendentry{$N = 20$}
\addplot [line width=0.24pt, midnightblue263565, forget plot]
table {%
0.5 0.729485988616943
0.799999952316284 0.686794996261597
1.10000002384186 0.653900146484375
1.39999997615814 0.647153854370117
1.70000004768372 0.607811093330383
2 0.58252739906311
2.29999995231628 0.512459278106689
2.59999990463257 0.459176421165466
2.90000009536743 0.398085117340088
3.20000004768372 0.337982654571533
3.5 0.258954048156738
3.79999995231628 0.202508926391602
4.09999990463257 0.166673541069031
4.40000009536743 0.115568161010742
4.69999980926514 0.0843895673751831
5 0.0581542253494263
};
\addplot [line width=0.24pt, darkslategray279768, forget plot]
table {%
0.5 0.480141639709473
0.799999952316284 0.436210155487061
1.10000002384186 0.41448438167572
1.39999997615814 0.386221885681152
1.70000004768372 0.350464820861816
2 0.328454494476318
2.29999995231628 0.285124540328979
2.59999990463257 0.22403347492218
2.90000009536743 0.181278109550476
3.20000004768372 0.132907748222351
3.5 0.0918769836425781
4.09999990463257 0.0381274223327637
4.40000009536743 0.0249022245407104
4.69999980926514 0.0148550271987915
5 0.00758802890777588
};
\addplot [very thick, darkslategray279768, dash pattern=on 1pt off 10pt]
table {%
0.5 0.49399995803833
0.799999952316284 0.450000047683716
1.10000002384186 0.428200006484985
1.39999997615814 0.399800062179565
1.70000004768372 0.363800048828125
2 0.341599941253662
2.29999995231628 0.297800064086914
2.59999990463257 0.23580002784729
2.90000009536743 0.192199945449829
3.20000004768372 0.142600059509277
3.5 0.100199937820435
4.09999990463257 0.0437999963760376
4.40000009536743 0.0296000242233276
4.69999980926514 0.0185999870300293
5 0.0104000568389893
};
\addlegendentry{$N = 30$}
\addplot [line width=0.24pt, darkslategray279768, forget plot]
table {%
0.5 0.507858276367188
0.799999952316284 0.463789820671082
1.10000002384186 0.44191563129425
1.39999997615814 0.413378119468689
1.70000004768372 0.377135157585144
2 0.354745388031006
2.29999995231628 0.310475468635559
2.59999990463257 0.24756646156311
2.90000009536743 0.203121900558472
3.20000004768372 0.152292251586914
3.5 0.108523011207581
4.09999990463257 0.0494725704193115
4.40000009536743 0.0342978239059448
4.69999980926514 0.0223449468612671
5 0.0132119655609131
};
\addplot [line width=0.24pt, olivedrab10412248, forget plot]
table {%
0.5 0.339951515197754
0.799999952316284 0.295796394348145
1.10000002384186 0.257891178131104
1.39999997615814 0.247644662857056
1.70000004768372 0.215781688690186
2 0.170718669891357
2.29999995231628 0.15256679058075
2.59999990463257 0.117770433425903
2.90000009536743 0.0820674896240234
3.20000004768372 0.0517104864120483
3.5 0.0321407318115234
3.79999995231628 0.0223234891891479
4.09999990463257 0.0118087530136108
4.40000009536743 0.00353157520294189
4.69999980926514 0.00240743160247803
5 0.000492095947265625
};
\addplot [very thick, olivedrab10412248, dashed]
table {%
0.5 0.35319995880127
0.799999952316284 0.308599948883057
1.10000002384186 0.27020001411438
1.39999997615814 0.259799957275391
1.70000004768372 0.227400064468384
2 0.181400060653687
2.29999995231628 0.162799954414368
2.59999990463257 0.126999974250793
2.90000009536743 0.0900000333786011
3.20000004768372 0.0582000017166138
3.5 0.0374000072479248
3.79999995231628 0.0268000364303589
4.09999990463257 0.0152000188827515
4.40000009536743 0.0055999755859375
4.69999980926514 0.00419998168945312
5 0.00160002708435059
};
\addlegendentry{$N = 40$}
\addplot [line width=0.24pt, olivedrab10412248, forget plot]
table {%
0.5 0.366448521614075
0.799999952316284 0.321403622627258
1.10000002384186 0.282508850097656
1.39999997615814 0.271955251693726
1.70000004768372 0.239018321037292
2 0.192081332206726
2.29999995231628 0.173033237457275
2.59999990463257 0.136229515075684
2.90000009536743 0.0979325771331787
3.20000004768372 0.0646895170211792
3.5 0.0426592826843262
3.79999995231628 0.0312764644622803
4.09999990463257 0.0185912847518921
4.40000009536743 0.00766849517822266
4.69999980926514 0.00599265098571777
5 0.002707839012146
};
\addplot [line width=0.24pt, rosybrown199122124, forget plot]
table {%
0.5 0.234258890151978
0.799999952316284 0.203808784484863
1.10000002384186 0.170132398605347
1.39999997615814 0.146915912628174
1.70000004768372 0.118351817131042
2 0.105770587921143
2.29999995231628 0.0836043357849121
2.59999990463257 0.0593082904815674
3.20000004768372 0.0265660285949707
3.5 0.00915706157684326
3.79999995231628 0.00672519207000732
4.09999990463257 0.00240743160247803
4.69999980926514 0.000240325927734375
5 -0.000154256820678711
};
\addplot [very thick, rosybrown199122124, dash pattern=on 3pt off 5pt on 1pt off 5pt on 1pt off 5pt]
table {%
0.5 0.24619996547699
0.799999952316284 0.215199947357178
1.10000002384186 0.180799961090088
1.39999997615814 0.157000064849854
1.70000004768372 0.127599954605103
2 0.114599943161011
2.29999995231628 0.0915999412536621
2.59999990463257 0.0662000179290771
3.20000004768372 0.0313999652862549
3.5 0.0121999979019165
3.79999995231628 0.00940001010894775
4.09999990463257 0.00419998168945312
4.69999980926514 0.00119996070861816
5 0.000399947166442871
};
\addlegendentry{$N = 50$}
\addplot [line width=0.24pt, rosybrown199122124, forget plot]
table {%
0.5 0.258141040802002
0.799999952316284 0.226591229438782
1.10000002384186 0.191467523574829
1.39999997615814 0.167083978652954
1.70000004768372 0.136848092079163
2 0.123429417610168
2.29999995231628 0.0995956659317017
2.59999990463257 0.0730917453765869
3.20000004768372 0.0362340211868286
3.5 0.0152429342269897
3.79999995231628 0.0120747089385986
4.09999990463257 0.00599265098571777
4.69999980926514 0.00215959548950195
5 0.000954270362854004
};
\end{axis}

\end{tikzpicture}\label{fig:con_l6}}
\subfloat[Level of discordance with $l=6$]{
\begin{tikzpicture}[scale = 0.95]

\definecolor{darkgray176}{RGB}{176,176,176}
\definecolor{darkslategray279768}{RGB}{27,97,68}
\definecolor{lightgray204}{RGB}{204,204,204}
\definecolor{midnightblue263565}{RGB}{26,35,65}
\definecolor{olivedrab10412248}{RGB}{104,122,48}
\definecolor{rosybrown199122124}{RGB}{199,122,124}

\begin{axis}[
legend cell align={left},
legend style={
  fill opacity=0.8,
  draw opacity=1,
  text opacity=1,
  at={(0.03,0.97)},
  anchor=north west,
  draw=lightgray204
},
tick align=outside,
tick pos=left,
x grid style={darkgray176},
xlabel={\(\displaystyle \kappa\)},
xmin=0.275, xmax=5.225,
xtick style={color=black},
y grid style={darkgray176},
ylabel={\(\displaystyle \delta(\kappa)\)},
ymin=0, ymax=0.35,
ytick style={color=black}
]
\addplot [line width=0.24pt, midnightblue263565, forget plot]
table {%
0.5 0.0574203729629517
0.799999952316284 0.0679017305374146
1.10000002384186 0.0768803358078003
1.39999997615814 0.0808984041213989
1.70000004768372 0.0929099321365356
2 0.102151989936829
2.29999995231628 0.123405814170837
2.59999990463257 0.141277074813843
2.90000009536743 0.160162091255188
3.20000004768372 0.177996158599854
3.5 0.204137802124023
3.79999995231628 0.223002433776855
4.09999990463257 0.239891886711121
4.40000009536743 0.2623291015625
4.69999980926514 0.279457688331604
5 0.296033382415771
};
\addplot [very thick, midnightblue263565, dotted]
table {%
0.5 0.0601166486740112
0.799999952316284 0.0707733631134033
1.10000002384186 0.079906702041626
1.39999997615814 0.0840467214584351
1.70000004768372 0.0962433815002441
2 0.10564661026001
2.29999995231628 0.127093315124512
2.59999990463257 0.145126700401306
2.90000009536743 0.164086699485779
3.20000004768372 0.181949973106384
3.5 0.208050012588501
3.79999995231628 0.226840019226074
4.09999990463257 0.243696689605713
4.40000009536743 0.265990018844604
4.69999980926514 0.283003330230713
5 0.299423336982727
};
\addlegendentry{$N= 20$}
\addplot [line width=0.24pt, midnightblue263565, forget plot]
table {%
0.5 0.0628130435943604
0.799999952316284 0.0736449956893921
1.10000002384186 0.0829330682754517
1.39999997615814 0.0871949195861816
1.70000004768372 0.0995767116546631
2 0.10914134979248
2.29999995231628 0.130780816078186
2.59999990463257 0.14897632598877
2.90000009536743 0.16801130771637
3.20000004768372 0.185903787612915
3.5 0.211962223052979
3.79999995231628 0.230677485466003
4.09999990463257 0.247501492500305
4.40000009536743 0.269650936126709
4.69999980926514 0.286548972129822
5 0.302813291549683
};
\addplot [line width=0.24pt, darkslategray279768, forget plot]
table {%
0.5 0.0825035572052002
0.799999952316284 0.0922156572341919
1.10000002384186 0.0990726947784424
1.39999997615814 0.107747673988342
1.70000004768372 0.119629621505737
2 0.128375291824341
2.29999995231628 0.14240825176239
2.59999990463257 0.163695573806763
2.90000009536743 0.182723760604858
3.20000004768372 0.202239871025085
3.5 0.223884463310242
4.09999990463257 0.258884310722351
4.40000009536743 0.275587320327759
5 0.306318640708923
};
\addplot [very thick, darkslategray279768, dash pattern=on 1pt off 10pt]
table {%
0.5 0.0849578380584717
0.799999952316284 0.0947467088699341
1.10000002384186 0.101699948310852
1.39999997615814 0.110464453697205
1.70000004768372 0.122451066970825
2 0.131304502487183
2.29999995231628 0.145402193069458
2.59999990463257 0.166746616363525
2.90000009536743 0.185842275619507
3.20000004768372 0.20533561706543
3.5 0.226873397827148
3.79999995231628 0.244446635246277
4.09999990463257 0.261733293533325
4.40000009536743 0.278351068496704
5 0.308913350105286
};
\addlegendentry{$N= 30$}
\addplot [line width=0.24pt, darkslategray279768, forget plot]
table {%
0.5 0.0874119997024536
0.799999952316284 0.0972776412963867
1.10000002384186 0.104327321052551
1.39999997615814 0.113181233406067
1.70000004768372 0.125272631645203
2 0.134233593940735
2.29999995231628 0.148396253585815
2.59999990463257 0.169797658920288
2.90000009536743 0.188960790634155
3.20000004768372 0.208431243896484
3.5 0.229862213134766
3.79999995231628 0.247435450553894
4.09999990463257 0.264582395553589
4.40000009536743 0.281114816665649
5 0.311508059501648
};
\addplot [line width=0.24pt, olivedrab10412248, forget plot]
table {%
0.5 0.0934960842132568
0.799999952316284 0.104216933250427
1.10000002384186 0.113372206687927
1.39999997615814 0.119863271713257
1.70000004768372 0.13258945941925
2 0.147217512130737
2.29999995231628 0.158140301704407
2.59999990463257 0.178972840309143
2.90000009536743 0.194276571273804
3.20000004768372 0.21379542350769
3.5 0.232650518417358
3.79999995231628 0.248614549636841
4.09999990463257 0.265839099884033
4.40000009536743 0.282196521759033
4.69999980926514 0.296158909797668
5 0.311532258987427
};
\addplot [very thick, olivedrab10412248, dashed]
table {%
0.5 0.0956716537475586
0.799999952316284 0.106456637382507
1.10000002384186 0.115638375282288
1.39999997615814 0.122220039367676
1.70000004768372 0.135031700134277
2 0.149685025215149
2.29999995231628 0.160665035247803
2.59999990463257 0.181581616401672
2.90000009536743 0.196813344955444
3.20000004768372 0.216286659240723
3.5 0.235103368759155
3.79999995231628 0.251073360443115
4.09999990463257 0.268208265304565
4.40000009536743 0.284463405609131
4.69999980926514 0.298316717147827
5 0.313666701316833
};
\addlegendentry{$N= 40$}
\addplot [line width=0.24pt, olivedrab10412248, forget plot]
table {%
0.5 0.0978472232818604
0.799999952316284 0.108696460723877
1.10000002384186 0.117904424667358
1.39999997615814 0.124576687812805
1.70000004768372 0.137473821640015
2 0.152152419090271
2.29999995231628 0.163189649581909
2.59999990463257 0.184190511703491
2.90000009536743 0.199350118637085
3.20000004768372 0.218777894973755
3.5 0.237556219100952
3.79999995231628 0.25353217124939
4.09999990463257 0.270577549934387
4.40000009536743 0.286730051040649
4.69999980926514 0.300474405288696
5 0.31580114364624
};
\addplot [line width=0.24pt, rosybrown199122124, forget plot]
table {%
0.5 0.103131413459778
0.799999952316284 0.112255454063416
1.10000002384186 0.122450113296509
1.39999997615814 0.131719470024109
1.70000004768372 0.143835186958313
2 0.153734683990479
2.29999995231628 0.167783856391907
2.59999990463257 0.183183312416077
2.90000009536743 0.201000213623047
3.20000004768372 0.220195651054382
3.5 0.238153100013733
3.79999995231628 0.255581974983215
4.09999990463257 0.269907116889954
4.40000009536743 0.285382151603699
4.69999980926514 0.301861643791199
5 0.313813805580139
};
\addplot [very thick, rosybrown199122124, dash pattern=on 3pt off 5pt on 1pt off 5pt on 1pt off 5pt]
table {%
0.5 0.105098724365234
0.799999952316284 0.114270687103271
1.10000002384186 0.12449061870575
1.39999997615814 0.133798718452454
1.70000004768372 0.145933389663696
2 0.155913352966309
2.29999995231628 0.170002698898315
2.59999990463257 0.185397386550903
2.90000009536743 0.203238725662231
3.20000004768372 0.222381353378296
3.5 0.240255951881409
3.79999995231628 0.257646679878235
4.09999990463257 0.271894693374634
4.40000009536743 0.287338733673096
4.69999980926514 0.303729295730591
5 0.315666675567627
};
\addlegendentry{$N= 50$}
\addplot [line width=0.24pt, rosybrown199122124, forget plot]
table {%
0.5 0.107065916061401
0.799999952316284 0.116285920143127
1.10000002384186 0.12653112411499
1.39999997615814 0.135877847671509
1.70000004768372 0.14803147315979
2 0.158092021942139
2.29999995231628 0.172221541404724
2.59999990463257 0.18761134147644
2.90000009536743 0.205477118492126
3.20000004768372 0.22456693649292
3.5 0.242358922958374
3.79999995231628 0.259711384773254
4.09999990463257 0.273882150650024
4.40000009536743 0.289295196533203
4.69999980926514 0.305597066879272
5 0.317519545555115
};
\end{axis}

\end{tikzpicture}\label{fig:dis_l6}}
\caption{Results of the sensitivity analysis inspecting the number of agents in the system with 6 nearest neighbours.}
\end{figure}
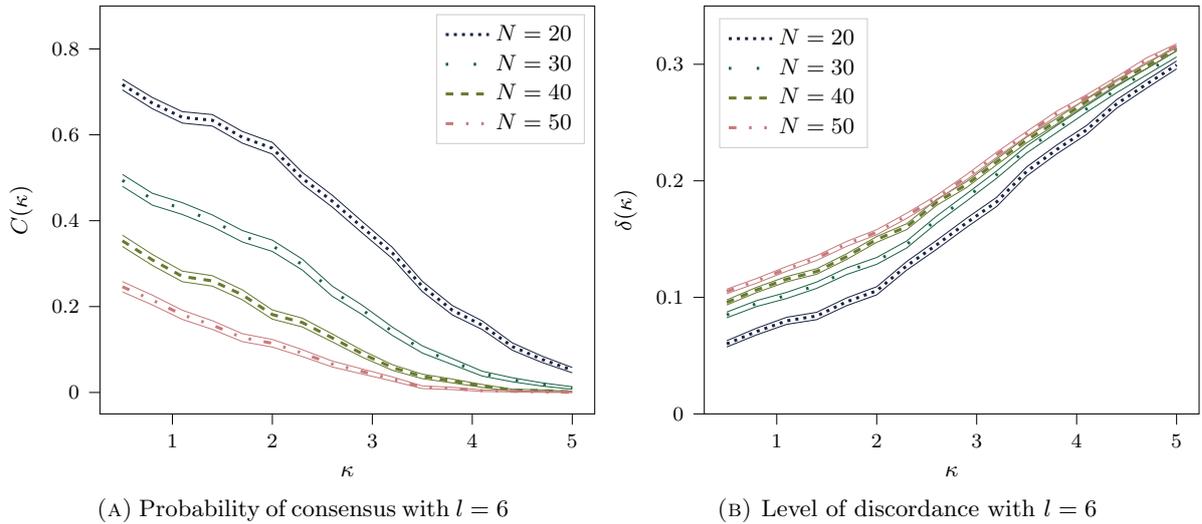

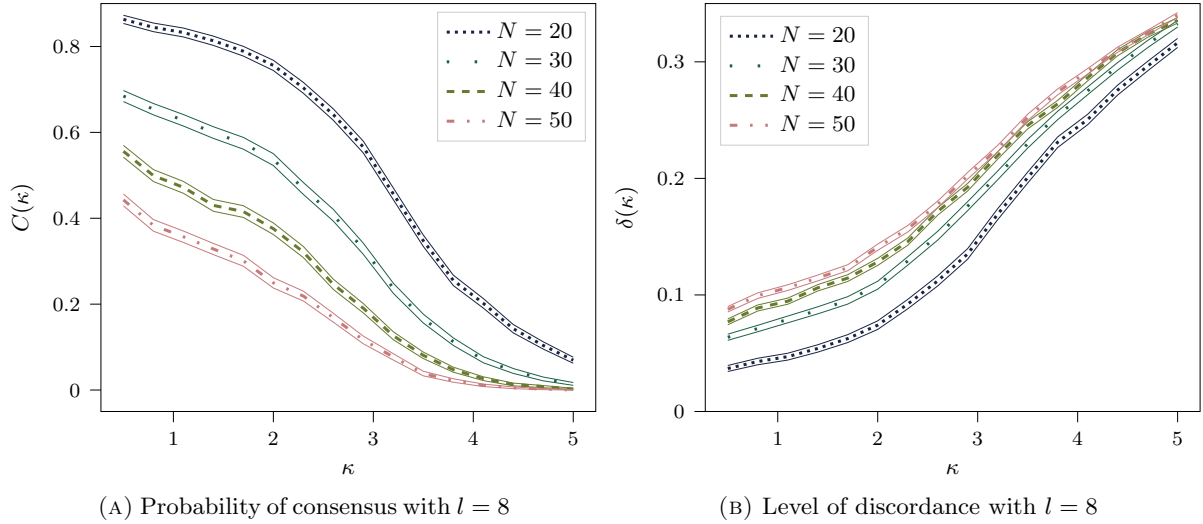
\begin{figure}
\centering
\subfloat[Probability of consensus with $l=8$]{
\begin{tikzpicture}[scale = 0.95]

\definecolor{darkgray176}{RGB}{176,176,176}
\definecolor{darkslategray279768}{RGB}{27,97,68}
\definecolor{lightgray204}{RGB}{204,204,204}
\definecolor{midnightblue263565}{RGB}{26,35,65}
\definecolor{olivedrab10412248}{RGB}{104,122,48}
\definecolor{rosybrown199122124}{RGB}{199,122,124}

\begin{axis}[
legend cell align={left},
legend style={fill opacity=0.8, draw opacity=1, text opacity=1, draw=lightgray204},
tick align=outside,
tick pos=left,
x grid style={darkgray176},
xlabel={\(\displaystyle \kappa\)},
xmin=0.275, xmax=5.225,
xtick style={color=black},
y grid style={darkgray176},
ylabel={\(\displaystyle C(\kappa)\)},
ymin=-0.05, ymax=0.9,
ytick style={color=black}
]
\addplot [line width=0.24pt, midnightblue263565, forget plot]
table {%
0.5 0.85367488861084
0.799999952316284 0.834558010101318
1.10000002384186 0.822251796722412
1.39999997615814 0.802601099014282
1.70000004768372 0.777690291404724
2 0.744095087051392
2.29999995231628 0.691144227981567
2.59999990463257 0.628106474876404
2.90000009536743 0.552061319351196
3.5 0.334400177001953
3.79999995231628 0.243312358856201
4.09999990463257 0.190871238708496
4.40000009536743 0.133102178573608
4.69999980926514 0.0953458547592163
5 0.0623557567596436
};
\addplot [very thick, midnightblue263565, dotted]
table {%
0.5 0.863199949264526
0.799999952316284 0.844599962234497
1.10000002384186 0.832599997520447
1.39999997615814 0.813400030136108
1.70000004768372 0.789000034332275
2 0.75600004196167
2.29999995231628 0.703799962997437
2.59999990463257 0.64139997959137
2.90000009536743 0.565799951553345
3.5 0.347599983215332
3.79999995231628 0.255399942398071
4.09999990463257 0.202000021934509
4.40000009536743 0.142799973487854
4.69999980926514 0.103800058364868
5 0.0693999528884888
};
\addlegendentry{$N = 20$}
\addplot [line width=0.24pt, midnightblue263565, forget plot]
table {%
0.5 0.872725129127502
0.799999952316284 0.854642033576965
1.10000002384186 0.842948198318481
1.39999997615814 0.824198961257935
1.70000004768372 0.800309658050537
2 0.767904996871948
2.29999995231628 0.716455698013306
2.59999990463257 0.654693603515625
2.90000009536743 0.579538822174072
3.20000004768372 0.47060751914978
3.5 0.360799789428711
3.79999995231628 0.267487645149231
4.09999990463257 0.213128805160522
4.40000009536743 0.152497887611389
4.69999980926514 0.11225414276123
5 0.076444149017334
};
\addplot [line width=0.24pt, darkslategray279768, forget plot]
table {%
0.5 0.671719789505005
0.799999952316284 0.640814542770386
1.10000002384186 0.614804029464722
1.39999997615814 0.586018443107605
1.70000004768372 0.560895919799805
2 0.522777915000916
2.29999995231628 0.455566644668579
2.59999990463257 0.393382549285889
2.90000009536743 0.31597638130188
3.20000004768372 0.225212931632996
3.5 0.156076550483704
3.79999995231628 0.103838086128235
4.09999990463257 0.0634996891021729
4.40000009536743 0.038690447807312
4.69999980926514 0.0214055776596069
5 0.010920524597168
};
\addplot [very thick, darkslategray279768, dash pattern=on 1pt off 10pt]
table {%
0.5 0.684599995613098
0.799999952316284 0.654000043869019
1.10000002384186 0.628200054168701
1.39999997615814 0.599600076675415
1.70000004768372 0.574599981307983
2 0.53659999370575
2.29999995231628 0.469399929046631
2.59999990463257 0.407000064849854
2.90000009536743 0.328999996185303
3.20000004768372 0.236999988555908
3.5 0.166399955749512
3.79999995231628 0.112599968910217
4.09999990463257 0.0706000328063965
4.40000009536743 0.0443999767303467
4.69999980926514 0.0257999897003174
5 0.01419997215271
};
\addlegendentry{$N = 30$}
\addplot [line width=0.24pt, darkslategray279768, forget plot]
table {%
0.5 0.697480201721191
0.799999952316284 0.667185544967651
1.10000002384186 0.641595959663391
1.39999997615814 0.613181591033936
1.70000004768372 0.588304162025452
2 0.550422191619873
2.29999995231628 0.483233332633972
2.59999990463257 0.420617461204529
2.90000009536743 0.342023611068726
3.20000004768372 0.24878716468811
3.5 0.176723480224609
3.79999995231628 0.121361970901489
4.09999990463257 0.0777002573013306
4.40000009536743 0.0501095056533813
4.69999980926514 0.0301944017410278
5 0.0174795389175415
};
\addplot [line width=0.24pt, olivedrab10412248, forget plot]
table {%
0.5 0.542227983474731
0.799999952316284 0.485540747642517
1.10000002384186 0.458162426948547
1.39999997615814 0.416277170181274
1.70000004768372 0.401939630508423
2 0.362176537513733
2.29999995231628 0.308454990386963
2.59999990463257 0.234258890151978
2.90000009536743 0.179517269134521
3.20000004768372 0.117189168930054
3.5 0.0732458829879761
3.79999995231628 0.0415099859237671
4.09999990463257 0.0219560861587524
4.40000009536743 0.00986015796661377
5 0.00104367733001709
};
\addplot [very thick, olivedrab10412248, dashed]
table {%
0.5 0.555999994277954
0.799999952316284 0.499400019645691
1.10000002384186 0.472000002861023
1.39999997615814 0.430000066757202
1.70000004768372 0.415600061416626
2 0.375599980354309
2.29999995231628 0.321399927139282
2.59999990463257 0.24619996547699
2.90000009536743 0.190400004386902
3.20000004768372 0.126399993896484
3.5 0.0808000564575195
3.79999995231628 0.0473999977111816
4.09999990463257 0.0263999700546265
4.40000009536743 0.0130000114440918
4.69999980926514 0.00800001621246338
5 0.00240004062652588
};
\addlegendentry{$N = 40$}
\addplot [line width=0.24pt, olivedrab10412248, forget plot]
table {%
0.5 0.569772005081177
0.799999952316284 0.513259291648865
1.10000002384186 0.485837578773499
1.39999997615814 0.44372284412384
1.70000004768372 0.42926037311554
2 0.389023542404175
2.29999995231628 0.334344983100891
2.59999990463257 0.258141040802002
2.90000009536743 0.201282739639282
3.20000004768372 0.135610818862915
3.5 0.0883541107177734
3.79999995231628 0.0532900094985962
4.09999990463257 0.0308438539505005
4.40000009536743 0.0161397457122803
4.69999980926514 0.0104693174362183
5 0.00375628471374512
};
\addplot [line width=0.24pt, rosybrown199122124, forget plot]
table {%
0.5 0.429031729698181
0.799999952316284 0.369922876358032
1.10000002384186 0.343719601631165
1.39999997615814 0.315382480621338
1.70000004768372 0.288285732269287
2 0.237603902816772
2.29999995231628 0.207340240478516
2.59999990463257 0.158027172088623
2.90000009536743 0.107703924179077
3.20000004768372 0.0698016881942749
3.5 0.0330735445022583
3.79999995231628 0.0181161165237427
4.09999990463257 0.00810885429382324
4.40000009536743 0.00320637226104736
4.69999980926514 0.000901341438293457
5 -0.000191926956176758
};
\addplot [very thick, rosybrown199122124, dash pattern=on 3pt off 5pt on 1pt off 5pt on 1pt off 5pt]
table {%
0.5 0.442800045013428
0.799999952316284 0.383399963378906
1.10000002384186 0.35699999332428
1.39999997615814 0.328400015830994
1.70000004768372 0.300999999046326
2 0.249599933624268
2.29999995231628 0.218799948692322
2.59999990463257 0.168400049209595
2.90000009536743 0.116600036621094
3.20000004768372 0.0772000551223755
3.5 0.0384000539779663
3.79999995231628 0.0221999883651733
4.09999990463257 0.0110000371932983
4.40000009536743 0.00520002841949463
4.69999980926514 0.00220000743865967
5 0.000200033187866211
};
\addlegendentry{$N = 50$}
\addplot [line width=0.24pt, rosybrown199122124, forget plot]
table {%
0.5 0.456568241119385
0.799999952316284 0.39687716960907
1.10000002384186 0.370280385017395
1.39999997615814 0.341417551040649
1.70000004768372 0.313714265823364
2 0.261596083641052
2.29999995231628 0.230259776115417
2.59999990463257 0.178772926330566
2.90000009536743 0.12549614906311
3.5 0.0437264442443848
3.79999995231628 0.026283860206604
4.09999990463257 0.0138911008834839
4.40000009536743 0.00719356536865234
4.69999980926514 0.00349867343902588
5 0.00059199333190918
};
\end{axis}

\end{tikzpicture}\label{fig:con_l8}}
\subfloat[Level of discordance with $l=8$]{
\begin{tikzpicture}[scale = 0.95]

\definecolor{darkgray176}{RGB}{176,176,176}
\definecolor{darkslategray279768}{RGB}{27,97,68}
\definecolor{lightgray204}{RGB}{204,204,204}
\definecolor{midnightblue263565}{RGB}{26,35,65}
\definecolor{olivedrab10412248}{RGB}{104,122,48}
\definecolor{rosybrown199122124}{RGB}{199,122,124}

\begin{axis}[
legend cell align={left},
legend style={
  fill opacity=0.8,
  draw opacity=1,
  text opacity=1,
  at={(0.03,0.97)},
  anchor=north west,
  draw=lightgray204
},
tick align=outside,
tick pos=left,
x grid style={darkgray176},
xlabel={\(\displaystyle \kappa\)},
xmin=0.275, xmax=5.225,
xtick style={color=black},
y grid style={darkgray176},
ylabel={\(\displaystyle \delta(\kappa)\)},
ymin=0, ymax=0.35,
ytick style={color=black}
]
\addplot [line width=0.24pt, midnightblue263565, forget plot]
table {%
0.5 0.0343161821365356
0.799999952316284 0.040099024772644
1.10000002384186 0.0441269874572754
1.39999997615814 0.050931453704834
1.70000004768372 0.0591408014297485
2 0.0704272985458374
2.29999995231628 0.0880963802337646
2.59999990463257 0.107708811759949
2.90000009536743 0.13068699836731
3.20000004768372 0.164571762084961
3.5 0.196001291275024
3.79999995231628 0.226496696472168
4.09999990463257 0.246065139770508
4.40000009536743 0.271521210670471
4.69999980926514 0.292253017425537
5 0.312362194061279
};
\addplot [very thick, midnightblue263565, dotted]
table {%
0.5 0.0369025468826294
0.799999952316284 0.0428899526596069
1.10000002384186 0.0470600128173828
1.39999997615814 0.0540950298309326
1.70000004768372 0.0625375509262085
2 0.0741075277328491
2.29999995231628 0.0921299457550049
2.59999990463257 0.112025022506714
2.90000009536743 0.13522744178772
3.20000004768372 0.169275045394897
3.5 0.200685024261475
3.79999995231628 0.231037497520447
4.09999990463257 0.250522494316101
4.40000009536743 0.275772571563721
4.69999980926514 0.296324968338013
5 0.316244959831238
};
\addlegendentry{$N= 20$}
\addplot [line width=0.24pt, midnightblue263565, forget plot]
table {%
0.5 0.0394887924194336
0.799999952316284 0.0456808805465698
1.10000002384186 0.0499930381774902
1.39999997615814 0.0572584867477417
1.70000004768372 0.0659341812133789
2 0.0777877569198608
2.29999995231628 0.0961636304855347
2.59999990463257 0.116341233253479
2.90000009536743 0.139768004417419
3.20000004768372 0.173978209495544
3.5 0.205368757247925
3.79999995231628 0.235578298568726
4.09999990463257 0.254979848861694
4.40000009536743 0.280023813247681
4.69999980926514 0.300396919250488
5 0.320127725601196
};
\addplot [line width=0.24pt, darkslategray279768, forget plot]
table {%
0.5 0.061058521270752
0.799999952316284 0.0684311389923096
1.39999997615814 0.0837516784667969
1.70000004768372 0.0921523571014404
2 0.104966402053833
2.29999995231628 0.125545382499695
2.59999990463257 0.147053480148315
2.90000009536743 0.172350406646729
3.20000004768372 0.198777675628662
3.5 0.226024389266968
3.79999995231628 0.250379920005798
4.09999990463257 0.271786332130432
4.40000009536743 0.292878985404968
4.69999980926514 0.312157154083252
5 0.329838395118713
};
\addplot [very thick, darkslategray279768, dash pattern=on 1pt off 10pt]
table {%
0.5 0.0636999607086182
0.799999952316284 0.0711983442306519
1.39999997615814 0.0867816209793091
1.70000004768372 0.095341682434082
2 0.10835337638855
2.29999995231628 0.129143357276917
2.59999990463257 0.150820016860962
2.90000009536743 0.176221609115601
3.20000004768372 0.202556610107422
3.5 0.22974169254303
3.79999995231628 0.253986597061157
4.09999990463257 0.27517831325531
4.40000009536743 0.296133279800415
4.69999980926514 0.315225005149841
5 0.332705020904541
};
\addlegendentry{$N= 30$}
\addplot [line width=0.24pt, darkslategray279768, forget plot]
table {%
0.5 0.0663414001464844
0.799999952316284 0.0739655494689941
1.39999997615814 0.0898116827011108
1.70000004768372 0.0985310077667236
2 0.111740350723267
2.29999995231628 0.132741212844849
2.59999990463257 0.154586553573608
2.90000009536743 0.180092930793762
3.20000004768372 0.206335663795471
3.5 0.233458876609802
3.79999995231628 0.257593393325806
4.40000009536743 0.299387693405151
4.69999980926514 0.318292856216431
5 0.335571646690369
};
\addplot [line width=0.24pt, olivedrab10412248, forget plot]
table {%
0.5 0.0744258165359497
0.799999952316284 0.0862195491790771
1.10000002384186 0.0920190811157227
1.39999997615814 0.103364706039429
1.70000004768372 0.111569881439209
2 0.125303983688354
2.29999995231628 0.142575860023499
2.59999990463257 0.168447136878967
2.90000009536743 0.189224362373352
3.20000004768372 0.216488003730774
3.5 0.242408871650696
3.79999995231628 0.260945558547974
4.09999990463257 0.28365421295166
4.40000009536743 0.305217504501343
4.69999980926514 0.321033954620361
5 0.333257436752319
};
\addplot [very thick, olivedrab10412248, dashed]
table {%
0.5 0.076912522315979
0.799999952316284 0.0888062715530396
1.10000002384186 0.0946462154388428
1.39999997615814 0.106098771095276
1.70000004768372 0.114477515220642
2 0.128356218338013
2.29999995231628 0.145701289176941
2.59999990463257 0.171673774719238
2.90000009536743 0.192445039749146
3.20000004768372 0.219687461853027
3.5 0.245506286621094
3.79999995231628 0.263924956321716
4.09999990463257 0.286456227302551
4.40000009536743 0.307814955711365
4.69999980926514 0.323514938354492
5 0.335581302642822
};
\addlegendentry{$N= 40$}
\addplot [line width=0.24pt, olivedrab10412248, forget plot]
table {%
0.5 0.0793992280960083
0.799999952316284 0.0913928747177124
1.10000002384186 0.0972733497619629
1.39999997615814 0.108832836151123
1.70000004768372 0.117385149002075
2 0.131408452987671
2.29999995231628 0.148826718330383
2.59999990463257 0.174900412559509
2.90000009536743 0.195665597915649
3.20000004768372 0.22288703918457
3.5 0.248603582382202
3.79999995231628 0.266904473304749
4.09999990463257 0.289258360862732
4.40000009536743 0.310412526130676
4.69999980926514 0.325996041297913
5 0.337905168533325
};
\addplot [line width=0.24pt, rosybrown199122124, forget plot]
table {%
0.5 0.0853885412216187
0.799999952316284 0.0970290899276733
1.10000002384186 0.103846907615662
1.39999997615814 0.112063050270081
1.70000004768372 0.120772957801819
2 0.138318300247192
2.29999995231628 0.153940916061401
2.59999990463257 0.174878001213074
2.90000009536743 0.199479222297668
3.20000004768372 0.222775220870972
3.5 0.249700307846069
3.79999995231628 0.271644353866577
4.09999990463257 0.289460897445679
4.40000009536743 0.308196544647217
4.69999980926514 0.322907209396362
5 0.338226675987244
};
\addplot [very thick, rosybrown199122124, dash pattern=on 3pt off 5pt on 1pt off 5pt on 1pt off 5pt]
table {%
0.5 0.0877159833908081
0.799999952316284 0.0994110107421875
1.10000002384186 0.106274008750916
1.39999997615814 0.11456298828125
1.70000004768372 0.123340964317322
2 0.140979051589966
2.29999995231628 0.156741976737976
2.59999990463257 0.177711963653564
2.90000009536743 0.202326059341431
3.20000004768372 0.225574970245361
3.5 0.252372026443481
3.79999995231628 0.274201989173889
4.09999990463257 0.291837930679321
4.40000009536743 0.31044602394104
4.69999980926514 0.325032949447632
5 0.340220928192139
};
\addlegendentry{$N= 50$}
\addplot [line width=0.24pt, rosybrown199122124, forget plot]
table {%
0.5 0.0900435447692871
0.799999952316284 0.101792812347412
1.10000002384186 0.10870099067688
1.39999997615814 0.117062926292419
1.70000004768372 0.125908970832825
2 0.14363968372345
2.29999995231628 0.15954315662384
2.59999990463257 0.180546045303345
2.90000009536743 0.205172777175903
3.20000004768372 0.228374838829041
3.5 0.255043745040894
3.79999995231628 0.276759624481201
4.09999990463257 0.294215083122253
4.40000009536743 0.312695384025574
4.69999980926514 0.327158808708191
5 0.342215299606323
};
\end{axis}

\end{tikzpicture}\label{fig:dis_l8}}
\caption{Results of the sensitivity analysis inspecting the number of agents in the system with 8 nearest neighbours.}
\end{figure}

We vary the total number of agents $N\in\{20,30,40,50\}$. Additionally, for each of these values of $N$, we vary the nearest number of neighbours taking values $l\in \{4,6,8\}$. We present the results grouped according to the number of nearest neighbours. Within Figures~\ref{fig:con_l4},~\ref{fig:con_l6}, and~\ref{fig:con_l8}, depicting the probability of consensus, we observe that as the population grows, the probability of consensus decreases. By comparing between these figures we see that as the number of neighbours increases the probability of consensus increases. Both of these results are conceivable. A larger population (keeping the number of neighbours constant) is likely to make it difficult for an opinion to spread throughout the entire network. Similarly, the more neighbours the agents have (keeping the population size constant), the easier it should be for an opinion to spread throughout the population. Quite logically, we observe the inverse effect in Figures~\ref{fig:dis_l4},~\ref{fig:dis_l6}, and~\ref{fig:dis_l8}, on the level of discordance. We would like to draw the reader's attention to the fact that both of these effects (comparing between population size) becomes smaller as the stubbornness $\kappa$ increases to $5$. We believe this is conceivable because a greater $\kappa$ means that the agents in the population become increasingly independent of one another and thus are less effected by the network they form a part of.

\subsection{Probability of rewiring}
\begin{figure}
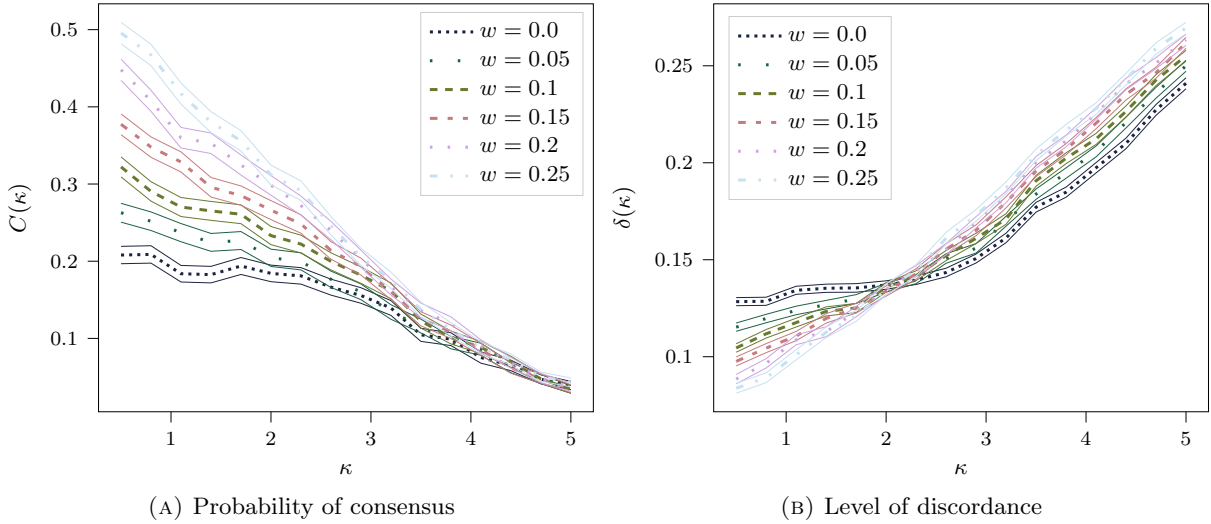

\centering
\subfloat[Probability of consensus]{\input{fig_con_rewire}\label{fig:con_rew}}
\subfloat[Level of discordance]{\input{fig_dis_rewire}\label{fig:dis_rew}}
\caption{Results of the sensitivity analysis inspecting the probability of rewiring.}
\end{figure}

In order to see how the probability of rewiring affects the outcome of the model we take the probability of rewiring each edge from the set $w\in\{0,$ $0.05,$ $0.10,$ $0.15,$ $0.20,$ $0.25\}$.
We present the probability of consensus for these probabilities of rewiring in Figure~\ref{fig:con_rew} and the level of discordance in Figure~\ref{fig:dis_rew}. Again we can see the effect of the network decreasing as $\kappa$ increases in Figure~\ref{fig:con_rew} by the fact that the probabilities of consensus seem to merge at roughly $\kappa=4$. Sensibly, we see that the probability of consensus is higher for networks with more rewiring. We posit that this because of the greater number of cross population connections which make the formation of two (or more) polarised groups more difficult.

We also observe an interesting phenomenon in Figure~\ref{fig:dis_rew}. Namely, that level of convergence for different probabilities of rewiring all seem to intersect just after $\kappa=2$. Furthermore, these have an inverse ordering before and after this crossing point. Our explanation for this effect follows: Before the crossing point, a higher probability of rewiring makes polarisation harder, and so the population tends toward consensus which has the lowest level of discordance. The setting with low level of rewiring has much more structure which allows more easily for polarisation. After the crossing point, a high enough $\kappa$ creates more room not only for polarisation but also fragmentation. As the agents become more independent, the cross population connections at greater $w$ create a higher level of discordance. More structured populations with low $w$ have more of a balance between polarisation (with relatively low discordance) and some fragmentation.

\subsection{Prior belief}
The parameters $\alpha$ and $\beta$ of the Beta-distributed belief are a measure of optimism in the agents in which a larger ratio $\alpha/\beta$ indicating greater optimism. To identify the effect of this prior belief distribution we choose values $(\alpha,\beta)\in\{(4,4), (4,2), (3,3), (3,2), (2,2), (2,1), (1,1)\}$. This takes into consideration that $\alpha\geq \beta$ must hold. This is required to ensure that $\hat{\theta}_0>\TC=0.5$ and the agents do not switch immediately away from an opinion they have just switched to. First we consider those prior beliefs in which $\alpha=\beta$. Thereafter we consider the prior belief combinations in which $\alpha>\beta$. 
\subsubsection{Priors with \texorpdfstring{$\alpha = \beta$}{alpha=beta}}
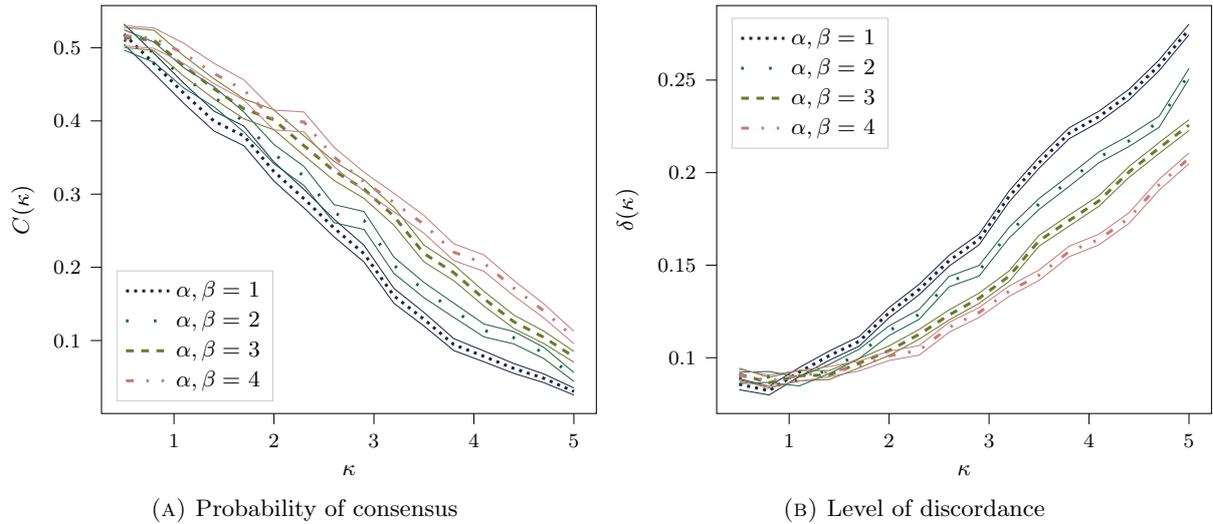
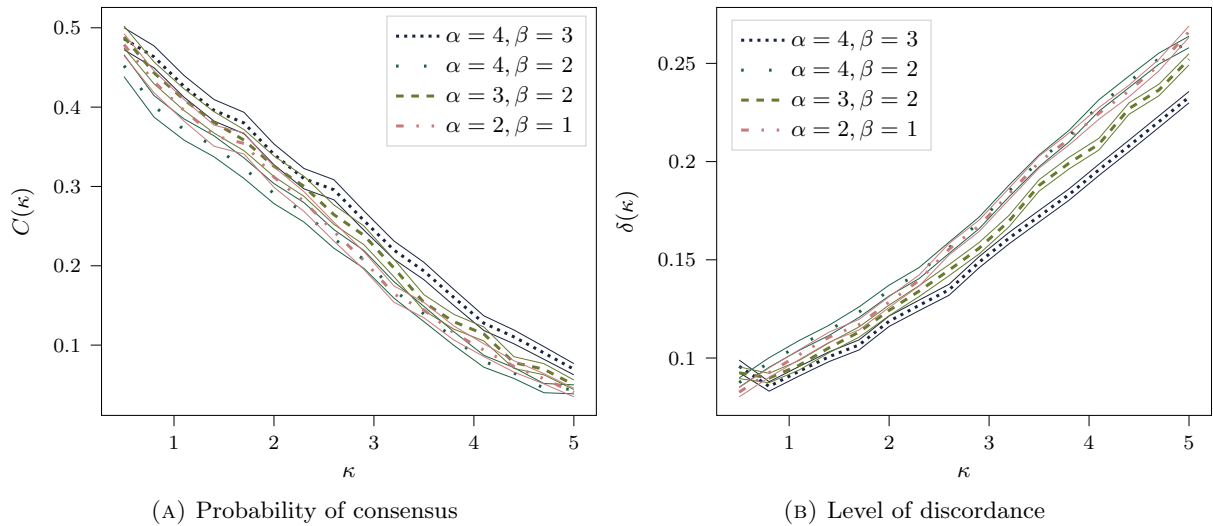
\begin{figure}
\centering
\subfloat[Probability of consensus]{
\begin{tikzpicture}[scale = 0.95]

\definecolor{darkgray176}{RGB}{176,176,176}
\definecolor{darkslategray279768}{RGB}{27,97,68}
\definecolor{lightgray204}{RGB}{204,204,204}
\definecolor{midnightblue263565}{RGB}{26,35,65}
\definecolor{olivedrab10412248}{RGB}{104,122,48}
\definecolor{rosybrown199122124}{RGB}{199,122,124}

\begin{axis}[
legend cell align={left},
legend style={
  fill opacity=0.8,
  draw opacity=1,
  text opacity=1,
  at={(0.03,0.03)},
  anchor=south west,
  draw=lightgray204
},
tick align=outside,
tick pos=left,
x grid style={darkgray176},
xlabel={\(\displaystyle \kappa\)},
xmin=0.275, xmax=5.225,
xtick style={color=black},
y grid style={darkgray176},
ylabel={\(\displaystyle C(\kappa)\)},
ymin=0.000310687514518963, ymax=0.557580344270979,
ytick style={color=black}
]
\addplot [line width=0.24pt, midnightblue263565, forget plot]
table {%
0.5 0.504550099372864
0.799999952316284 0.464553594589233
1.10000002384186 0.423649787902832
1.39999997615814 0.386221885681152
1.70000004768372 0.365949869155884
2 0.317758321762085
2.29999995231628 0.281174182891846
2.59999990463257 0.241737365722656
2.90000009536743 0.207340240478516
3.20000004768372 0.15081262588501
3.5 0.119902610778809
3.79999995231628 0.086295485496521
4.09999990463257 0.0709491968154907
4.40000009536743 0.0551255941390991
4.69999980926514 0.0424512624740601
5 0.0256410837173462
};
\addplot [very thick, midnightblue263565, dotted]
table {%
0.5 0.518399953842163
0.799999952316284 0.478399991989136
1.10000002384186 0.437399983406067
1.39999997615814 0.399800062179565
1.70000004768372 0.379400014877319
2 0.33080005645752
2.29999995231628 0.293799996376038
2.59999990463257 0.25380003452301
2.90000009536743 0.218799948692322
3.20000004768372 0.16100001335144
3.5 0.129199981689453
3.79999995231628 0.0944000482559204
4.09999990463257 0.0784000158309937
4.40000009536743 0.0618000030517578
4.69999980926514 0.0484000444412231
5 0.0304000377655029
};
\addlegendentry{$\alpha, \beta =$ 1}
\addplot [line width=0.24pt, midnightblue263565, forget plot]
table {%
0.5 0.532249927520752
0.799999952316284 0.492246389389038
1.10000002384186 0.451150178909302
1.39999997615814 0.413378119468689
1.70000004768372 0.392850160598755
2 0.343841671943665
2.29999995231628 0.306425809860229
2.59999990463257 0.265862703323364
2.90000009536743 0.230259776115417
3.20000004768372 0.171187400817871
3.5 0.138497352600098
3.79999995231628 0.10250449180603
4.09999990463257 0.085850715637207
4.40000009536743 0.0684744119644165
4.69999980926514 0.0543487071990967
5 0.0351588726043701
};
\addplot [line width=0.24pt, darkslategray279768, forget plot]
table {%
0.5 0.49674379825592
0.799999952316284 0.479941844940186
1.10000002384186 0.444788336753845
1.39999997615814 0.419265747070312
1.70000004768372 0.384829878807068
2 0.341339588165283
2.29999995231628 0.311819314956665
2.59999990463257 0.261045694351196
2.90000009536743 0.251781702041626
3.20000004768372 0.191654801368713
3.5 0.158612370491028
3.79999995231628 0.131547570228577
4.09999990463257 0.104997396469116
4.40000009536743 0.0953458547592163
4.69999980926514 0.0753529071807861
5 0.0441474914550781
};
\addplot [very thick, darkslategray279768, dash pattern=on 1pt off 10pt]
table {%
0.5 0.510599970817566
0.799999952316284 0.493800044059753
1.10000002384186 0.458600044250488
1.39999997615814 0.432999968528748
1.70000004768372 0.398400068283081
2 0.354599952697754
2.29999995231628 0.32480001449585
2.59999990463257 0.273400068283081
2.90000009536743 0.263999938964844
3.20000004768372 0.202800035476685
3.5 0.169000029563904
3.79999995231628 0.141200065612793
4.09999990463257 0.113800048828125
4.40000009536743 0.103800058364868
4.69999980926514 0.0829999446868896
5 0.0501999855041504
};
\addlegendentry{$\alpha, \beta =$ 2}
\addplot [line width=0.24pt, darkslategray279768, forget plot]
table {%
0.5 0.524456262588501
0.799999952316284 0.507658243179321
1.10000002384186 0.472411632537842
1.39999997615814 0.446734309196472
1.70000004768372 0.411970138549805
2 0.367860317230225
2.29999995231628 0.337780594825745
2.59999990463257 0.285754323005676
2.90000009536743 0.276218295097351
3.20000004768372 0.213945269584656
3.5 0.17938756942749
4.09999990463257 0.122602581977844
4.40000009536743 0.11225414276123
4.69999980926514 0.0906471014022827
5 0.0562525987625122
};
\addplot [line width=0.24pt, olivedrab10412248, forget plot]
table {%
0.5 0.500146150588989
0.799999952316284 0.496143460273743
1.10000002384186 0.460958361625671
1.39999997615814 0.430028557777405
1.70000004768372 0.402735829353333
2 0.387812852859497
2.29999995231628 0.352846145629883
2.59999990463257 0.318748474121094
2.90000009536743 0.294214844703674
3.20000004768372 0.257496953010559
3.5 0.207536458969116
3.79999995231628 0.181669473648071
4.09999990463257 0.1471107006073
4.40000009536743 0.116414070129395
4.69999980926514 0.0959242582321167
5 0.0703754425048828
};
\addplot [very thick, olivedrab10412248, dashed]
table {%
0.5 0.513999938964844
0.799999952316284 0.509999990463257
1.10000002384186 0.474799990653992
1.39999997615814 0.44379997253418
1.70000004768372 0.416399955749512
2 0.401399970054626
2.29999995231628 0.366199970245361
2.59999990463257 0.331799983978271
2.90000009536743 0.307000041007996
3.20000004768372 0.269799947738647
3.5 0.218999981880188
3.79999995231628 0.192600011825562
4.09999990463257 0.15719997882843
4.40000009536743 0.125599980354309
4.69999980926514 0.104400038719177
5 0.0778000354766846
};
\addlegendentry{$\alpha, \beta =$ 3}
\addplot [line width=0.24pt, olivedrab10412248, forget plot]
table {%
0.5 0.527853846549988
0.799999952316284 0.523856520652771
1.10000002384186 0.488641738891602
1.39999997615814 0.457571506500244
1.70000004768372 0.43006420135498
2 0.414987087249756
2.29999995231628 0.37955379486084
2.59999990463257 0.344851493835449
2.90000009536743 0.319785118103027
3.20000004768372 0.282103061676025
3.5 0.23046350479126
3.79999995231628 0.203530550003052
4.09999990463257 0.167289257049561
4.40000009536743 0.134785890579224
4.69999980926514 0.112875699996948
5 0.0852246284484863
};
\addplot [line width=0.24pt, rosybrown199122124, forget plot]
table {%
0.5 0.502347946166992
0.799999952316284 0.49914538860321
1.10000002384186 0.478142499923706
1.39999997615814 0.450975060462952
1.70000004768372 0.428433656692505
2 0.387614011764526
2.29999995231628 0.385028719902039
2.59999990463257 0.33757209777832
2.90000009536743 0.30430018901825
3.20000004768372 0.276040315628052
3.5 0.246659994125366
3.79999995231628 0.2091064453125
4.09999990463257 0.194593787193298
4.40000009536743 0.161149144172668
4.69999980926514 0.130770564079285
5 0.0957313776016235
};
\addplot [very thick, rosybrown199122124, dash pattern=on 3pt off 5pt on 1pt off 5pt on 1pt off 5pt]
table {%
0.5 0.516200065612793
0.799999952316284 0.513000011444092
1.10000002384186 0.491999983787537
1.39999997615814 0.464800000190735
1.70000004768372 0.442199945449829
2 0.40120005607605
2.29999995231628 0.398599982261658
2.59999990463257 0.350800037384033
2.90000009536743 0.317199945449829
3.20000004768372 0.288599967956543
3.5 0.258800029754639
3.79999995231628 0.220600008964539
4.09999990463257 0.20580005645752
4.40000009536743 0.171599984169006
4.69999980926514 0.140400052070618
5 0.104200005531311
};
\addlegendentry{$\alpha, \beta =$ 4}
\addplot [line width=0.24pt, rosybrown199122124, forget plot]
table {%
0.5 0.530051946640015
0.799999952316284 0.526854634284973
1.10000002384186 0.505857467651367
1.39999997615814 0.478624939918518
1.70000004768372 0.455966353416443
2 0.414785981178284
2.29999995231628 0.412171363830566
2.59999990463257 0.364027857780457
2.90000009536743 0.330099821090698
3.20000004768372 0.301159620285034
3.5 0.270940065383911
3.79999995231628 0.232093572616577
4.09999990463257 0.217006206512451
4.40000009536743 0.182050824165344
4.69999980926514 0.150029420852661
5 0.112668514251709
};
\end{axis}

\end{tikzpicture}\label{fig:con_unit_prior}}
\subfloat[Level of discordance]{
\begin{tikzpicture}[scale = 0.95]

\definecolor{darkgray176}{RGB}{176,176,176}
\definecolor{darkslategray279768}{RGB}{27,97,68}
\definecolor{lightgray204}{RGB}{204,204,204}
\definecolor{midnightblue263565}{RGB}{26,35,65}
\definecolor{olivedrab10412248}{RGB}{104,122,48}
\definecolor{rosybrown199122124}{RGB}{199,122,124}

\begin{axis}[
legend cell align={left},
legend style={
  fill opacity=0.8,
  draw opacity=1,
  text opacity=1,
  at={(0.03,0.97)},
  anchor=north west,
  draw=lightgray204
},
tick align=outside,
tick pos=left,
x grid style={darkgray176},
xlabel={\(\displaystyle \kappa\)},
xmin=0.275, xmax=5.225,
xtick style={color=black},
y grid style={darkgray176},
ylabel={\(\displaystyle \delta(\kappa)\)},
ymin=0.0698479148086679, ymax=0.290270352806771,
ytick style={color=black}
]
\addplot [line width=0.24pt, midnightblue263565, forget plot]
table {%
0.5 0.0826984643936157
0.799999952316284 0.0798671245574951
1.10000002384186 0.0897704362869263
1.39999997615814 0.0985625982284546
1.70000004768372 0.106229662895203
2 0.121625185012817
2.29999995231628 0.134018778800964
2.59999990463257 0.149518251419067
2.90000009536743 0.160985589027405
3.20000004768372 0.184152960777283
3.5 0.202366590499878
3.79999995231628 0.218262553215027
4.09999990463257 0.227355122566223
4.40000009536743 0.239273071289062
4.69999980926514 0.254988431930542
5 0.274488806724548
};
\addplot [very thick, midnightblue263565, dotted]
table {%
0.5 0.0856499671936035
0.799999952316284 0.0822149515151978
1.10000002384186 0.092210054397583
1.39999997615814 0.101070046424866
1.70000004768372 0.108839988708496
2 0.124364972114563
2.29999995231628 0.136834979057312
2.59999990463257 0.152430057525635
2.90000009536743 0.163915038108826
3.20000004768372 0.187119960784912
3.5 0.20536994934082
3.79999995231628 0.221235036849976
4.09999990463257 0.230280041694641
4.40000009536743 0.242164969444275
4.69999980926514 0.257910013198853
5 0.277369976043701
};
\addlegendentry{$\alpha, \beta =$ 1}
\addplot [line width=0.24pt, midnightblue263565, forget plot]
table {%
0.5 0.0886015892028809
0.799999952316284 0.0845628976821899
1.10000002384186 0.0946495532989502
1.39999997615814 0.103577375411987
1.70000004768372 0.11145031452179
2 0.127104878425598
2.29999995231628 0.139651298522949
2.59999990463257 0.155341744422913
2.90000009536743 0.166844367980957
3.20000004768372 0.190087080001831
3.5 0.208373427391052
3.79999995231628 0.224207401275635
4.09999990463257 0.233204960823059
4.40000009536743 0.245056986808777
4.69999980926514 0.260831594467163
5 0.280251145362854
};
\addplot [line width=0.24pt, darkslategray279768, forget plot]
table {%
0.5 0.0860685110092163
0.799999952316284 0.0866835117340088
1.10000002384186 0.0847327709197998
1.39999997615814 0.0914018154144287
1.70000004768372 0.0995343923568726
2 0.112000942230225
2.29999995231628 0.120736956596375
2.59999990463257 0.138327360153198
2.90000009536743 0.144129991531372
3.20000004768372 0.164640188217163
3.5 0.180120229721069
3.79999995231628 0.192540764808655
4.09999990463257 0.205359935760498
4.40000009536743 0.214116215705872
4.69999980926514 0.224445104598999
5 0.250517725944519
};
\addplot [very thick, darkslategray279768, dash pattern=on 1pt off 10pt]
table {%
0.5 0.0891150236129761
0.799999952316284 0.0896099805831909
1.10000002384186 0.0871399641036987
1.39999997615814 0.0938700437545776
1.70000004768372 0.102059960365295
2 0.114624977111816
2.29999995231628 0.123404979705811
2.59999990463257 0.141129970550537
2.90000009536743 0.146994948387146
3.20000004768372 0.167549967765808
3.5 0.183079957962036
3.79999995231628 0.195510029792786
4.09999990463257 0.208315014839172
4.40000009536743 0.217100024223328
4.69999980926514 0.227414965629578
5 0.253414988517761
};
\addlegendentry{$\alpha, \beta =$ 2}
\addplot [line width=0.24pt, darkslategray279768, forget plot]
table {%
0.5 0.0921614170074463
0.799999952316284 0.0925365686416626
1.10000002384186 0.0895472764968872
1.39999997615814 0.096338152885437
1.70000004768372 0.104585647583008
2 0.117249011993408
2.29999995231628 0.126073122024536
2.59999990463257 0.143932580947876
2.90000009536743 0.149860024452209
3.20000004768372 0.170459866523743
3.5 0.186039805412292
3.79999995231628 0.198479175567627
4.09999990463257 0.211270093917847
4.40000009536743 0.220083713531494
4.69999980926514 0.230384826660156
5 0.256312370300293
};
\addplot [line width=0.24pt, olivedrab10412248, forget plot]
table {%
0.5 0.0879542827606201
0.799999952316284 0.0836055278778076
1.10000002384186 0.0874879360198975
1.39999997615814 0.0881998538970947
1.70000004768372 0.0945230722427368
2 0.101217865943909
2.29999995231628 0.109981060028076
2.59999990463257 0.120393753051758
2.90000009536743 0.129355192184448
3.20000004768372 0.141186952590942
3.5 0.160282850265503
4.09999990463257 0.181885480880737
4.40000009536743 0.197458505630493
4.69999980926514 0.210293173789978
5 0.222750902175903
};
\addplot [very thick, olivedrab10412248, dashed]
table {%
0.5 0.0911550521850586
0.799999952316284 0.0865099430084229
1.10000002384186 0.0902600288391113
1.39999997615814 0.090630054473877
1.70000004768372 0.0970000028610229
2 0.103814959526062
2.29999995231628 0.112625002861023
2.59999990463257 0.123115062713623
2.90000009536743 0.132135033607483
3.20000004768372 0.144029974937439
3.5 0.163210034370422
3.79999995231628 0.174085021018982
4.09999990463257 0.184785008430481
4.40000009536743 0.200394988059998
4.69999980926514 0.213259935379028
5 0.225664973258972
};
\addlegendentry{$\alpha, \beta =$ 3}
\addplot [line width=0.24pt, olivedrab10412248, forget plot]
table {%
0.5 0.0943557024002075
0.799999952316284 0.0894144773483276
1.10000002384186 0.0930321216583252
1.39999997615814 0.0930601358413696
1.70000004768372 0.0994769334793091
2 0.106412172317505
2.29999995231628 0.11526894569397
2.59999990463257 0.125836253166199
2.90000009536743 0.134914875030518
3.20000004768372 0.146872997283936
3.5 0.166137218475342
3.79999995231628 0.177057504653931
4.09999990463257 0.187684535980225
4.40000009536743 0.203331470489502
4.69999980926514 0.216226816177368
5 0.228579044342041
};
\addplot [line width=0.24pt, rosybrown199122124, forget plot]
table {%
0.5 0.0871880054473877
0.799999952316284 0.0845805406570435
1.10000002384186 0.0872913599014282
1.39999997615814 0.0896799564361572
1.70000004768372 0.0928163528442383
2 0.0984926223754883
2.29999995231628 0.101385951042175
2.59999990463257 0.11375904083252
2.90000009536743 0.121763944625854
3.20000004768372 0.133159399032593
3.5 0.141784071922302
3.79999995231628 0.15456748008728
4.09999990463257 0.16101861000061
4.40000009536743 0.172574758529663
4.69999980926514 0.190470695495605
5 0.20485258102417
};
\addplot [very thick, rosybrown199122124, dash pattern=on 3pt off 5pt on 1pt off 5pt on 1pt off 5pt]
table {%
0.5 0.0903999805450439
0.799999952316284 0.0875749588012695
1.10000002384186 0.0902299880981445
1.39999997615814 0.092460036277771
1.70000004768372 0.0954949855804443
2 0.101009964942932
2.29999995231628 0.103970050811768
2.59999990463257 0.116415023803711
2.90000009536743 0.124434947967529
3.20000004768372 0.135930061340332
3.5 0.144585013389587
3.79999995231628 0.157389998435974
4.09999990463257 0.163895010948181
4.40000009536743 0.175449967384338
4.69999980926514 0.193410038948059
5 0.207739949226379
};
\addlegendentry{$\alpha, \beta =$ 4}
\addplot [line width=0.24pt, rosybrown199122124, forget plot]
table {%
0.5 0.0936119556427002
0.799999952316284 0.0905694961547852
1.10000002384186 0.0931686162948608
1.39999997615814 0.0952399969100952
1.70000004768372 0.0981736183166504
2 0.103527426719666
2.29999995231628 0.10655403137207
2.59999990463257 0.119071006774902
2.90000009536743 0.127106070518494
3.20000004768372 0.138700604438782
3.5 0.147385954856873
3.79999995231628 0.160212516784668
4.09999990463257 0.166771411895752
4.40000009536743 0.178325176239014
4.69999980926514 0.196349382400513
5 0.210627317428589
};
\end{axis}

\end{tikzpicture}\label{fig:dis_unit_prior}}
\caption{Results of the sensitivity analysis inspecting the prior belief distribution of the agents with $\alpha = \beta$.}
\end{figure}

In these cases switching early is quite likely as the initial estimate is on the cusp of the critical level. The general trend we observe in Figure~\ref{fig:con_unit_prior}, is that the lower $\alpha$ and $\beta$ lead to lower probability of consensus than higher vales of $\alpha$ and $\beta$. This is explained by the fact that greater values of $\alpha$ and $\beta$ mean that the change in their estimate from one round to the next is comparatively small at the start of a run with an opinion. As an illustration, consider an agent with 5 affirming experiences during the warm up. If this agent has the $\alpha = \beta =1$ prior, their estimate at the end of the warm up is $\hat{\theta}=0.86$ which may allow them to retain this opinion in the face of a disagreeing neighbourhood. If instead this agent had the $\alpha=\beta=4$ prior, their estimate at the end of the warm up would be $\hat{\theta}=0.69$. This lower estimate can withstand less disagreement and so it makes sense that greater initial values of $\alpha=\beta$ lead to more consensus. We see the opposite effect on the level of discordance in Figure~\ref{fig:dis_unit_prior}.

\subsubsection{Priors with \texorpdfstring{$\alpha>\beta$}{alpha > beta}}
The case with $\alpha>\beta$ exemplifies a greater optimism of an agent in their opinion at the start of a run. The greater level of optimism is represented by a ratio of 2 to 1 in the settings of $(\alpha,\beta)\in \{(2,1),(4,2)\}$. The remaining settings $(\alpha,\beta)\in\{(3,2),(4,3)\}$ also exhibit optimism but to a lesser extent. In Figures~\ref{fig:con_gen_prior} and~\ref{fig:dis_gen_prior} respectively, we see that the more optimistic settings lead to less consensus and more discordance than the somewhat less optimistic settings. Furthermore, we see that the more optimistic settings (both having a ratio of 2:1) do not differ greatly from one another. Within the less optimistic settings we note that the slightly more optimistic of the two ($\alpha=3, \beta=2$) leads to less consensus and more discordance than ($\alpha = 4,\beta = 3$). In summary, as optimism increases, the probability of consensus decreases and the proportion of discordance increases.
\begin{figure}
\centering
\subfloat[Probability of consensus]{
\begin{tikzpicture}[scale = 0.95]

\definecolor{darkgray176}{RGB}{176,176,176}
\definecolor{darkslategray279768}{RGB}{27,97,68}
\definecolor{lightgray204}{RGB}{204,204,204}
\definecolor{midnightblue263565}{RGB}{26,35,65}
\definecolor{olivedrab10412248}{RGB}{104,122,48}
\definecolor{rosybrown199122124}{RGB}{199,122,124}

\begin{axis}[
legend cell align={left},
legend style={fill opacity=0.8, draw opacity=1, text opacity=1, draw=lightgray204},
tick align=outside,
tick pos=left,
x grid style={darkgray176},
xlabel={\(\displaystyle \kappa\)},
xmin=0.275, xmax=5.225,
xtick style={color=black},
y grid style={darkgray176},
ylabel={\(\displaystyle C(\kappa)\)},
ymin=0.0113802801149888, ymax=0.525630576047822,
ytick style={color=black}
]
\addplot [line width=0.24pt, midnightblue263565, forget plot]
table {%
0.5 0.472345948219299
0.799999952316284 0.449577927589417
1.10000002384186 0.412492513656616
1.39999997615814 0.382443785667419
1.70000004768372 0.366545796394348
2 0.327265739440918
2.29999995231628 0.296982645988464
2.59999990463257 0.283149242401123
2.90000009536743 0.244887590408325
3.20000004768372 0.208321452140808
3.5 0.182452201843262
3.79999995231628 0.149838209152222
4.09999990463257 0.118351817131042
4.40000009536743 0.101713299751282
5 0.0625463724136353
};
\addplot [very thick, midnightblue263565, dotted]
table {%
0.5 0.486199975013733
0.799999952316284 0.46340000629425
1.10000002384186 0.426200032234192
1.39999997615814 0.396000027656555
1.70000004768372 0.379999995231628
2 0.340399980545044
2.29999995231628 0.309800028800964
2.59999990463257 0.295799970626831
2.90000009536743 0.256999969482422
3.20000004768372 0.219799995422363
3.5 0.193400025367737
3.79999995231628 0.159999966621399
4.09999990463257 0.127599954605103
4.40000009536743 0.110399961471558
4.69999980926514 0.0901999473571777
5 0.069599986076355
};
\addlegendentry{$\alpha = 4, \beta =$ 3}
\addplot [line width=0.24pt, midnightblue263565, forget plot]
table {%
0.5 0.500054001808167
0.799999952316284 0.477222084999084
1.10000002384186 0.439907550811768
1.39999997615814 0.409556150436401
1.70000004768372 0.393454194068909
2 0.35353422164917
2.29999995231628 0.322617411613464
2.59999990463257 0.308450818061829
2.90000009536743 0.269112467765808
3.20000004768372 0.231278538703918
3.5 0.204347848892212
3.79999995231628 0.170161843299866
4.09999990463257 0.136848092079163
4.40000009536743 0.119086623191833
4.69999980926514 0.0981404781341553
5 0.0766535997390747
};
\addplot [line width=0.24pt, darkslategray279768, forget plot]
table {%
0.5 0.438404202461243
0.799999952316284 0.387415051460266
1.10000002384186 0.358404040336609
1.39999997615814 0.337373852729797
1.70000004768372 0.310038089752197
2 0.278409600257874
2.29999995231628 0.255131721496582
2.59999990463257 0.222068190574646
2.90000009536743 0.196749687194824
3.20000004768372 0.159197807312012
3.5 0.129605054855347
3.79999995231628 0.0995897054672241
4.09999990463257 0.0720973014831543
4.40000009536743 0.0579764842987061
4.69999980926514 0.0400054454803467
5 0.0385026931762695
};
\addplot [very thick, darkslategray279768, dash pattern=on 1pt off 10pt]
table {%
0.5 0.452199935913086
0.799999952316284 0.401000022888184
1.10000002384186 0.371799945831299
1.39999997615814 0.350600004196167
1.70000004768372 0.322999954223633
2 0.291000008583069
2.29999995231628 0.267400026321411
2.59999990463257 0.233799934387207
2.90000009536743 0.20799994468689
3.20000004768372 0.169600009918213
3.5 0.13919997215271
3.79999995231628 0.108199954032898
4.09999990463257 0.0795999765396118
4.40000009536743 0.0648000240325928
4.69999980926514 0.0457999706268311
5 0.0441999435424805
};
\addlegendentry{$\alpha = 4, \beta =$ 2}
\addplot [line width=0.24pt, darkslategray279768, forget plot]
table {%
0.5 0.465995788574219
0.799999952316284 0.414584875106812
1.10000002384186 0.385195970535278
1.39999997615814 0.363826155662537
1.70000004768372 0.335961818695068
2 0.303590416908264
2.29999995231628 0.27966833114624
2.59999990463257 0.245531797409058
2.90000009536743 0.219250321388245
3.20000004768372 0.180002212524414
3.5 0.148794889450073
3.79999995231628 0.116810321807861
4.09999990463257 0.0871026515960693
4.40000009536743 0.0716235637664795
4.69999980926514 0.051594614982605
5 0.049897313117981
};
\addplot [line width=0.24pt, olivedrab10412248, forget plot]
table {%
0.5 0.474544405937195
0.799999952316284 0.429430484771729
1.10000002384186 0.395770072937012
1.39999997615814 0.367340326309204
1.70000004768372 0.345108032226562
2 0.312413215637207
2.29999995231628 0.287495374679565
2.59999990463257 0.252175807952881
2.90000009536743 0.22658896446228
3.20000004768372 0.186171174049377
3.5 0.143605709075928
3.79999995231628 0.120096445083618
4.09999990463257 0.105190753936768
4.40000009536743 0.0705666542053223
4.69999980926514 0.0625463724136353
5 0.0439589023590088
};
\addplot [very thick, olivedrab10412248, dashed]
table {%
0.5 0.488399982452393
0.799999952316284 0.443199992179871
1.10000002384186 0.40939998626709
1.39999997615814 0.380800008773804
1.70000004768372 0.358399987220764
2 0.325399994850159
2.29999995231628 0.30019998550415
2.59999990463257 0.264400005340576
2.90000009536743 0.238399982452393
3.20000004768372 0.197200059890747
3.5 0.153599977493286
3.79999995231628 0.129400014877319
4.09999990463257 0.113999962806702
4.40000009536743 0.0779999494552612
4.69999980926514 0.069599986076355
5 0.0499999523162842
};
\addlegendentry{$\alpha = 3, \beta =$ 2}
\addplot [line width=0.24pt, olivedrab10412248, forget plot]
table {%
0.5 0.50225555896759
0.799999952316284 0.456969618797302
1.10000002384186 0.423029899597168
1.39999997615814 0.394259691238403
1.70000004768372 0.371691942214966
2 0.33838677406311
2.29999995231628 0.312904596328735
2.59999990463257 0.276624202728271
2.90000009536743 0.250211000442505
3.20000004768372 0.208228826522827
3.5 0.163594365119934
3.79999995231628 0.138703584671021
4.09999990463257 0.122809290885925
4.40000009536743 0.0854333639144897
4.69999980926514 0.0766535997390747
5 0.0560411214828491
};
\addplot [line width=0.24pt, rosybrown199122124, forget plot]
table {%
0.5 0.464953184127808
0.799999952316284 0.418269515037537
1.10000002384186 0.382443785667419
1.39999997615814 0.350663185119629
1.70000004768372 0.340943098068237
2 0.298959970474243
2.29999995231628 0.268540859222412
2.59999990463257 0.231111645698547
2.90000009536743 0.195377707481384
3.20000004768372 0.154126405715942
3.5 0.134462714195251
3.79999995231628 0.106350421905518
4.09999990463257 0.0855263471603394
4.40000009536743 0.0659803152084351
4.69999980926514 0.0511419773101807
5 0.0347553491592407
};
\addplot [very thick, rosybrown199122124, dash pattern=on 3pt off 5pt on 1pt off 5pt on 1pt off 5pt]
table {%
0.5 0.478800058364868
0.799999952316284 0.432000041007996
1.10000002384186 0.396000027656555
1.39999997615814 0.363999962806702
1.70000004768372 0.354200005531311
2 0.311800003051758
2.29999995231628 0.281000018119812
2.59999990463257 0.243000030517578
2.90000009536743 0.206599950790405
3.20000004768372 0.164399981498718
3.5 0.144199967384338
3.79999995231628 0.115200042724609
4.09999990463257 0.0936000347137451
4.40000009536743 0.073199987411499
4.69999980926514 0.0576000213623047
5 0.0401999950408936
};
\addlegendentry{$\alpha = 2, \beta =$ 1}
\addplot [line width=0.24pt, rosybrown199122124, forget plot]
table {%
0.5 0.492646813392639
0.799999952316284 0.445730566978455
1.10000002384186 0.409556150436401
1.39999997615814 0.377336740493774
1.70000004768372 0.367456912994385
2 0.324640035629272
2.29999995231628 0.293459177017212
2.59999990463257 0.254888296127319
2.90000009536743 0.217822313308716
3.20000004768372 0.174673557281494
3.5 0.153937339782715
3.79999995231628 0.124049544334412
4.09999990463257 0.101673603057861
4.40000009536743 0.080419659614563
4.69999980926514 0.0640580654144287
5 0.0456447601318359
};
\end{axis}

\end{tikzpicture}\label{fig:con_gen_prior}}
\subfloat[Level of discordance]{
\begin{tikzpicture}[scale = 0.95]

\definecolor{darkgray176}{RGB}{176,176,176}
\definecolor{darkslategray279768}{RGB}{27,97,68}
\definecolor{lightgray204}{RGB}{204,204,204}
\definecolor{midnightblue263565}{RGB}{26,35,65}
\definecolor{olivedrab10412248}{RGB}{104,122,48}
\definecolor{rosybrown199122124}{RGB}{199,122,124}

\begin{axis}[
legend cell align={left},
legend style={
  fill opacity=0.8,
  draw opacity=1,
  text opacity=1,
  at={(0.03,0.97)},
  anchor=north west,
  draw=lightgray204
},
tick align=outside,
tick pos=left,
x grid style={darkgray176},
xlabel={\(\displaystyle \kappa\)},
xmin=0.275, xmax=5.225,
xtick style={color=black},
y grid style={darkgray176},
ylabel={\(\displaystyle \delta(\kappa)\)},
ymin=0.0707860494579103, ymax=0.278467504969025,
ytick style={color=black}
]
\addplot [line width=0.24pt, midnightblue263565, forget plot]
table {%
0.5 0.0926030874252319
0.799999952316284 0.0831252336502075
1.10000002384186 0.09084153175354
1.39999997615814 0.0982069969177246
1.70000004768372 0.104115009307861
2 0.116081953048706
2.29999995231628 0.124136328697205
2.59999990463257 0.131957054138184
2.90000009536743 0.145925641059875
3.20000004768372 0.158146262168884
3.79999995231628 0.180325388908386
4.09999990463257 0.192976951599121
4.40000009536743 0.205055594444275
4.69999980926514 0.217416167259216
5 0.229916214942932
};
\addplot [very thick, midnightblue263565, dotted]
table {%
0.5 0.0957850217819214
0.799999952316284 0.0855350494384766
1.10000002384186 0.0932650566101074
1.39999997615814 0.100679993629456
1.70000004768372 0.106670022010803
2 0.118729948997498
2.29999995231628 0.126804947853088
2.59999990463257 0.13475501537323
2.90000009536743 0.148769974708557
3.20000004768372 0.161044955253601
3.79999995231628 0.183220028877258
4.09999990463257 0.195850014686584
4.40000009536743 0.207990050315857
5 0.232805013656616
};
\addlegendentry{$\alpha = 4, \beta =$ 3}
\addplot [line width=0.24pt, midnightblue263565, forget plot]
table {%
0.5 0.0989668369293213
0.799999952316284 0.0879447460174561
1.10000002384186 0.0956884622573853
1.39999997615814 0.103152990341187
1.70000004768372 0.109225034713745
2 0.121377944946289
2.59999990463257 0.137552976608276
2.90000009536743 0.151614427566528
3.20000004768372 0.163943767547607
3.5 0.17512845993042
3.79999995231628 0.18611466884613
4.09999990463257 0.198723077774048
4.40000009536743 0.210924386978149
5 0.2356938123703
};
\addplot [line width=0.24pt, darkslategray279768, forget plot]
table {%
0.5 0.0851080417633057
0.799999952316284 0.0953185558319092
1.10000002384186 0.103909015655518
1.39999997615814 0.111511468887329
1.70000004768372 0.120725035667419
2 0.131611943244934
2.29999995231628 0.140447616577148
2.59999990463257 0.153277277946472
2.90000009536743 0.165846228599548
3.20000004768372 0.182092428207397
3.5 0.197365045547485
3.79999995231628 0.210044026374817
4.09999990463257 0.225698947906494
4.69999980926514 0.249627232551575
5 0.257985234260559
};
\addplot [very thick, darkslategray279768, dash pattern=on 1pt off 10pt]
table {%
0.5 0.087494969367981
0.799999952316284 0.0977449417114258
1.10000002384186 0.106415033340454
1.39999997615814 0.114099979400635
1.70000004768372 0.123394966125488
2 0.134354948997498
2.29999995231628 0.143254995346069
2.59999990463257 0.156144976615906
2.90000009536743 0.168784976005554
3.20000004768372 0.185070037841797
3.5 0.20036506652832
3.79999995231628 0.212985038757324
4.09999990463257 0.228620052337646
4.40000009536743 0.240615010261536
4.69999980926514 0.252444982528687
5 0.260915040969849
};
\addlegendentry{$\alpha = 4, \beta =$ 2}
\addplot [line width=0.24pt, darkslategray279768, forget plot]
table {%
0.5 0.0898820161819458
0.799999952316284 0.100171446800232
1.10000002384186 0.108920931816101
1.39999997615814 0.11668848991394
1.70000004768372 0.126065015792847
2 0.137098073959351
2.29999995231628 0.14606237411499
2.59999990463257 0.159012675285339
2.90000009536743 0.17172384262085
3.20000004768372 0.188047528266907
3.5 0.203364968299866
3.79999995231628 0.215925931930542
4.09999990463257 0.231541037559509
4.40000009536743 0.243552446365356
4.69999980926514 0.255262732505798
5 0.263844728469849
};
\addplot [line width=0.24pt, olivedrab10412248, forget plot]
table {%
0.5 0.0895092487335205
0.799999952316284 0.0874092578887939
1.10000002384186 0.0939034223556519
1.39999997615814 0.102706432342529
1.70000004768372 0.11062490940094
2 0.121541500091553
2.29999995231628 0.131281614303589
2.59999990463257 0.141923666000366
2.90000009536743 0.152981281280518
3.20000004768372 0.16684103012085
3.5 0.184946537017822
3.79999995231628 0.196264982223511
4.09999990463257 0.205947160720825
4.40000009536743 0.224095106124878
4.69999980926514 0.233525991439819
5 0.249208569526672
};
\addplot [very thick, olivedrab10412248, dashed]
table {%
0.5 0.0924949645996094
0.799999952316284 0.0898150205612183
1.10000002384186 0.0963200330734253
1.39999997615814 0.105234980583191
1.70000004768372 0.113240003585815
2 0.124250054359436
2.29999995231628 0.134070038795471
2.59999990463257 0.144734978675842
2.90000009536743 0.155889987945557
3.20000004768372 0.169734954833984
3.5 0.187855005264282
3.79999995231628 0.199219942092896
4.09999990463257 0.208925008773804
4.40000009536743 0.226984977722168
4.69999980926514 0.236479997634888
5 0.252074956893921
};
\addlegendentry{$\alpha = 3, \beta =$ 2}
\addplot [line width=0.24pt, olivedrab10412248, forget plot]
table {%
0.5 0.0954807996749878
0.799999952316284 0.092220664024353
1.10000002384186 0.0987365245819092
1.39999997615814 0.107763648033142
1.70000004768372 0.115855097770691
2 0.12695848941803
2.29999995231628 0.136858344078064
2.59999990463257 0.147546291351318
2.90000009536743 0.158798694610596
3.20000004768372 0.172628998756409
3.5 0.190763473510742
3.79999995231628 0.20217502117157
4.09999990463257 0.211902737617493
4.40000009536743 0.229874849319458
4.69999980926514 0.239434003829956
5 0.254941463470459
};
\addplot [line width=0.24pt, rosybrown199122124, forget plot]
table {%
0.5 0.0802260637283325
0.799999952316284 0.0901663303375244
1.10000002384186 0.0991815328598022
1.39999997615814 0.108054399490356
1.70000004768372 0.114336133003235
2 0.125773668289185
2.29999995231628 0.136645317077637
2.59999990463257 0.152498483657837
2.90000009536743 0.164384841918945
3.20000004768372 0.180640459060669
3.5 0.196652889251709
3.79999995231628 0.208491086959839
4.09999990463257 0.221639633178711
4.40000009536743 0.233025789260864
4.69999980926514 0.246200561523438
5 0.263282537460327
};
\addplot [very thick, rosybrown199122124, dash pattern=on 3pt off 5pt on 1pt off 5pt on 1pt off 5pt]
table {%
0.5 0.0826050043106079
0.799999952316284 0.0925849676132202
1.10000002384186 0.101680040359497
1.39999997615814 0.110625028610229
1.70000004768372 0.11701500415802
2 0.128499984741211
2.29999995231628 0.139445066452026
2.59999990463257 0.155420064926147
2.90000009536743 0.167284965515137
3.20000004768372 0.183560013771057
3.5 0.199694991111755
3.79999995231628 0.21150004863739
4.09999990463257 0.224604964256287
4.40000009536743 0.235975027084351
4.69999980926514 0.249165058135986
5 0.266155004501343
};
\addlegendentry{$\alpha = 2, \beta =$ 1}
\addplot [line width=0.24pt, rosybrown199122124, forget plot]
table {%
0.5 0.0849838256835938
0.799999952316284 0.095003604888916
1.10000002384186 0.104178428649902
1.39999997615814 0.113195657730103
1.70000004768372 0.119693875312805
2 0.131226301193237
2.29999995231628 0.142244696617126
2.59999990463257 0.158341407775879
2.90000009536743 0.170185208320618
3.5 0.202737092971802
3.79999995231628 0.214509010314941
4.09999990463257 0.227570414543152
4.40000009536743 0.238924145698547
4.69999980926514 0.252129435539246
5 0.269027471542358
};
\end{axis}

\end{tikzpicture}\label{fig:dis_gen_prior}}
\caption{Results of the sensitivity analysis inspecting the prior belief distribution of the agents with $\alpha > \beta$.}
\end{figure}

\subsection{Opinion reliability}
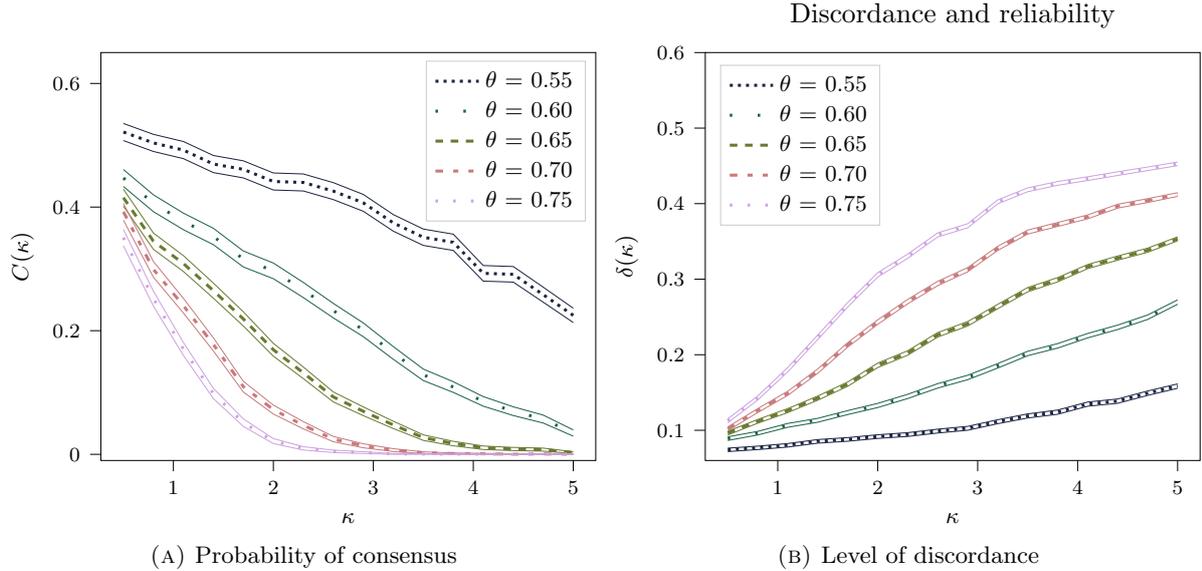
\begin{figure}
    \centering
    \subfloat[Probability of consensus]{
\begin{tikzpicture}[scale = 0.95]

\definecolor{darkgray176}{RGB}{176,176,176}
\definecolor{darkslategray279768}{RGB}{27,97,68}
\definecolor{lightgray204}{RGB}{204,204,204}
\definecolor{midnightblue263565}{RGB}{26,35,65}
\definecolor{olivedrab10412248}{RGB}{104,122,48}
\definecolor{plum205162224}{RGB}{205,162,224}
\definecolor{rosybrown199122124}{RGB}{199,122,124}

\begin{axis}[
legend cell align={left},
legend style={fill opacity=0.5, draw opacity=1, text opacity=1, draw=lightgray204},
tick align=outside,
tick pos=left,
x grid style={darkgray176},
xlabel={\(\displaystyle \kappa\)},
xmin=0.275, xmax=5.225,
xtick style={color=black},
y grid style={darkgray176},
ylabel={\(\displaystyle C(\kappa)\)},
ymin=-0.01, ymax=0.65,
ytick style={color=black}
]
\addplot [line width=0.24pt, midnightblue263565, forget plot]
table {%
0.5 0.507553406910983
0.8 0.489941107350903
1.1 0.478342393590118
1.4 0.455766347214535
1.7 0.447182931378907
2 0.427636220727838
2.3 0.426041527061945
2.6 0.412094165232457
2.9 0.393183605907466
3.2 0.360985102482508
3.5 0.33796866262167
3.8 0.329841668786658
4.1 0.280186728866341
4.4 0.278607017235809
4.7 0.246069111732309
5 0.213228861159454
};
\addplot [very thick, midnightblue263565, dotted]
table {%
0.5 0.5214
0.8 0.5038
1.1 0.4922
1.4 0.4696
1.7 0.461
2 0.4414
2.3 0.4398
2.6 0.4258
2.9 0.4068
3.2 0.3744
3.5 0.3512
3.8 0.343
4.1 0.2928
4.4 0.2912
4.7 0.2582
5 0.2248
};
\addlegendentry{$\theta$ = 0.55}
\addplot [line width=0.24pt, midnightblue263565, forget plot]
table {%
0.5 0.535246593089016
0.8 0.517658892649097
1.1 0.506057606409882
1.4 0.483433652785465
1.7 0.474817068621093
2 0.455163779272162
2.3 0.453558472938055
2.6 0.439505834767543
2.9 0.420416394092534
3.2 0.387814897517491
3.5 0.36443133737833
3.8 0.356158331213342
4.1 0.305413271133659
4.4 0.30379298276419
4.7 0.270330888267691
5 0.236371138840546
};
\addplot [line width=0.24pt, darkslategray279768, forget plot]
table {%
0.5 0.433218788546721
0.8 0.392586771779655
1.1 0.363368013749143
1.4 0.339158339799074
1.7 0.30311327201808
2 0.284136829670134
2.3 0.253555085766115
2.6 0.220692703591486
2.9 0.190283566988786
3.2 0.15373648475034
3.5 0.119321018020666
3.8 0.100168761728512
4.1 0.0778451025303489
4.4 0.0627370367247733
4.7 0.0505736504670225
5 0.0293481677734904
};
\addplot [very thick, darkslategray279768, dash pattern=on 1pt off 10pt]
table {%
0.5 0.447
0.8 0.4062
1.1 0.3768
1.4 0.3524
1.7 0.316
2 0.2968
2.3 0.2658
2.6 0.2324
2.9 0.2014
3.2 0.164
3.5 0.1286
3.8 0.1088
4.1 0.0856
4.4 0.0698
4.7 0.057
5 0.0344
};
\addlegendentry{$\theta$ = 0.60}
\addplot [line width=0.24pt, darkslategray279768, forget plot]
table {%
0.5 0.460781211453279
0.8 0.419813228220345
1.1 0.390231986250857
1.4 0.365641660200926
1.7 0.32888672798192
2 0.309463170329866
2.3 0.278044914233885
2.6 0.244107296408514
2.9 0.212516433011214
3.2 0.17426351524966
3.5 0.137878981979334
3.8 0.117431238271488
4.1 0.093354897469651
4.4 0.0768629632752266
4.7 0.0634263495329774
5 0.0394518322265095
};
\addplot [line width=0.24pt, olivedrab10412248, forget plot]
table {%
0.5 0.401541486898377
0.8 0.331625272472374
1.1 0.29579635224427
1.4 0.252569821130204
1.7 0.207536462828603
2 0.158807512938897
2.3 0.122423545185775
2.6 0.0847573979900034
2.9 0.0625464046764221
3.2 0.0416982033856799
3.5 0.0225072798084011
3.8 0.0137756170957251
4.1 0.0077613623752229
4.4 0.0060405611840092
4.7 0.0051927497353204
5 0.0013353251653694
};
\addplot [very thick, olivedrab10412248, dashed]
table {%
0.5 0.4152
0.8 0.3448
1.1 0.3086
1.4 0.2648
1.7 0.219
2 0.1692
2.3 0.1318
2.6 0.0928
2.9 0.0696
3.2 0.0476
3.5 0.027
3.8 0.0174
4.1 0.0106
4.4 0.0086
4.7 0.0076
5 0.0028
};
\addlegendentry{$\theta$ = 0.65}
\addplot [line width=0.24pt, olivedrab10412248, forget plot]
table {%
0.5 0.428858513101623
0.8 0.357974727527626
1.1 0.32140364775573
1.4 0.277030178869796
1.7 0.230463537171397
2 0.179592487061103
2.3 0.141176454814225
2.6 0.100842602009996
2.9 0.0766535953235778
3.2 0.05350179661432
3.5 0.0314927201915988
3.8 0.0210243829042748
4.1 0.013438637624777
4.4 0.0111594388159907
4.7 0.0100072502646796
5 0.0042646748346305
};
\addplot [line width=0.24pt, rosybrown199122124, forget plot]
table {%
0.5 0.378865430356424
0.8 0.286507449197454
1.1 0.228555116901007
1.4 0.16720195684236
1.7 0.100747909059378
2 0.0652167542881508
2.3 0.0411336705581769
2.6 0.0203063167945457
2.9 0.0105663492552225
3.2 0.0045220516065547
3.5 0.001188462484806
3.8 0.0002403763137562
4.1 1.63136627451006e-05
4.4 -0.0001919607980398
4.7 -7.87601968294841e-05
5 -0.0001919607980398
};
\addplot [very thick, rosybrown199122124, dash pattern=on 3pt off 4pt on 2pt off 4pt on 2pt off 4pt]
table {%
0.5 0.3924
0.8 0.2992
1.1 0.2404
1.4 0.1778
1.7 0.1094
2 0.0724
2.3 0.047
2.6 0.0246
2.9 0.0138
3.2 0.0068
3.5 0.0026
3.8 0.0012
4.1 0.0008
4.4 0.0002
4.7 0.0006
5 0.0002
};
\addlegendentry{$\theta$ = 0.70}
\addplot [line width=0.24pt, rosybrown199122124, forget plot]
table {%
0.5 0.405934569643576
0.8 0.311892550802546
1.1 0.252244883098993
1.4 0.18839804315764
1.7 0.118052090940622
2 0.0795832457118492
2.3 0.052866329441823
2.6 0.0288936832054542
2.9 0.0170336507447774
3.2 0.0090779483934452
3.5 0.0040115375151939
3.8 0.0021596236862437
4.1 0.0015836863372548
4.4 0.0005919607980398
4.7 0.0012787601968294
5 0.0005919607980398
};
\addplot [line width=0.24pt, plum205162224, forget plot]
table {%
0.5 0.33737385842338
0.8 0.239768832199524
1.1 0.159978259955804
1.4 0.090143888112386
1.7 0.0450906727686921
2 0.0173887423203189
2.3 0.0068972118786704
2.6 0.0028842145456236
2.9 0.0013353251653694
3.2 -7.87601968294841e-05
3.5 -0.0001542608310173
3.8 -0.0001919607980398
4.1 -0.0001919607980398
4.4 0
4.7 -0.0001542608310173
5 -0.0001919607980398
};
\addplot [very thick, plum205162224, dash pattern=on 1pt off 6pt]
table {%
0.5 0.3506
0.8 0.2518
1.1 0.1704
1.4 0.0984
1.7 0.0512
2 0.0214
2.3 0.0096
2.6 0.0048
2.9 0.0028
3.2 0.0006
3.5 0.0004
3.8 0.0002
4.1 0.0002
4.4 0
4.7 0.0004
5 0.0002
};
\addlegendentry{$\theta$ = 0.75}
\addplot [line width=0.24pt, plum205162224, forget plot]
table {%
0.5 0.36382614157662
0.8 0.263831167800476
1.1 0.180821740044196
1.4 0.106656111887614
1.7 0.0573093272313078
2 0.025411257679681
2.3 0.0123027881213295
2.6 0.0067157854543763
2.9 0.0042646748346305
3.2 0.0012787601968294
3.5 0.0009542608310173
3.8 0.0005919607980398
4.1 0.0005919607980398
4.4 0
4.7 0.0009542608310173
5 0.0005919607980398
};
\end{axis}

\end{tikzpicture}\label{fig:con_rel}}
\subfloat[Level of discordance]{
\begin{tikzpicture}[scale = 0.95]

\definecolor{darkgray176}{RGB}{176,176,176}
\definecolor{darkslategray279768}{RGB}{27,97,68}
\definecolor{lightgray204}{RGB}{204,204,204}
\definecolor{midnightblue263565}{RGB}{26,35,65}
\definecolor{olivedrab10412248}{RGB}{104,122,48}
\definecolor{plum205162224}{RGB}{205,162,224}
\definecolor{rosybrown199122124}{RGB}{199,122,124}

\begin{axis}[
legend cell align={left},
legend style={
  fill opacity=0.8,
  draw opacity=1,
  text opacity=1,
  at={(0.03,0.97)},
  anchor=north west,
  draw=lightgray204
},
tick align=outside,
tick pos=left,
title={Discordance and reliability},
x grid style={darkgray176},
xlabel={\(\displaystyle \kappa\)},
xmin=0.275, xmax=5.225,
xtick style={color=black},
y grid style={darkgray176},
ylabel={\(\displaystyle \delta(\kappa)\)},
ymin=0.06, ymax=0.6,
ytick style={color=black}
]
\addplot [line width=0.24pt, midnightblue263565, forget plot]
table {%
0.5 0.0717638731002808
0.799999952316284 0.0744373798370361
1.10000002384186 0.0777894258499146
1.39999997615814 0.0829442739486694
1.70000004768372 0.0854583978652954
2 0.0892065763473511
2.29999995231628 0.0918335914611816
2.59999990463257 0.0963326692581177
2.90000009536743 0.100042939186096
3.20000004768372 0.108678221702576
3.5 0.116244196891785
3.79999995231628 0.12116539478302
4.09999990463257 0.131837964057922
4.40000009536743 0.135681629180908
4.69999980926514 0.146157026290894
5 0.15556263923645
};
\addplot [very thick, midnightblue263565, dotted]
table {%
0.5 0.0740699768066406
0.799999952316284 0.0767250061035156
1.10000002384186 0.0801249742507935
1.39999997615814 0.0853350162506104
1.70000004768372 0.0878900289535522
2 0.0916500091552734
2.29999995231628 0.0943449735641479
2.59999990463257 0.0989099740982056
2.90000009536743 0.102625012397766
3.20000004768372 0.111330032348633
3.5 0.118970036506653
3.79999995231628 0.123975038528442
4.09999990463257 0.134609937667847
4.40000009536743 0.138565063476562
4.69999980926514 0.149075031280518
5 0.158480048179626
};
\addlegendentry{$\theta$ = 0.55}
\addplot [line width=0.24pt, midnightblue263565, forget plot]
table {%
0.5 0.0763760805130005
0.799999952316284 0.0790125131607056
1.10000002384186 0.0824606418609619
1.39999997615814 0.0877257585525513
1.70000004768372 0.0903216600418091
2 0.0940933227539062
2.29999995231628 0.0968564748764038
2.59999990463257 0.101487278938293
2.90000009536743 0.105207085609436
3.20000004768372 0.1139817237854
3.5 0.121695756912231
3.79999995231628 0.126784563064575
4.09999990463257 0.137382030487061
4.40000009536743 0.141448259353638
4.69999980926514 0.151993036270142
5 0.161397457122803
};
\addplot [line width=0.24pt, darkslategray279768, forget plot]
table {%
0.5 0.0865178108215332
0.799999952316284 0.0937660932540894
1.10000002384186 0.103972554206848
1.39999997615814 0.110636591911316
1.70000004768372 0.120800614356995
2 0.130080580711365
2.29999995231628 0.141998410224915
2.59999990463257 0.15585732460022
2.90000009536743 0.166937232017517
3.20000004768372 0.182347297668457
3.5 0.198715567588806
3.79999995231628 0.208827018737793
4.09999990463257 0.222300291061401
4.40000009536743 0.233340501785278
4.69999980926514 0.246692776679993
5 0.266806840896606
};
\addplot [very thick, darkslategray279768, dash pattern=on 1pt off 10pt]
table {%
0.5 0.0889400243759155
0.799999952316284 0.0961649417877197
1.10000002384186 0.106505036354065
1.39999997615814 0.11321496963501
1.70000004768372 0.123425006866455
2 0.132830023765564
2.29999995231628 0.144824981689453
2.59999990463257 0.158779978752136
2.90000009536743 0.169875025749207
3.20000004768372 0.185289978981018
3.5 0.201640009880066
3.79999995231628 0.211770057678223
4.09999990463257 0.22522497177124
4.40000009536743 0.236269950866699
4.69999980926514 0.249639987945557
5 0.269705057144165
};
\addlegendentry{$\theta$ = 0.60}
\addplot [line width=0.24pt, darkslategray279768, forget plot]
table {%
0.5 0.0913622379302979
0.799999952316284 0.0985639095306396
1.10000002384186 0.109037518501282
1.39999997615814 0.115793347358704
1.70000004768372 0.126049399375916
2 0.135579466819763
2.29999995231628 0.147651553153992
2.59999990463257 0.161702752113342
2.90000009536743 0.172812700271606
3.20000004768372 0.188232779502869
3.5 0.204564452171326
3.79999995231628 0.214712977409363
4.09999990463257 0.228149652481079
4.40000009536743 0.23919951915741
4.69999980926514 0.252587199211121
5 0.272603034973145
};
\addplot [line width=0.24pt, olivedrab10412248, forget plot]
table {%
0.5 0.0937927961349487
0.799999952316284 0.109345436096191
1.10000002384186 0.123399019241333
1.39999997615814 0.139783382415771
1.70000004768372 0.158194184303284
2 0.182974100112915
2.29999995231628 0.199358344078064
2.59999990463257 0.223380923271179
2.90000009536743 0.238003253936768
3.20000004768372 0.260830163955688
3.5 0.283243179321289
3.79999995231628 0.296461462974548
4.09999990463257 0.31390905380249
4.40000009536743 0.325154542922974
4.69999980926514 0.335617303848267
5 0.350788831710815
};
\addplot [very thick, olivedrab10412248, dashed]
table {%
0.5 0.0962749719619751
0.799999952316284 0.11183500289917
1.10000002384186 0.126055002212524
1.39999997615814 0.142549991607666
1.70000004768372 0.161059975624084
2 0.185950040817261
2.29999995231628 0.202314972877502
2.59999990463257 0.226359963417053
2.90000009536743 0.240949988365173
3.20000004768372 0.263749957084656
3.5 0.28608500957489
3.79999995231628 0.299279928207397
4.09999990463257 0.316650032997131
4.40000009536743 0.32790994644165
4.69999980926514 0.338364958763123
5 0.353404998779297
};
\addlegendentry{$\theta$ = 0.65}
\addplot [line width=0.24pt, olivedrab10412248, forget plot]
table {%
0.5 0.098757266998291
0.799999952316284 0.114324569702148
1.10000002384186 0.128710985183716
1.39999997615814 0.145316600799561
1.70000004768372 0.163925766944885
2 0.188925862312317
2.29999995231628 0.205271601676941
2.59999990463257 0.229339122772217
2.90000009536743 0.243896722793579
3.20000004768372 0.266669750213623
3.5 0.288926839828491
3.79999995231628 0.302098512649536
4.09999990463257 0.319390892982483
4.40000009536743 0.330665469169617
4.69999980926514 0.341112613677979
5 0.356021165847778
};
\addplot [line width=0.24pt, rosybrown199122124, forget plot]
table {%
0.5 0.0994566679000854
0.799999952316284 0.123502373695374
1.10000002384186 0.146190881729126
1.39999997615814 0.175413489341736
1.70000004768372 0.2100830078125
2 0.240235447883606
2.29999995231628 0.267345786094666
2.59999990463257 0.291424751281738
2.90000009536743 0.309791922569275
3.20000004768372 0.33799934387207
3.5 0.359714388847351
4.09999990463257 0.379887580871582
4.40000009536743 0.394063949584961
4.69999980926514 0.401486992835999
5 0.409619569778442
};
\addplot [very thick, rosybrown199122124, dash pattern=on 3pt off 4pt on 2pt off 4pt on 2pt off 4pt]
table {%
0.5 0.101954936981201
0.799999952316284 0.126100063323975
1.10000002384186 0.148929953575134
1.39999997615814 0.178315043449402
1.70000004768372 0.213034987449646
2 0.243209958076477
2.29999995231628 0.270305037498474
2.59999990463257 0.294309973716736
2.90000009536743 0.312559962272644
3.20000004768372 0.340754985809326
3.5 0.362360000610352
4.09999990463257 0.382465004920959
4.40000009536743 0.396630048751831
4.69999980926514 0.404034972190857
5 0.412085056304932
};
\addlegendentry{$\theta$ = 0.70}
\addplot [line width=0.24pt, rosybrown199122124, forget plot]
table {%
0.5 0.104453325271606
0.799999952316284 0.128697633743286
1.10000002384186 0.151669144630432
1.39999997615814 0.181216478347778
1.70000004768372 0.215986967086792
2 0.246184587478638
2.29999995231628 0.273264169692993
2.59999990463257 0.297195196151733
2.90000009536743 0.315328121185303
3.20000004768372 0.343510627746582
3.5 0.365005612373352
4.09999990463257 0.385042428970337
4.40000009536743 0.399196147918701
4.69999980926514 0.406583070755005
5 0.414550542831421
};
\addplot [line width=0.24pt, plum205162224, forget plot]
table {%
0.5 0.10994017124176
0.799999952316284 0.140107154846191
1.10000002384186 0.177322506904602
1.39999997615814 0.220106482505798
1.70000004768372 0.263160109519958
2 0.302819728851318
2.29999995231628 0.327584862709045
2.59999990463257 0.355931282043457
2.90000009536743 0.368016481399536
3.20000004768372 0.399762272834778
3.5 0.415764570236206
3.79999995231628 0.424283504486084
4.09999990463257 0.430603861808777
4.40000009536743 0.437280654907227
5 0.450294375419617
};
\addplot [very thick, plum205162224, dash pattern=on 1pt off 6pt]
table {%
0.5 0.112529993057251
0.799999952316284 0.142794966697693
1.10000002384186 0.18020498752594
1.70000004768372 0.266149997711182
2 0.305709958076477
2.29999995231628 0.330430030822754
2.59999990463257 0.358615040779114
2.90000009536743 0.370700001716614
3.20000004768372 0.402330040931702
3.5 0.418305039405823
3.79999995231628 0.42679500579834
4.09999990463257 0.433084964752197
4.40000009536743 0.439729928970337
4.69999980926514 0.446069955825806
5 0.452679991722107
};
\addlegendentry{$\theta$ = 0.75}
\addplot [line width=0.24pt, plum205162224, forget plot]
table {%
0.5 0.115119934082031
0.799999952316284 0.145482897758484
1.10000002384186 0.183087468147278
1.70000004768372 0.269139885902405
2 0.308600187301636
2.29999995231628 0.333275079727173
2.59999990463257 0.361298799514771
2.90000009536743 0.373383522033691
3.20000004768372 0.404897689819336
3.5 0.42084538936615
3.79999995231628 0.429306507110596
4.09999990463257 0.435566186904907
4.40000009536743 0.442179441452026
4.69999980926514 0.448456287384033
5 0.455065608024597
};
\end{axis}

\end{tikzpicture}\label{fig:dis_rel}}
    \caption{Results of the sensitivity analysis inspecting the effect of the reliability of the opinions $\theta_0 = \theta_1$.}
    \label{fig:SA_rel}
\end{figure}

To elucidate the effect of the level of reliability of the opinions we choose these $\theta_0=\theta_1\in\{0.55, 0.60, 0.65,0.70, 0.75\}$. We plot the probability of consensus for these settings in Figure~\ref{fig:con_rel} and the level of discordance in Figure~\ref{fig:dis_rel}. We see that a greater reliability of both opinions lead to less consensus and more discordance. This is conceivable as when the reliability is greater, convergence of belief to this true parameter allows for neighbourhoods with more disagreement (and thus greater $\TTC$). As the reliability increases agents are less dependent on their network to agree with them (thus decreasing $\TTC$) in order for their belief of an opinion to converge. 

\subsection{Warm-up length}
\begin{figure}[h]
\subfloat[Probability of consensus]{
\begin{tikzpicture}[scale = 0.95]

\definecolor{darkgray176}{RGB}{176,176,176}
\definecolor{darkslategray279768}{RGB}{27,97,68}
\definecolor{lightgray204}{RGB}{204,204,204}
\definecolor{midnightblue263565}{RGB}{26,35,65}
\definecolor{olivedrab10412248}{RGB}{104,122,48}
\definecolor{rosybrown199122124}{RGB}{199,122,124}

\begin{axis}[
legend cell align={left},
legend style={fill opacity=0.8, draw opacity=1, text opacity=1, draw=lightgray204},
tick align=outside,
tick pos=left,
x grid style={darkgray176},
xlabel={\(\displaystyle \kappa\)},
xmin=0.275, xmax=5.225,
xtick style={color=black},
y grid style={darkgray176},
ylabel={\(\displaystyle C(\kappa)\)},
ymin=-0.0145699256893526, ymax=0.492030856788483,
ytick style={color=black}
]
\addplot [line width=0.24pt, midnightblue263565, forget plot]
table {%
0.5 0.441396474838257
0.799999952316284 0.414285182952881
1.10000002384186 0.398755073547363
1.39999997615814 0.35959529876709
1.70000004768372 0.339554905891418
2 0.309642314910889
2.29999995231628 0.282556653022766
2.59999990463257 0.250993609428406
2.90000009536743 0.237997531890869
3.20000004768372 0.200082421302795
3.5 0.152176856994629
3.79999995231628 0.128439903259277
4.09999990463257 0.115445494651794
4.40000009536743 0.0901439189910889
4.69999980926514 0.0631183385848999
5 0.044901967048645
};
\addplot [semithick, midnightblue263565, dotted]
table {%
0.5 0.455199956893921
0.799999952316284 0.427999973297119
1.10000002384186 0.412400007247925
1.39999997615814 0.373000025749207
1.70000004768372 0.352800011634827
2 0.32260000705719
2.29999995231628 0.295199990272522
2.59999990463257 0.263200044631958
2.90000009536743 0.25
3.20000004768372 0.211400032043457
3.5 0.162400007247925
3.79999995231628 0.138000011444092
4.09999990463257 0.124600052833557
4.40000009536743 0.0983999967575073
4.69999980926514 0.0701999664306641
5 0.0509999990463257
};
\addlegendentry{$t_s = 0$}
\addplot [line width=0.24pt, midnightblue263565, forget plot]
table {%
0.5 0.469003558158875
0.799999952316284 0.441714882850647
1.10000002384186 0.426044940948486
1.39999997615814 0.386404752731323
1.70000004768372 0.366045117378235
2 0.335557699203491
2.29999995231628 0.307843327522278
2.59999990463257 0.275406360626221
2.90000009536743 0.262002468109131
3.20000004768372 0.222717523574829
3.5 0.172623157501221
3.79999995231628 0.147560119628906
4.09999990463257 0.13375449180603
4.40000009536743 0.106656074523926
4.69999980926514 0.0772815942764282
5 0.0570980310440063
};
\addplot [line width=0.24pt, darkslategray279768, forget plot]
table {%
0.5 0.429031729698181
0.799999952316284 0.377871513366699
1.10000002384186 0.345703125
1.39999997615814 0.31696629524231
1.70000004768372 0.29638946056366
2 0.252963900566101
2.29999995231628 0.225802659988403
2.59999990463257 0.195573687553406
2.90000009536743 0.172086715698242
3.20000004768372 0.123393416404724
3.5 0.101134061813354
3.79999995231628 0.0774616003036499
4.09999990463257 0.0636904239654541
4.40000009536743 0.0469790697097778
4.69999980926514 0.0351294279098511
5 0.0256410837173462
};
\addplot [semithick, darkslategray279768, dash pattern=on 1pt off 10pt]
table {%
0.5 0.442800045013428
0.799999952316284 0.39139997959137
1.10000002384186 0.358999967575073
1.39999997615814 0.330000042915344
1.70000004768372 0.309200048446655
2 0.265200018882751
2.29999995231628 0.237599968910217
2.59999990463257 0.206799983978271
2.90000009536743 0.182800054550171
3.20000004768372 0.132799983024597
3.5 0.109799981117249
3.79999995231628 0.0851999521255493
4.09999990463257 0.0707999467849731
4.40000009536743 0.0532000064849854
4.69999980926514 0.040600061416626
5 0.0304000377655029
};
\addlegendentry{$t_s = 10$}
\addplot [line width=0.24pt, darkslategray279768, forget plot]
table {%
0.5 0.456568241119385
0.799999952316284 0.40492844581604
1.10000002384186 0.372296810150146
1.39999997615814 0.343033671379089
1.70000004768372 0.322010517120361
2 0.277436017990112
2.29999995231628 0.249397397041321
2.59999990463257 0.218026280403137
2.90000009536743 0.19351327419281
3.20000004768372 0.14220654964447
3.5 0.118465900421143
3.79999995231628 0.0929384231567383
4.09999990463257 0.0779095888137817
4.40000009536743 0.0594209432601929
4.69999980926514 0.0460705757141113
5 0.0351588726043701
};
\addplot [line width=0.24pt, olivedrab10412248, forget plot]
table {%
0.5 0.419664144515991
0.799999952316284 0.363566637039185
1.10000002384186 0.346893429756165
1.39999997615814 0.293028831481934
1.70000004768372 0.260059833526611
2 0.229931592941284
2.29999995231628 0.196357607841492
2.59999990463257 0.166811347007751
2.90000009536743 0.140102505683899
3.20000004768372 0.0992037057876587
3.5 0.0805313587188721
3.79999995231628 0.060450553894043
4.09999990463257 0.0409455299377441
4.40000009536743 0.0347553491592407
4.69999980926514 0.0210388898849487
5 0.0132375955581665
};
\addplot [semithick, olivedrab10412248, dashed]
table {%
0.5 0.43340003490448
0.799999952316284 0.376999974250793
1.10000002384186 0.360199928283691
1.39999997615814 0.305799961090088
1.70000004768372 0.27240002155304
2 0.24179995059967
2.29999995231628 0.207599997520447
2.59999990463257 0.17739999294281
2.90000009536743 0.149999976158142
3.20000004768372 0.107800006866455
3.5 0.0884000062942505
3.79999995231628 0.0673999786376953
4.09999990463257 0.0468000173568726
4.40000009536743 0.0401999950408936
4.69999980926514 0.0254000425338745
5 0.0168000459671021
};
\addlegendentry{$t_s = 20$}
\addplot [line width=0.24pt, olivedrab10412248, forget plot]
table {%
0.5 0.447135806083679
0.799999952316284 0.390433430671692
1.10000002384186 0.373506546020508
1.39999997615814 0.318571209907532
1.70000004768372 0.284740209579468
2 0.253668308258057
2.29999995231628 0.218842387199402
2.59999990463257 0.187988638877869
2.90000009536743 0.159897565841675
3.20000004768372 0.116396307945251
3.5 0.0962686538696289
4.09999990463257 0.052654504776001
4.40000009536743 0.0456447601318359
4.69999980926514 0.0297611951828003
5 0.020362377166748
};
\addplot [line width=0.24pt, rosybrown199122124, forget plot]
table {%
0.5 0.413488388061523
0.799999952316284 0.359396696090698
1.10000002384186 0.324690222740173
1.39999997615814 0.276830077171326
1.70000004768372 0.249417543411255
2 0.211461901664734
2.29999995231628 0.178147912025452
2.59999990463257 0.146721124649048
2.90000009536743 0.111379623413086
3.20000004768372 0.0886039733886719
3.5 0.0640718936920166
3.79999995231628 0.0422630310058594
4.09999990463257 0.0280483961105347
4.40000009536743 0.0199404954910278
5 0.00845742225646973
};
\addplot [semithick, rosybrown199122124, dash pattern=on 3pt off 4pt on 2pt off 4pt on 2pt off 4pt]
table {%
0.5 0.427199959754944
0.799999952316284 0.37279999256134
1.10000002384186 0.337800025939941
1.39999997615814 0.289399981498718
1.70000004768372 0.261600017547607
2 0.223000049591064
2.29999995231628 0.189000010490417
2.59999990463257 0.156800031661987
2.90000009536743 0.120399951934814
3.20000004768372 0.0967999696731567
3.5 0.0712000131607056
3.79999995231628 0.0482000112533569
4.09999990463257 0.0329999923706055
4.40000009536743 0.0241999626159668
4.69999980926514 0.0180000066757202
5 0.0113999843597412
};
\addlegendentry{$t_s = 30$}
\addplot [line width=0.24pt, rosybrown199122124, forget plot]
table {%
0.5 0.440911531448364
0.799999952316284 0.386203289031982
1.10000002384186 0.35090970993042
1.39999997615814 0.3019700050354
1.70000004768372 0.27378249168396
2 0.234538078308105
2.29999995231628 0.199852108955383
2.59999990463257 0.166878819465637
2.90000009536743 0.129420399665833
3.20000004768372 0.104995965957642
3.5 0.0783281326293945
3.79999995231628 0.0541369915008545
4.09999990463257 0.0379515886306763
4.40000009536743 0.0284595489501953
4.69999980926514 0.0216852426528931
5 0.0143426656723022
};
\end{axis}

\end{tikzpicture}\label{fig:con_WU}}
\subfloat[Level of discordance]{
\begin{tikzpicture}[scale = 0.95]

\definecolor{darkgray176}{RGB}{176,176,176}
\definecolor{darkslategray279768}{RGB}{27,97,68}
\definecolor{lightgray204}{RGB}{204,204,204}
\definecolor{midnightblue263565}{RGB}{26,35,65}
\definecolor{olivedrab10412248}{RGB}{104,122,48}
\definecolor{rosybrown199122124}{RGB}{199,122,124}

\begin{axis}[
legend cell align={left},
legend style={
  fill opacity=0.8,
  draw opacity=1,
  text opacity=1,
  at={(0.03,0.97)},
  anchor=north west,
  draw=lightgray204
},
tick align=outside,
tick pos=left,
x grid style={darkgray176},
xlabel={\(\displaystyle \kappa\)},
xmin=0.275, xmax=5.225,
xtick style={color=black},
y grid style={darkgray176},
ylabel={\(\displaystyle \delta(\kappa)\)},
ymin=0.0734784539720718, ymax=0.334537407116523,
ytick style={color=black}
]
\addplot [line width=0.24pt, midnightblue263565, forget plot]
table {%
0.5 0.0853447914123535
0.799999952316284 0.0898114442825317
1.10000002384186 0.0940989255905151
1.39999997615814 0.104550957679749
1.70000004768372 0.111563444137573
2 0.120487332344055
2.29999995231628 0.130328774452209
2.59999990463257 0.142777919769287
2.90000009536743 0.149377822875977
3.20000004768372 0.16299033164978
3.5 0.18084716796875
3.79999995231628 0.188945651054382
4.09999990463257 0.201463580131531
4.40000009536743 0.216299533843994
4.69999980926514 0.231646060943604
5 0.244739890098572
};
\addplot [very thick, midnightblue263565, dotted]
table {%
0.5 0.0877749919891357
1.10000002384186 0.0965399742126465
1.39999997615814 0.107079982757568
1.70000004768372 0.11416494846344
2 0.123139977455139
2.29999995231628 0.133080005645752
2.59999990463257 0.145614981651306
2.90000009536743 0.152264952659607
3.20000004768372 0.16592001914978
3.5 0.183794975280762
3.79999995231628 0.191844940185547
4.09999990463257 0.20441997051239
4.40000009536743 0.219274997711182
4.69999980926514 0.234560012817383
5 0.24762499332428
};
\addlegendentry{$t_s = 0$}
\addplot [line width=0.24pt, midnightblue263565, forget plot]
table {%
0.5 0.090205192565918
1.10000002384186 0.0989811420440674
1.39999997615814 0.109609127044678
1.70000004768372 0.116766571998596
2 0.125792622566223
2.29999995231628 0.135831236839294
2.59999990463257 0.148452043533325
2.90000009536743 0.155152201652527
3.20000004768372 0.168849587440491
3.5 0.186742782592773
3.79999995231628 0.194744348526001
4.09999990463257 0.207376480102539
4.40000009536743 0.222250461578369
4.69999980926514 0.237473964691162
5 0.250510096549988
};
\addplot [line width=0.24pt, darkslategray279768, forget plot]
table {%
0.5 0.0874271392822266
0.799999952316284 0.0984716415405273
1.10000002384186 0.106120347976685
1.39999997615814 0.116323471069336
1.70000004768372 0.125488638877869
2 0.140403032302856
2.29999995231628 0.149804472923279
2.59999990463257 0.16381561756134
2.90000009536743 0.175900340080261
3.20000004768372 0.196629643440247
3.5 0.213377475738525
3.79999995231628 0.225793480873108
4.09999990463257 0.23554253578186
4.40000009536743 0.24803352355957
4.69999980926514 0.263640880584717
5 0.277959823608398
};
\addplot [very thick, darkslategray279768, dash pattern=on 1pt off 10pt]
table {%
0.5 0.0898799896240234
0.799999952316284 0.100924968719482
1.10000002384186 0.108615040779114
1.39999997615814 0.118924975395203
1.70000004768372 0.12819504737854
2 0.143180012702942
2.29999995231628 0.152629971504211
2.59999990463257 0.166705012321472
2.90000009536743 0.1788649559021
3.20000004768372 0.199574947357178
3.5 0.216354966163635
3.79999995231628 0.228744983673096
4.09999990463257 0.238510012626648
4.40000009536743 0.250930070877075
4.69999980926514 0.266540050506592
5 0.280849933624268
};
\addlegendentry{$t_s = 10$}
\addplot [line width=0.24pt, darkslategray279768, forget plot]
table {%
0.5 0.0923328399658203
0.799999952316284 0.103378415107727
1.10000002384186 0.111109614372253
1.39999997615814 0.121526479721069
1.70000004768372 0.130901336669922
2 0.145956993103027
2.29999995231628 0.155455589294434
2.59999990463257 0.169594407081604
2.90000009536743 0.181829690933228
3.20000004768372 0.202520370483398
3.5 0.219332456588745
3.79999995231628 0.231696605682373
4.09999990463257 0.241477489471436
4.40000009536743 0.253826498985291
4.69999980926514 0.269439101219177
5 0.283740282058716
};
\addplot [line width=0.24pt, olivedrab10412248, forget plot]
table {%
0.5 0.0897181034088135
0.799999952316284 0.101224422454834
1.10000002384186 0.108378887176514
1.39999997615814 0.122557401657104
1.70000004768372 0.135172843933105
2 0.149186730384827
2.29999995231628 0.162156581878662
2.59999990463257 0.178186774253845
2.90000009536743 0.194465160369873
3.20000004768372 0.21632719039917
3.5 0.22989022731781
3.79999995231628 0.244478106498718
4.09999990463257 0.26056432723999
4.40000009536743 0.273540616035461
4.69999980926514 0.287638425827026
5 0.302013635635376
};
\addplot [very thick, olivedrab10412248, dashed]
table {%
0.5 0.0921750068664551
0.799999952316284 0.103675007820129
1.10000002384186 0.110934972763062
1.39999997615814 0.125174999237061
1.70000004768372 0.137899994850159
2 0.152024984359741
2.29999995231628 0.165019989013672
2.59999990463257 0.181139945983887
2.90000009536743 0.197499990463257
3.20000004768372 0.219334959983826
3.5 0.23291003704071
3.79999995231628 0.24745500087738
4.09999990463257 0.263484954833984
4.40000009536743 0.276520013809204
4.69999980926514 0.290509939193726
5 0.304834961891174
};
\addlegendentry{$t_s = 20$}
\addplot [line width=0.24pt, olivedrab10412248, forget plot]
table {%
0.5 0.0946319103240967
0.799999952316284 0.106125593185425
1.10000002384186 0.113491177558899
1.39999997615814 0.127792596817017
1.70000004768372 0.140627145767212
2 0.154863357543945
2.29999995231628 0.167883396148682
2.59999990463257 0.184093236923218
2.90000009536743 0.200534820556641
3.20000004768372 0.222342848777771
3.5 0.235929727554321
3.79999995231628 0.250431895256042
4.09999990463257 0.266405701637268
4.40000009536743 0.279499411582947
4.69999980926514 0.293381571769714
5 0.307656288146973
};
\addplot [line width=0.24pt, rosybrown199122124, forget plot]
table {%
0.5 0.0915148258209229
0.799999952316284 0.101216673851013
1.10000002384186 0.111532807350159
1.39999997615814 0.125951051712036
1.70000004768372 0.138873100280762
2 0.154826879501343
2.29999995231628 0.17229151725769
2.59999990463257 0.189432978630066
2.90000009536743 0.207094669342041
3.20000004768372 0.224441051483154
3.5 0.243963122367859
3.79999995231628 0.260833859443665
4.09999990463257 0.279519557952881
4.40000009536743 0.290388822555542
4.69999980926514 0.305586576461792
5 0.317048907279968
};
\addplot [very thick, rosybrown199122124, dash pattern=on 3pt off 4pt on 2pt off 4pt on 2pt off 4pt]
table {%
0.5 0.093999981880188
0.799999952316284 0.103649973869324
1.10000002384186 0.114065051078796
1.39999997615814 0.128555059432983
1.70000004768372 0.14162003993988
2 0.157660007476807
2.29999995231628 0.175230026245117
2.59999990463257 0.192425012588501
2.90000009536743 0.210085034370422
3.20000004768372 0.227479934692383
3.5 0.246994972229004
3.79999995231628 0.263790011405945
4.09999990463257 0.28245997428894
4.40000009536743 0.293290019035339
4.69999980926514 0.308480024337769
5 0.319859981536865
};
\addlegendentry{$t_s = 30$}
\addplot [line width=0.24pt, rosybrown199122124, forget plot]
table {%
0.5 0.0964851379394531
0.799999952316284 0.106083273887634
1.10000002384186 0.116597294807434
1.39999997615814 0.131158947944641
1.70000004768372 0.144366979598999
2 0.160493135452271
2.29999995231628 0.178168416023254
2.59999990463257 0.195416927337646
2.90000009536743 0.213075399398804
3.20000004768372 0.230518937110901
3.5 0.250026941299438
3.79999995231628 0.266746163368225
4.09999990463257 0.28540050983429
4.40000009536743 0.296191215515137
4.69999980926514 0.311373353004456
5 0.322671055793762
};
\end{axis}

\end{tikzpicture}\label{fig:dis_WU}}
\caption{Results of the sensitivity analysis inspecting the effect of the warm up period.}
\end{figure}
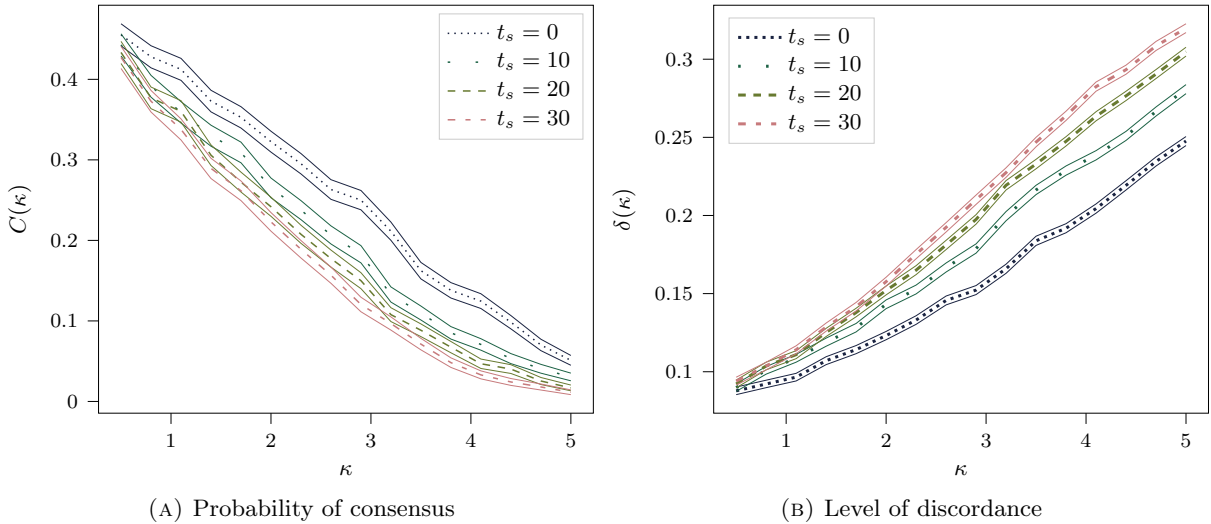

We vary the warm-up length $t_s$ taking values $t_s \in \{0,10,20,30\}$. In Figure~\ref{fig:con_WU}, which plots the probability of consensus for differing warm-up lengths, we see that a shorter warm-up period leads to more consensus. Similarly in Figure~\ref{fig:dis_WU} we see that longer warm-up periods lead to more discordance. This result is conceivable as in the early stages of an interaction with an opinion, the estimated reliability is changing a lot more than toward the end. In other words, having a longer warm-up period allows for an agent's belief distribution to `settle' before having to compete with a network adjusted threshold $\TTC$.

\subsection{Discussion}
We highlight the fact that in the model, the agents are exposed only to attractive forces between each other. That is we follow the assumption of assimilative social influence between agents and yet we have rich results illustrating a range of possible outcomes from polarisation to consensus. We believe that especially for agent-based simulation modellers who would like to include an opinion dynamic within a greater context, this model may prove to be useful because of the diversity of its outcomes which interact in a conceivable way with the parameter settings.

\section{Experiments on framework instance}\label{sec:Experiments}
The sensitivity analysis conducted in \S\ref{sec:SensAn} showcases that the model is capable of a variety of end states and that these interact with the model parameters in a logical way. A strength of agent-based modelling is describing micro-behaviour rules of agents resulting from the combination of their characteristics and their environment (possibly interactions between agents) and subsequently observing the resulting macro behaviour of the population. The strength of our model then is the possibility of modelling agents with different parameters; prior belief distributions, stubbornness parameters, or interpretation of agreement and disagreement with an opinion ($p_j$ and $l_j$ for $j\in V$). This section is devoted to the results of two experiments in which we introduce heterogeneity to the model. We do this by drawing the stubbornness parameter for different agents from a distribution in one experiment, and defining the reliability of one opinion to be greater than the other in another experiment.

As a result of the sensitivity analysis a suitable set of parameters were chosen as the baseline for the experiments presented in \S\ref{sec:Experiments}. The definition of suitable parameters we use are those which present a richness in the type of results that may be obtained in the steady state. We chose a population of $N=30$ agents, connected to their 6 nearest neighbours with a rewiring probability of $w=0.2$. As prior belief parameters we chose $\alpha=4$ and $\beta=2$. This was done in order to keep frivolous switching back and forth between opinions to a minimum which is more likely when the agent's prior estimate is closer to typical threshold values. The opinion's reliability is kept at $\theta=0.6$ as this setting provides a relatively high probability of consensus at low $\kappa$ and reaches a minimum toward the end of our chosen range at $\kappa= 5$. The warm-up period is set to 10 rounds which we believe to strike a good balance between allowing the agents' belief to settle without blocking dynamics completely.
\subsection{Heterogeneous agent stubbornness}
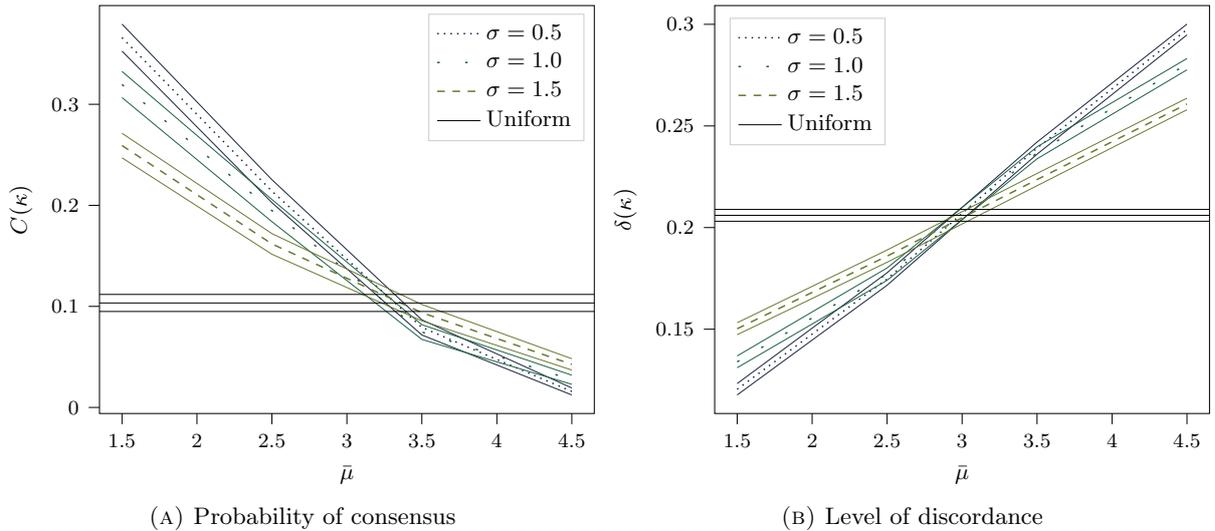
\begin{figure}
    \centering
    \subfloat[Probability of consensus]{
\begin{tikzpicture}[scale = 0.95]

\definecolor{darkgray176}{RGB}{176,176,176}
\definecolor{darkslategray279768}{RGB}{27,97,68}
\definecolor{lightgray204}{RGB}{204,204,204}
\definecolor{midnightblue263565}{RGB}{26,35,65}
\definecolor{olivedrab10412248}{RGB}{104,122,48}

\begin{axis}[
legend cell align={left},
legend style={fill opacity=0.8, draw opacity=1, text opacity=1, draw=lightgray204},
tick align=outside,
tick pos=left,
x grid style={darkgray176},
xlabel={\(\displaystyle \bar{\mu}\)},
xmin=1.35, xmax=4.65,
xtick style={color=black},
y grid style={darkgray176},
ylabel={\(\displaystyle C(\kappa)\)},
ymin=-0.00600697978791352, ymax=0.397702747025542,
ytick style={color=black}
]
\addplot [line width=0.28pt, midnightblue263565, forget plot]
table {%
1.5 0.352647695102343
2.5 0.20263186482839
3.5 0.0717145775513201
4.5 0.0123434623399708
};
\addplot [semithick, midnightblue263565, dotted]
table {%
1.5 0.366
2.5 0.214
3.5 0.0792
4.5 0.0158
};
\addlegendentry{$\sigma = 0.5$}
\addplot [line width=0.28pt, midnightblue263565, forget plot]
table {%
1.5 0.379352304897657
2.5 0.22536813517161
3.5 0.0866854224486798
4.5 0.0192565376600291
};
\addplot [line width=0.28pt, darkslategray279768, forget plot]
table {%
1.5 0.306674224549807
2.5 0.183822146556489
3.5 0.0673170851804514
4.5 0.0228750531409971
};
\addplot [semithick, darkslategray279768, dash pattern=on 1pt off 10pt]
table {%
1.5 0.3196
2.5 0.1948
3.5 0.0746
4.5 0.0274
};
\addlegendentry{$\sigma = 1.0$}
\addplot [line width=0.28pt, darkslategray279768, forget plot]
table {%
1.5 0.332525775450193
2.5 0.205777853443511
3.5 0.0818829148195485
4.5 0.0319249468590028
};
\addplot [line width=0.28pt, olivedrab10412248, forget plot]
table {%
1.5 0.247053838408979
2.5 0.151592152904593
3.5 0.0857186388164369
4.5 0.0370021329422002
};
\addplot [semithick, olivedrab10412248, dashed]
table {%
1.5 0.2592
2.5 0.1618
3.5 0.0938
4.5 0.0426
};
\addlegendentry{$\sigma = 1.5$}
\addplot [line width=0.28pt, olivedrab10412248, forget plot]
table {%
1.5 0.271346161591021
2.5 0.172007847095407
3.5 0.101881361183563
4.5 0.0481978670577997
};
\addplot [very thin, black]
table {%
1.35 0.0949602281653589
4.65 0.0949602281653589
};
\addlegendentry{Uniform}
\addplot [very thin, black, forget plot]
table {%
1.35 0.1034
4.65 0.1034
};
\addplot [very thin, black, forget plot]
table {%
1.35 0.111839771834641
4.65 0.111839771834641
};
\end{axis}

\end{tikzpicture}\label{fig:C_het_kappa}}
    \subfloat[Level of discordance]{
\begin{tikzpicture}[scale = 0.95]

\definecolor{darkgray176}{RGB}{176,176,176}
\definecolor{darkslategray279768}{RGB}{27,97,68}
\definecolor{lightgray204}{RGB}{204,204,204}
\definecolor{midnightblue263565}{RGB}{26,35,65}
\definecolor{olivedrab10412248}{RGB}{104,122,48}

\begin{axis}[
legend cell align={left},
legend style={
  fill opacity=0.8,
  draw opacity=1,
  text opacity=1,
  at={(0.03,0.97)},
  anchor=north west,
  draw=lightgray204
},
tick align=outside,
tick pos=left,
x grid style={darkgray176},
xlabel={\(\displaystyle \bar{\mu}\)},
xmin=1.35, xmax=4.65,
xtick style={color=black},
y grid style={darkgray176},
ylabel={\(\displaystyle \delta(\kappa)\)},
ymin=0.108521173674606, ymax=0.309223056704886,
ytick style={color=black}
]
\addplot [line width=0.28pt, midnightblue263565, forget plot]
table {%
1.5 0.117643986539619
2.5 0.171629648882298
3.5 0.236111504471734
4.5 0.29479308949346
};
\addplot [semithick, midnightblue263565, dotted]
table {%
1.5 0.12044
2.5 0.174697777777778
3.5 0.239097777777778
4.5 0.297446666666667
};
\addlegendentry{$\sigma = 0.5$}
\addplot [line width=0.28pt, midnightblue263565, forget plot]
table {%
1.5 0.123236013460381
2.5 0.177765906673257
3.5 0.242084051083822
4.5 0.300100243839873
};
\addplot [line width=0.28pt, darkslategray279768, forget plot]
table {%
1.5 0.131035252294776
2.5 0.17405747123054
3.5 0.233823629006046
4.5 0.277611879895787
};
\addplot [semithick, darkslategray279768, dash pattern=on 1pt off 10pt]
table {%
1.5 0.13394
2.5 0.177097777777778
3.5 0.236775555555555
4.5 0.280397777777778
};
\addlegendentry{$\sigma = 1.0$}
\addplot [line width=0.28pt, darkslategray279768, forget plot]
table {%
1.5 0.136844747705224
2.5 0.180138084325016
3.5 0.239727482105065
4.5 0.283183675659769
};
\addplot [line width=0.28pt, olivedrab10412248, forget plot]
table {%
1.5 0.147278626791016
2.5 0.182940491174928
3.5 0.220592541811706
4.5 0.25796567570746
};
\addplot [semithick, olivedrab10412248, dashed]
table {%
1.5 0.150224444444444
2.5 0.185951111111111
3.5 0.223611111111111
4.5 0.260815555555556
};
\addlegendentry{$\sigma = 1.5$}
\addplot [line width=0.28pt, olivedrab10412248, forget plot]
table {%
1.5 0.153170262097872
2.5 0.188961731047294
3.5 0.226629680410516
4.5 0.263665435403651
};
\addplot [very thin, black]
table {%
1.35 0.203114987248141
4.65 0.203114987248141
};
\addlegendentry{Uniform}
\addplot [very thin, black, forget plot]
table {%
1.35 0.20602
4.65 0.20602
};
\addplot [very thin, black, forget plot]
table {%
1.35 0.208925012751859
4.65 0.208925012751859
};
\end{axis}

\end{tikzpicture}\label{fig:dis_het_kappa}}
    \caption{Results of experiment with agent stubbornness drawn from a Gaussian distribution with mean depicted on the horizontal axis and standard deviation shown in the legend.}  
\end{figure}

The value of $\kappa$ plays an important role in determining how sensitive the agents are to the opinions held by their neighbours. We conducted simulation runs in which the stubbornness of each agent was drawn from a Gaussian distribution centered on $\mu\in\{1.5, 2.5, 3.5, 4.5\}$. The standard deviation  of this distribution was varied taking values $\sigma\in \{0.5, 1.0, 1.5\}$. In fact we use the truncated Gaussian distribution on the support $\{0.5, 5.5\}$. We present the results from a set of simulation runs in which the stubbornness parameter for each agent was taken from the uniform distribution $\mathcal{U}_{[0.5, 5.5]}$. The resulting probability of consensus is depicted in Figure~\ref{fig:C_het_kappa} and the proportion of discordance is depicted in Figure~\ref{fig:dis_het_kappa}.

In Figure~\ref{fig:C_het_kappa} and Figure~\ref{fig:dis_het_kappa} we see that simulations with greater $\sigma$ behave more like the simulation in which the stubbornness is uniformly distributed than the other simulation runs. This means that when $\mu$ is relatively low, there is more discordance and less consensus for these runs compared to runs with lower $\sigma.$  The opposite holds for greater $\mu$. In words: A population which has, in general, a lower stubbornness, greater variability between agents helps diversify opinions. Conversely, in population with, in general, a greater stubbornness (stronger sense of individuality), greater variability between agents hinders diversity of opinions. This showcases a subtle (though possibly expected) paradox: In populations which are in general individualistic (greater $\mu$), a greater diversity of agent characteristics (greater $\sigma$) results in lower diversity of agent opinion.

\subsection{Opinions with different reliability}
This experiment bears the flavour of models of learning in populations. The agents in question are given a homogeneous $\kappa$ but the true reliability of the opinions are set unequal: $\theta_0>\theta_1$. Specifically we simulated the pairs $(\theta_0,\theta_1)\in\{(0.65,0.60),(0.70,0.60),(0.75, 0.60)\}$ which showcase a growing difference between the more reliable opinion and the other. We also experiment on the effect of a constant difference in the reliability between  two opinion's reliabilities by the pairs $(\theta_0,\theta_1)\in\{(0.65,0.60),(0.70,0.65),(0.75, 0.70)\}$, which highlight the effect of a greater general reliability while keeping the nominal difference between the two opinion's reliability constant.

\subsubsection{Growing difference}
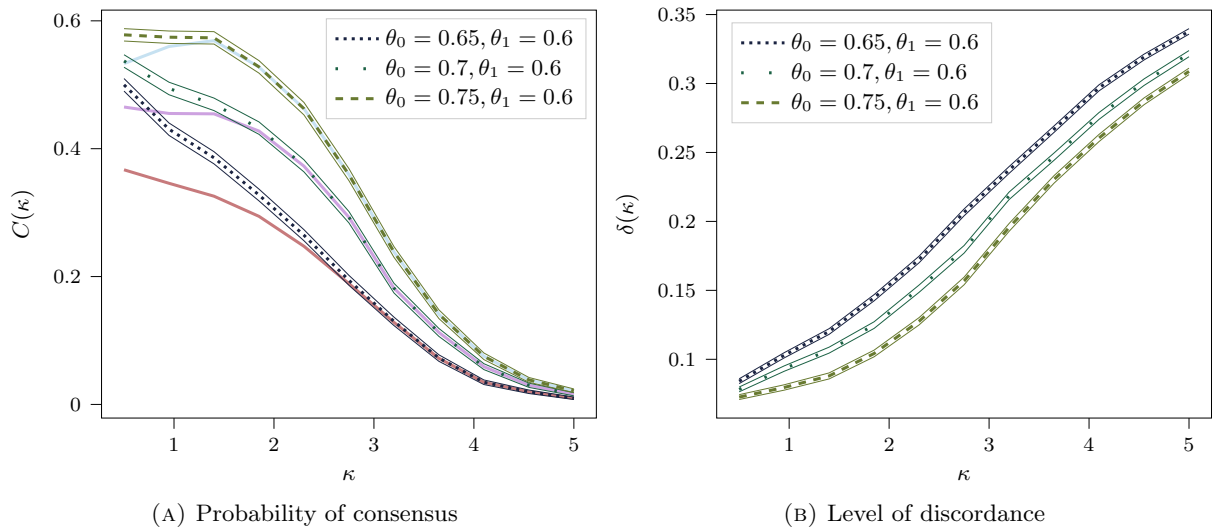
\begin{figure}[h]
    \centering
    \subfloat[Probability of consensus]{
\begin{tikzpicture}[scale = 0.95]

\definecolor{darkgray176}{RGB}{176,176,176}
\definecolor{darkslategray279768}{RGB}{27,97,68}
\definecolor{lightgray204}{RGB}{204,204,204}
\definecolor{midnightblue263565}{RGB}{26,35,65}
\definecolor{olivedrab10412248}{RGB}{104,122,48}
\definecolor{plum205162224}{RGB}{205,162,224}
\definecolor{powderblue198225241}{RGB}{198,225,241}
\definecolor{rosybrown199122124}{RGB}{199,122,124}

\begin{axis}[
legend cell align={left},
legend style={fill opacity=0.8, draw opacity=1, text opacity=1, draw=lightgray204},
tick align=outside,
tick pos=left,
x grid style={darkgray176},
xlabel={\(\displaystyle \kappa\)},
xmin=0.275, xmax=5.225,
xtick style={color=black},
y grid style={darkgray176},
ylabel={\(\displaystyle C(\kappa)\)},
ymin=-0.0213157032685023, ymax=0.61678425284709,
ytick style={color=black}
]
\addplot [very thick, rosybrown199122124, forget plot]
table {%
0.5 0.366999983787537
0.950000047683716 0.345900058746338
1.39999997615814 0.325700044631958
1.85000002384186 0.29419994354248
2.29999995231628 0.247400045394897
2.75 0.189000010490417
3.20000004768372 0.128399968147278
3.65000009536743 0.072100043296814
4.09999990463257 0.0346000194549561
4.55000019073486 0.0197999477386475
5 0.00960004329681396
};
\addplot [very thick, plum205162224, forget plot]
table {%
0.5 0.465199947357178
0.950000047683716 0.455100059509277
1.39999997615814 0.454499959945679
1.85000002384186 0.427600026130676
2.29999995231628 0.372699975967407
2.75 0.292700052261353
3.20000004768372 0.181900024414062
3.65000009536743 0.112599968910217
4.09999990463257 0.0591000318527222
4.55000019073486 0.0299999713897705
5 0.0159000158309937
};
\addplot [very thick, powderblue198225241, forget plot]
table {%
0.5 0.533100008964539
0.950000047683716 0.559900045394897
1.39999997615814 0.569000005722046
1.85000002384186 0.527699947357178
2.29999995231628 0.462599992752075
2.75 0.359400033950806
3.20000004768372 0.238600015640259
3.65000009536743 0.14110004901886
4.09999990463257 0.0752999782562256
4.55000019073486 0.0384000539779663
5 0.021399974822998
};
\addplot [line width=0.28pt, midnightblue263565, forget plot]
table {%
0.5 0.490100026130676
0.950000047683716 0.420795202255249
1.39999997615814 0.376259088516235
1.85000002384186 0.318400979042053
2.29999995231628 0.25605297088623
2.75 0.186348080635071
3.20000004768372 0.123408436775208
3.65000009536743 0.0674173831939697
4.09999990463257 0.0311127901077271
4.55000019073486 0.0170694589614868
5 0.00768887996673584
};
\addplot [very thick, midnightblue263565, dotted]
table {%
0.5 0.499899983406067
0.950000047683716 0.430500030517578
1.39999997615814 0.385800004005432
1.85000002384186 0.327600002288818
2.29999995231628 0.264699935913086
2.75 0.194100022315979
3.20000004768372 0.129999995231628
3.65000009536743 0.0724999904632568
4.09999990463257 0.0347000360488892
4.55000019073486 0.0197999477386475
5 0.00960004329681396
};
\addlegendentry{$\theta_0 = 0.65, \theta_1 = 0.6$}
\addplot [line width=0.28pt, midnightblue263565, forget plot]
table {%
0.5 0.509700059890747
0.950000047683716 0.440204858779907
1.39999997615814 0.395340919494629
1.85000002384186 0.336799025535583
2.29999995231628 0.273347020149231
2.75 0.201851963996887
3.20000004768372 0.136591553688049
3.65000009536743 0.0775825977325439
4.09999990463257 0.0382871627807617
4.55000019073486 0.0225305557250977
5 0.0115112066268921
};
\addplot [line width=0.28pt, darkslategray279768, forget plot]
table {%
0.5 0.527427196502686
0.950000047683716 0.484400629997253
1.39999997615814 0.459618330001831
1.85000002384186 0.422490477561951
2.29999995231628 0.363720417022705
2.75 0.283881068229675
3.20000004768372 0.174437403678894
3.65000009536743 0.106404423713684
4.09999990463257 0.0544780492782593
4.55000019073486 0.0266565084457397
5 0.0134482383728027
};
\addplot [very thick, darkslategray279768, dash pattern=on 1pt off 10pt]
table {%
0.5 0.537199974060059
0.950000047683716 0.494199991226196
1.39999997615814 0.469399929046631
1.85000002384186 0.432199954986572
2.29999995231628 0.373199939727783
2.75 0.292799949645996
3.20000004768372 0.182000041007996
3.65000009536743 0.112599968910217
4.09999990463257 0.0591000318527222
4.55000019073486 0.0299999713897705
5 0.0159000158309937
};
\addlegendentry{$\theta_0 = 0.7, \theta_1 = 0.6$}
\addplot [line width=0.28pt, darkslategray279768, forget plot]
table {%
0.5 0.546972751617432
0.950000047683716 0.503999352455139
1.39999997615814 0.47918164730072
1.85000002384186 0.441909551620483
2.29999995231628 0.38267970085144
2.75 0.301718950271606
3.20000004768372 0.189562559127808
3.65000009536743 0.11879563331604
4.09999990463257 0.0637218952178955
4.55000019073486 0.0333435535430908
5 0.0183517932891846
};
\addplot [line width=0.28pt, olivedrab10412248, forget plot]
table {%
0.5 0.56842029094696
0.950000047683716 0.564709186553955
1.39999997615814 0.563706159591675
1.85000002384186 0.518215417861938
2.29999995231628 0.452827453613281
2.75 0.349995374679565
3.20000004768372 0.23024594783783
3.65000009536743 0.134276747703552
4.09999990463257 0.0701280832290649
4.55000019073486 0.0346336364746094
5 0.0185636281967163
};
\addplot [very thick, olivedrab10412248, dashed]
table {%
0.5 0.578099966049194
0.950000047683716 0.574399948120117
1.39999997615814 0.573400020599365
1.85000002384186 0.527999997138977
2.29999995231628 0.462599992752075
2.75 0.359400033950806
3.20000004768372 0.238600015640259
3.65000009536743 0.14110004901886
4.09999990463257 0.0752999782562256
4.55000019073486 0.0384000539779663
5 0.021399974822998
};
\addlegendentry{$\theta_0 = 0.75, \theta_1 = 0.6$}
\addplot [line width=0.28pt, olivedrab10412248, forget plot]
table {%
0.5 0.587779760360718
0.950000047683716 0.584090948104858
1.39999997615814 0.583093881607056
1.85000002384186 0.537784576416016
2.29999995231628 0.472372531890869
2.75 0.368804574012756
3.20000004768372 0.246954083442688
3.65000009536743 0.147923231124878
4.09999990463257 0.0804719924926758
4.55000019073486 0.0421663522720337
5 0.0242364406585693
};
\end{axis}

\end{tikzpicture}\label{fig:C_Grow_dif}}
    \subfloat[Level of discordance]{
\begin{tikzpicture}[scale = 0.95]

\definecolor{darkgray176}{RGB}{176,176,176}
\definecolor{darkslategray279768}{RGB}{27,97,68}
\definecolor{lightgray204}{RGB}{204,204,204}
\definecolor{midnightblue263565}{RGB}{26,35,65}
\definecolor{olivedrab10412248}{RGB}{104,122,48}

\begin{axis}[
legend cell align={left},
legend style={
  fill opacity=0.8,
  draw opacity=1,
  text opacity=1,
  at={(0.03,0.97)},
  anchor=north west,
  draw=lightgray204
},
tick align=outside,
tick pos=left,
x grid style={darkgray176},
xlabel={\(\displaystyle \kappa\)},
xmin=0.275, xmax=5.225,
xtick style={color=black},
y grid style={darkgray176},
ylabel={\(\displaystyle \delta(\kappa)\)},
ymin=0.0573924233032778, ymax=0.353231365552374,
ytick style={color=black}
]
\addplot [line width=0.28pt, midnightblue263565, forget plot]
table {%
0.5 0.0823736190795898
0.950000047683716 0.100993275642395
1.39999997615814 0.118106603622437
1.85000002384186 0.142402172088623
2.29999995231628 0.170010685920715
2.75 0.204047679901123
3.20000004768372 0.234823107719421
3.65000009536743 0.264567017555237
4.09999990463257 0.294670581817627
4.55000019073486 0.317070603370667
5 0.33580470085144
};
\addplot [very thick, midnightblue263565, dotted]
table {%
0.5 0.0841177701950073
0.950000047683716 0.102888941764832
1.39999997615814 0.120168924331665
1.85000002384186 0.144652247428894
2.29999995231628 0.172407746315002
2.75 0.206525564193726
3.20000004768372 0.237286686897278
3.65000009536743 0.26690673828125
4.09999990463257 0.296880006790161
4.55000019073486 0.319175481796265
5 0.337794423103333
};
\addlegendentry{$\theta_0 = 0.65, \theta_1 = 0.6$}
\addplot [line width=0.28pt, midnightblue263565, forget plot]
table {%
0.5 0.0858619213104248
0.950000047683716 0.104784607887268
1.39999997615814 0.122231125831604
1.85000002384186 0.146902322769165
2.29999995231628 0.17480480670929
2.75 0.209003448486328
3.20000004768372 0.239750146865845
3.65000009536743 0.269246339797974
4.09999990463257 0.299089312553406
4.55000019073486 0.321280479431152
5 0.339784145355225
};
\addplot [line width=0.28pt, darkslategray279768, forget plot]
table {%
0.5 0.0764086246490479
0.950000047683716 0.0913077592849731
1.39999997615814 0.104492425918579
1.85000002384186 0.122407674789429
2.29999995231628 0.148590683937073
2.75 0.177001476287842
3.20000004768372 0.215823650360107
3.65000009536743 0.244333624839783
4.09999990463257 0.274001598358154
4.55000019073486 0.298581123352051
5 0.319422960281372
};
\addplot [very thick, darkslategray279768, dash pattern=on 1pt off 10pt]
table {%
0.5 0.0781500339508057
0.950000047683716 0.0932555198669434
1.39999997615814 0.106655597686768
1.85000002384186 0.124812245368958
2.29999995231628 0.151229977607727
2.75 0.179748892784119
3.20000004768372 0.218536615371704
3.65000009536743 0.246977806091309
4.09999990463257 0.276484489440918
4.55000019073486 0.30095112323761
5 0.321654438972473
};
\addlegendentry{$\theta_0 = 0.7, \theta_1 = 0.6$}
\addplot [line width=0.28pt, darkslategray279768, forget plot]
table {%
0.5 0.0798914432525635
0.950000047683716 0.0952033996582031
1.39999997615814 0.108818650245667
1.85000002384186 0.127216815948486
2.29999995231628 0.153869390487671
2.75 0.182496309280396
3.20000004768372 0.22124969959259
3.65000009536743 0.249621868133545
4.09999990463257 0.278967380523682
4.55000019073486 0.303321123123169
5 0.323885917663574
};
\addplot [line width=0.28pt, olivedrab10412248, forget plot]
table {%
0.5 0.0708396434783936
0.950000047683716 0.0776189565658569
1.39999997615814 0.0857714414596558
1.85000002384186 0.101816773414612
2.29999995231628 0.124949812889099
2.75 0.15403950214386
3.20000004768372 0.192779183387756
3.65000009536743 0.227741241455078
4.09999990463257 0.257893919944763
4.55000019073486 0.284622669219971
5 0.306456446647644
};
\addplot [very thick, olivedrab10412248, dashed]
table {%
0.5 0.0725977420806885
0.950000047683716 0.0795677900314331
1.39999997615814 0.0879533290863037
1.85000002384186 0.104233384132385
2.29999995231628 0.127599954605103
2.75 0.156807780265808
3.20000004768372 0.195615530014038
3.65000009536743 0.230477809906006
4.09999990463257 0.260495543479919
4.55000019073486 0.287083387374878
5 0.308794498443604
};
\addlegendentry{$\theta_0 = 0.75, \theta_1 = 0.6$}
\addplot [line width=0.28pt, olivedrab10412248, forget plot]
table {%
0.5 0.0743559598922729
0.950000047683716 0.0815166234970093
1.39999997615814 0.0901352167129517
1.85000002384186 0.106649994850159
2.29999995231628 0.130250096321106
2.75 0.159576058387756
3.20000004768372 0.19845187664032
3.65000009536743 0.233214378356934
4.09999990463257 0.263097167015076
4.55000019073486 0.289544105529785
5 0.311132431030273
};
\end{axis}

\end{tikzpicture}\label{fig:Disc_Grow_dif}}
    \caption{The Results of the experiment in which the difference between $\theta_0$ and $\theta_1$ is growing. Additionally plotted is the probability of consensus on opinion $0$ keeping in mind that $\theta_0>\theta_1$.}
    \end{figure}

We depict the probability of consensus in the growing difference experiment in Figure~\ref{fig:C_Grow_dif}. The corresponding portion of discordance is depicted in Figure~\ref{fig:Disc_Grow_dif}. In Figure~\ref{fig:C_Grow_dif} we see that a greater difference in the opinion's reliability fosters a greater probability of consensus. We also note that as the stubbornness of the population grows, the less likely it becomes that the population reaches consensus on the `inferior' opinion. This becomes so extreme that if $\kappa$ is great enough, if there is consensus in the population, this is on the opinion with the greater reliability. Similarly in Figure~\ref{fig:Disc_Grow_dif} we see that a greater difference in reliability implies a lower portion of discordance in expectation. 

\subsubsection{Constant difference}
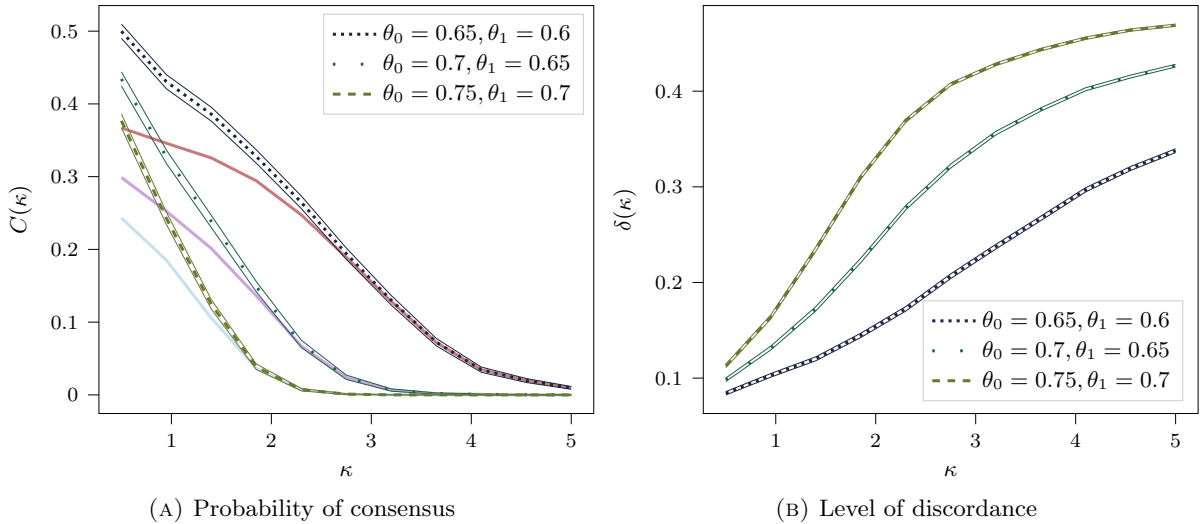
\begin{figure}
    \centering
    \subfloat[Probability of consensus]{
\begin{tikzpicture}[scale = 0.95]

\definecolor{darkgray176}{RGB}{176,176,176}
\definecolor{darkslategray279768}{RGB}{27,97,68}
\definecolor{lightgray204}{RGB}{204,204,204}
\definecolor{midnightblue263565}{RGB}{26,35,65}
\definecolor{olivedrab10412248}{RGB}{104,122,48}
\definecolor{plum205162224}{RGB}{205,162,224}
\definecolor{powderblue198225241}{RGB}{198,225,241}
\definecolor{rosybrown199122124}{RGB}{199,122,124}

\begin{axis}[
legend cell align={left},
legend style={fill opacity=0.8, draw opacity=1, text opacity=1, draw=lightgray204},
tick align=outside,
tick pos=left,
x grid style={darkgray176},
xlabel={\(\displaystyle \kappa\)},
xmin=0.275, xmax=5.225,
xtick style={color=black},
y grid style={darkgray176},
ylabel={\(\displaystyle C(\kappa)\)},
ymin=-0.0255857896999427, ymax=0.535189799304188,
ytick style={color=black}
]
\addplot [very thick, rosybrown199122124, forget plot]
table {%
0.5 0.366999983787537
0.950000047683716 0.345900058746338
1.39999997615814 0.325700044631958
1.85000002384186 0.29419994354248
2.29999995231628 0.247400045394897
2.75 0.189000010490417
3.20000004768372 0.128399968147278
3.65000009536743 0.072100043296814
4.09999990463257 0.0346000194549561
4.55000019073486 0.0197999477386475
5 0.00960004329681396
};
\addplot [very thick, plum205162224, forget plot]
table {%
0.5 0.298799991607666
0.950000047683716 0.252500057220459
1.39999997615814 0.201099991798401
1.85000002384186 0.136199951171875
2.29999995231628 0.0671999454498291
2.75 0.023900032043457
3.20000004768372 0.00670003890991211
3.65000009536743 0.00179994106292725
4.09999990463257 0.000699996948242188
4.55000019073486 0.000200033187866211
5 0.000100016593933105
};
\addplot [very thick, powderblue198225241, forget plot]
table {%
0.5 0.243100047111511
0.950000047683716 0.185199975967407
1.39999997615814 0.105399966239929
1.85000002384186 0.0364999771118164
2.29999995231628 0.00709998607635498
2.75 0.0010000467300415
3.20000004768372 0.000100016593933105
5 0
};
\addplot [line width=0.28pt, midnightblue263565, forget plot]
table {%
0.5 0.490100026130676
0.950000047683716 0.420795202255249
1.39999997615814 0.376259088516235
1.85000002384186 0.318400979042053
2.29999995231628 0.25605297088623
2.75 0.186348080635071
3.20000004768372 0.123408436775208
3.65000009536743 0.0674173831939697
4.09999990463257 0.0311127901077271
4.55000019073486 0.0170694589614868
5 0.00768887996673584
};
\addplot [very thick, midnightblue263565, dotted]
table {%
0.5 0.499899983406067
0.950000047683716 0.430500030517578
1.39999997615814 0.385800004005432
1.85000002384186 0.327600002288818
2.29999995231628 0.264699935913086
2.75 0.194100022315979
3.20000004768372 0.129999995231628
3.65000009536743 0.0724999904632568
4.09999990463257 0.0347000360488892
4.55000019073486 0.0197999477386475
5 0.00960004329681396
};
\addlegendentry{$\theta_0 = 0.65, \theta_1 = 0.6$}
\addplot [line width=0.28pt, midnightblue263565, forget plot]
table {%
0.5 0.509700059890747
0.950000047683716 0.440204858779907
1.39999997615814 0.395340919494629
1.85000002384186 0.336799025535583
2.29999995231628 0.273347020149231
2.75 0.201851963996887
3.20000004768372 0.136591553688049
3.65000009536743 0.0775825977325439
4.09999990463257 0.0382871627807617
4.55000019073486 0.0225305557250977
5 0.0115112066268921
};
\addplot [line width=0.28pt, darkslategray279768, forget plot]
table {%
0.5 0.424584984779358
0.950000047683716 0.318103075027466
1.85000002384186 0.141628384590149
2.29999995231628 0.0650957822799683
2.75 0.0211880207061768
3.20000004768372 0.00510108470916748
3.65000009536743 0.000969171524047852
4.09999990463257 0.00018155574798584
5 -9.59634780883789e-05
};
\addplot [very thick, darkslategray279768, dash pattern=on 1pt off 10pt]
table {%
0.5 0.434299945831299
0.950000047683716 0.327300071716309
1.39999997615814 0.238299965858459
1.85000002384186 0.148599982261658
2.29999995231628 0.070099949836731
2.75 0.0241999626159668
3.20000004768372 0.00670003890991211
3.65000009536743 0.00179994106292725
4.09999990463257 0.000699996948242188
4.55000019073486 0.000200033187866211
5 0.000100016593933105
};
\addlegendentry{$\theta_0 = 0.7, \theta_1 = 0.65$}
\addplot [line width=0.28pt, darkslategray279768, forget plot]
table {%
0.5 0.444015026092529
0.950000047683716 0.336496829986572
1.39999997615814 0.246650457382202
1.85000002384186 0.155571579933167
2.29999995231628 0.0751042366027832
2.75 0.0272119045257568
3.20000004768372 0.00829899311065674
3.65000009536743 0.00263082981109619
4.09999990463257 0.00121843814849854
4.55000019073486 0.000477194786071777
5 0.00029599666595459
};
\addplot [line width=0.28pt, olivedrab10412248, forget plot]
table {%
0.5 0.367302179336548
0.950000047683716 0.234297156333923
1.39999997615814 0.117833495140076
1.85000002384186 0.0348242521286011
2.29999995231628 0.00554287433624268
2.75 0.000380516052246094
3.20000004768372 -9.59634780883789e-05
5 0
};
\addplot [very thick, olivedrab10412248, dashed]
table {%
0.5 0.376800060272217
0.950000047683716 0.242699980735779
1.39999997615814 0.124300003051758
1.85000002384186 0.038599967956543
2.29999995231628 0.00720000267028809
2.75 0.0010000467300415
3.20000004768372 0.000100016593933105
5 0
};
\addlegendentry{$\theta_0 = 0.75, \theta_1 = 0.7$}
\addplot [line width=0.28pt, olivedrab10412248, forget plot]
table {%
0.5 0.386297821998596
0.950000047683716 0.251102805137634
1.39999997615814 0.13076651096344
1.85000002384186 0.0423756837844849
2.29999995231628 0.0088571310043335
2.75 0.00161945819854736
3.20000004768372 0.00029599666595459
4.09999990463257 0
5 0
};
\end{axis}

\end{tikzpicture}\label{fig:C_Const_dif}}
    \subfloat[Level of discordance]{
\begin{tikzpicture}[scale = 0.95]

\definecolor{darkgray176}{RGB}{176,176,176}
\definecolor{darkslategray279768}{RGB}{27,97,68}
\definecolor{lightgray204}{RGB}{204,204,204}
\definecolor{midnightblue263565}{RGB}{26,35,65}
\definecolor{olivedrab10412248}{RGB}{104,122,48}

\begin{axis}[
legend cell align={left},
legend style={
  fill opacity=0.8,
  draw opacity=1,
  text opacity=1,
  at={(0.97,0.03)},
  anchor=south east,
  draw=lightgray204
},
tick align=outside,
tick pos=left,
x grid style={darkgray176},
xlabel={\(\displaystyle \kappa\)},
xmin=0.275, xmax=5.225,
xtick style={color=black},
y grid style={darkgray176},
ylabel={\(\displaystyle \delta(\kappa)\)},
ymin=0.0629851048967294, ymax=0.489533474334249,
ytick style={color=black}
]
\addplot [line width=0.28pt, midnightblue263565, forget plot]
table {%
0.5 0.0823736190795898
0.950000047683716 0.100993275642395
1.39999997615814 0.118106603622437
1.85000002384186 0.142402172088623
2.29999995231628 0.170010685920715
2.75 0.204047679901123
3.20000004768372 0.234823107719421
3.65000009536743 0.264567017555237
4.09999990463257 0.294670581817627
4.55000019073486 0.317070603370667
5 0.33580470085144
};
\addplot [very thick, midnightblue263565, dotted]
table {%
0.5 0.0841177701950073
0.950000047683716 0.102888941764832
1.39999997615814 0.120168924331665
1.85000002384186 0.144652247428894
2.29999995231628 0.172407746315002
2.75 0.206525564193726
3.20000004768372 0.237286686897278
3.65000009536743 0.26690673828125
4.09999990463257 0.296880006790161
4.55000019073486 0.319175481796265
5 0.337794423103333
};
\addlegendentry{$\theta_0 = 0.65, \theta_1 = 0.6$}
\addplot [line width=0.28pt, midnightblue263565, forget plot]
table {%
0.5 0.0858619213104248
0.950000047683716 0.104784607887268
1.39999997615814 0.122231125831604
1.85000002384186 0.146902322769165
2.29999995231628 0.17480480670929
2.75 0.209003448486328
3.20000004768372 0.239750146865845
3.65000009536743 0.269246339797974
4.09999990463257 0.299089312553406
4.55000019073486 0.321280479431152
5 0.339784145355225
};
\addplot [line width=0.28pt, darkslategray279768, forget plot]
table {%
0.5 0.0964388847351074
0.950000047683716 0.129387378692627
1.39999997615814 0.1701740026474
1.85000002384186 0.220637440681458
2.29999995231628 0.275818824768066
2.75 0.319612264633179
3.20000004768372 0.354159116744995
3.65000009536743 0.379104137420654
4.09999990463257 0.400390863418579
4.55000019073486 0.413817882537842
5 0.425257325172424
};
\addplot [very thick, darkslategray279768, dash pattern=on 1pt off 10pt]
table {%
0.5 0.0982433557510376
0.950000047683716 0.131401062011719
1.39999997615814 0.172427773475647
1.85000002384186 0.223012208938599
2.29999995231628 0.278110027313232
2.75 0.321699976921082
3.20000004768372 0.356035590171814
3.65000009536743 0.380831122398376
4.09999990463257 0.401980042457581
4.55000019073486 0.415316700935364
5 0.426672220230103
};
\addlegendentry{$\theta_0 = 0.7, \theta_1 = 0.65$}
\addplot [line width=0.28pt, darkslategray279768, forget plot]
table {%
0.5 0.100047826766968
0.950000047683716 0.133414745330811
1.39999997615814 0.174681544303894
1.85000002384186 0.22538697719574
2.29999995231628 0.280401229858398
2.75 0.323787689208984
3.20000004768372 0.357911944389343
3.65000009536743 0.382558107376099
4.09999990463257 0.403569221496582
4.55000019073486 0.416815519332886
5 0.428087115287781
};
\addplot [line width=0.28pt, olivedrab10412248, forget plot]
table {%
0.5 0.110812783241272
0.950000047683716 0.161749720573425
1.39999997615814 0.232427716255188
1.85000002384186 0.307922005653381
2.29999995231628 0.367387294769287
2.75 0.405671119689941
3.20000004768372 0.426586508750916
3.65000009536743 0.442047238349915
4.09999990463257 0.454109072685242
4.55000019073486 0.462722778320312
5 0.467737317085266
};
\addplot [very thick, olivedrab10412248, dashed]
table {%
0.5 0.112684488296509
0.950000047683716 0.163910031318665
1.39999997615814 0.234750032424927
1.85000002384186 0.310073375701904
2.29999995231628 0.369222164154053
2.75 0.407265543937683
3.20000004768372 0.428047776222229
3.65000009536743 0.443403363227844
4.09999990463257 0.455393314361572
4.55000019073486 0.463932275772095
5 0.46894109249115
};
\addlegendentry{$\theta_0 = 0.75, \theta_1 = 0.7$}
\addplot [line width=0.28pt, olivedrab10412248, forget plot]
table {%
0.5 0.114556193351746
0.950000047683716 0.166070222854614
1.39999997615814 0.237072229385376
1.85000002384186 0.312224626541138
2.29999995231628 0.371057152748108
2.75 0.408859968185425
3.20000004768372 0.429509043693542
3.65000009536743 0.444759368896484
4.09999990463257 0.456677556037903
4.55000019073486 0.465141773223877
5 0.470144867897034
};
\end{axis}

\end{tikzpicture}\label{fig:Disc_Const_dif}}
    \caption{Results of the experiment in which the difference between $\theta_0$ and $\theta_1$ is constant. Additionally plotted is the probability of consensus on opinion $0$ keeping in mind that $\theta_0>\theta_1$.}
\end{figure}

We plot the probability of consensus in the sub-experiment with constant difference between $\theta_0$ and $\theta_1$ in Figure~\ref{fig:C_Const_dif}. We plot the corresponding portion of discordance in Figure~\ref{fig:Disc_Const_dif}. We see a similar trend in Figure~\ref{fig:C_Const_dif} to the one present in the sensitivity analysis: Lower reliability leads to more consensus of opinion. As in the experiment with a growing difference between opinion reliability, we see that greater stubbornness in the population leads to a greater chance of agreeing on the `better' opinion. In Figure~\ref{fig:Disc_Const_dif} we see again (as in the sensitivity analysis) that a greater reliability leads to more discordance. 

\section{Discussion}\label{sec:Discussion}
In this section we discuss the results of the experiments conducted. In doing so we also reflect on the merits of the model when interpreted as a heuristic for communication interpretation between agents. Subsequently we discuss differences between our framework and relevant literature. 
\subsection{Interpretation of the experiments}
As a result of the experiment with different $\kappa$ per agent we see that an increase in heterogeneity (greater standard deviation of the distribution from which we sample the stubbornness $\kappa$) decreases the differences which arise from shifting the mean $\mu$ of the distribution. From a modeller's perspective this may be intuitive, as a greater spread in the distribution should decrease the effect of shifting its mean. It does however also hint to the important difference between individualism and diversity. In this context: Populations with lower individualism tend toward more consensus. Furthermore, populations with diversity in the extent of its agent's individualism may increase or decrease the probability of consensus depending on the mean value of the population's individualism.

The experiment with opinions of different reliability show us that populations with greater stubbornness may be more sure that if they reach consensus, it is upon the better alternative. Furthermore, the greater the difference between two opinions, the less discordance one expects in the population. The more clear-cut the difference between two opinions, the easier it should be for the population to learn this and subsequently reach consensus on the better of the two. It should be said here that it is also true that enough stubbornness leads to general disagreement in the population. Thus balance may be important to the goal of reaching consensus on the better of two opinions. In general it seems that the agents in the model make good use of the information provided by their network: Consensus on the better opinion is more likely than on the worse opinion and this is increasingly the case the greater the difference between the two opinions. Though agents are not modelled `rationally' this outcome suggests that the heuristic method by which agents incorporate their neighbours opinions does aid the agents in making good decisions.

\subsection{Contributions and future work}
The framework we present in \S\ref{sec:solo}, and \S\ref{sec:many} addresses the current lack of models in the opinion dynamics literature which have sophisticated agents who may adjust their opinion in absence of network influence beyond the introduction of noise. Furthermore, it entails models from the same basic assumptions of assimilative forces between agents yet with a novel aspect: Opinions of alters do not affect an agent's belief of an opinion directly but rather the decision making process by which an agent chooses their opinion. This is an attempt towards modelling interacting agents as opposed to what Giardini \textit{et al.}\,\cite{Giardini2015} call interacting opinions. Furthermore, the framework for social influence we present generalises the majority rule dynamics~\cite{Galam2002} which were presented a step towards realism. As our model is a generalisation of majority rules we hope to take another such step. The framework also is computationally light despite the relatively high level of detail of the agents. This enables the modelling of agents with reasonable sophistication that is possible in general models using this framework.

The results of the framework instance and experiments we present in \S\ref{sec:ExModel}, and \S\ref{sec:Experiments} highlight that models from the framework have desirable and reasonable characteristics: An array of outcomes is possible entailing consensus, polarisation as well as fragmentation. That the model parameters adjust the probabilities of these outcomes in ways which align with what a modeller might expect is showcased in the sensitivity analysis. 

The definition of the opinions used in our framework is broad and allows for interesting future work in which agent behaviour may be coupled back to the reliability of an opinion. For example consider a population of agents who are faced with the choice of a means of transportation. The agent's belief on the reliability of the options available is likely to play a role in their decision making. Closing the feedback loop: The agent decision making (the number of people using each type) is likely to influence the reliability of the options available. The fact that our model is lightweight means that it may be straightforwardly implemented in agent-based models which investigate more than opinions but rather the interface between opinion dynamics and their effect on agent behaviour.
\bibliographystyle{ieeetr}
\bibliography{Biblio.bib}

\section*{Appendix: Watts-Strogatz network}
\begin{figure}[ht]
\centering
\subfloat[Step 1]{\includegraphics[width = 0.24\textwidth]{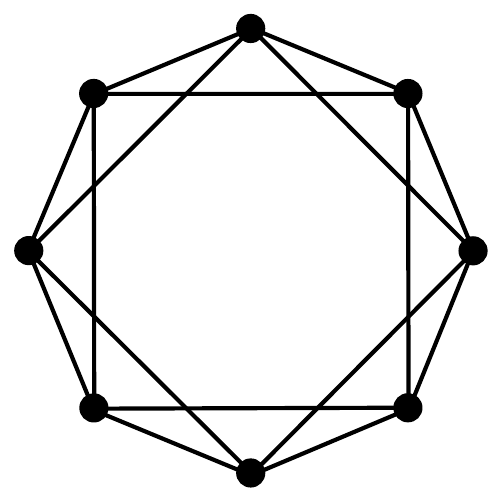}}
\subfloat[Step 2]{\includegraphics[width = 0.24\textwidth]{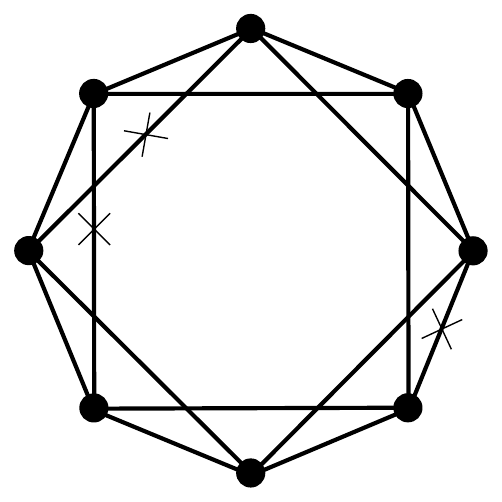}}
\subfloat[Step 2]{\includegraphics[width = 0.24\textwidth]{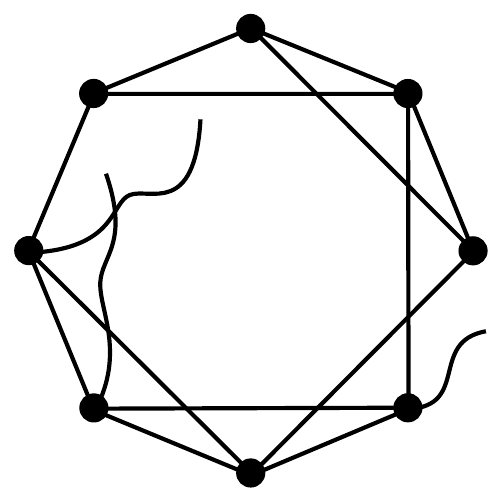}}
\subfloat[Step 3]{\includegraphics[width = 0.24\textwidth]{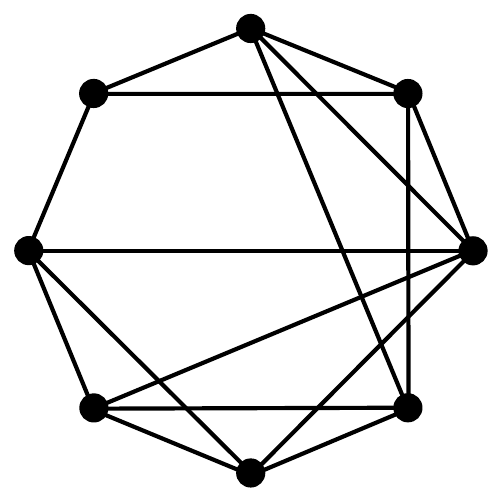}}
\caption{The steps to create a Watts-Strogatz random graph on $N=8$ agents with $l=4$ nearest neighbours.}\label{fig:WS_net}
\end{figure}

The agents in our example are embedded within a Watts-Strogatz random graph model~\cite{Watts1998}. The creation of a Watts-Strogatz random graph is illustrated in three steps. This uses, $N\in\mathbb{N}$ the number of agents in the population, $l\in 2\mathbb{N}$ the initial number of nearest neighbours to each agent, and $w\in (0,1)$, the rewiring probability. 
\begin{enumerate}
\item First, we arrange the population of $N$ agents, on a cycle graph and connected each agent to their $l$-nearest neighbours. 
\item Secondly, for each edge in the circulant created, we flip a coin which lands heads with probability $w$ and if it lands heads, we `cut' the edge off of one of its vertices.
\item Finally, each of the edges cut in this way is rewired to another vertex uniformly at random.
\end{enumerate}
This network structure has the property that average path lengths between vertices are short, yet there is still high clustering of vertices. We illustrate this process in Figure~\ref{fig:WS_net} for a network on $N=8$ agents, with $l=4$ nearest neighbours.

\end{document}